\documentclass[journal]{vgtc}                     

\onlineid{1786}

\vgtccategory{Research}

\usepackage[labelfont=sf]{subcaption}
\usepackage{makecell}
\usepackage{algorithm}
\usepackage[noend]{algpseudocode}
\algnewcommand\algorithmicforeach{\textbf{for each}}
\algdef{S}[FOR]{ForEach}[1]{\algorithmicforeach\ #1\ \algorithmicdo}
\MakeRobust{\Call} 

\newcommand{\edit}[1]{\textcolor{black}{#1}}
\newcommand{\new}[1]{\textcolor{black}{#1}}

\title{Approximate Puzzlepiece Compositing}

\author{%
  Xuan Huang, 
  Will Usher
  and 
  Valerio Pascucci
}

\authorfooter{
  \item
  	Xuan Huang and Valerio Pascucci are with the SCI Institute at the University of Utah.
  	E-mail: xuanhuang@sci.utah.edu
  \item
  	Will Usher is with Luminary Cloud.

}

\abstract{%
  \new{The increasing demand for larger and higher fidelity simulations has made}
  Adaptive Mesh Refinement (AMR) and unstructured mesh
  techniques essential to focus compute effort and memory cost on just the areas of interest in the simulation domain.
  The distribution of these meshes over the compute nodes is often determined by balancing compute, memory, and network
  costs, leading to distributions with jagged nonconvex boundaries that fit together much like puzzle pieces.
  \new{It is expensive, and sometimes impossible,}
  to re-partition the data posing a challenge
  for in situ and post hoc visualization as the data cannot be rendered
  using standard sort-last compositing techniques that require \new{a convex and disjoint data partitioning.}
  We present a new distributed volume rendering and compositing algorithm, Approximate Puzzlepiece Compositing, that enables fast and high-accuracy in-place rendering of AMR and unstructured meshes.
  Our approach builds on Moment-Based Ordered-Independent Transparency to achieve a scalable, order-independent compositing algorithm that requires little communication and does not impose requirements on the data partitioning.
  We evaluate the image quality and scalability of our approach on synthetic data and two large-scale unstructured meshes on HPC systems by comparing to state-of-the-art sort-last compositing techniques, highlighting our approach's minimal overhead at higher core counts.
  We demonstrate that Approximate Puzzlepiece Compositing provides a scalable, high-performance, and high-quality distributed rendering approach applicable to the complex data distributions encountered in large-scale CFD simulations.
}


\teaser{
  \centering
    \begin{subfigure}{0.32\linewidth}
         \centering
         \includegraphics[width=\linewidth]{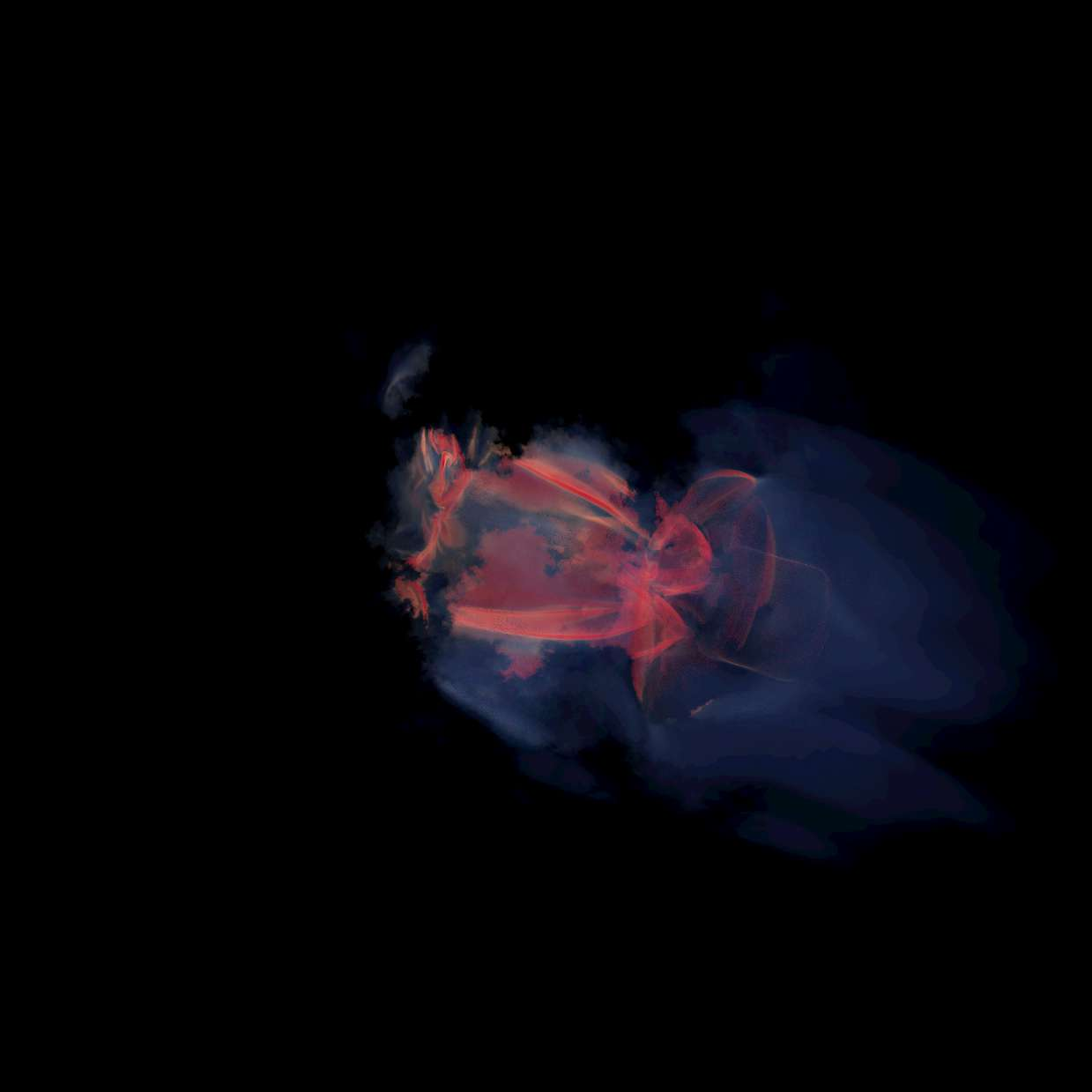}
         \caption{16 Distributed Domains (15.3s/frame)}
         \vspace{-0.5em}
    \end{subfigure}
    \begin{subfigure}{0.32\linewidth}
         \centering
         \includegraphics[width=\linewidth]{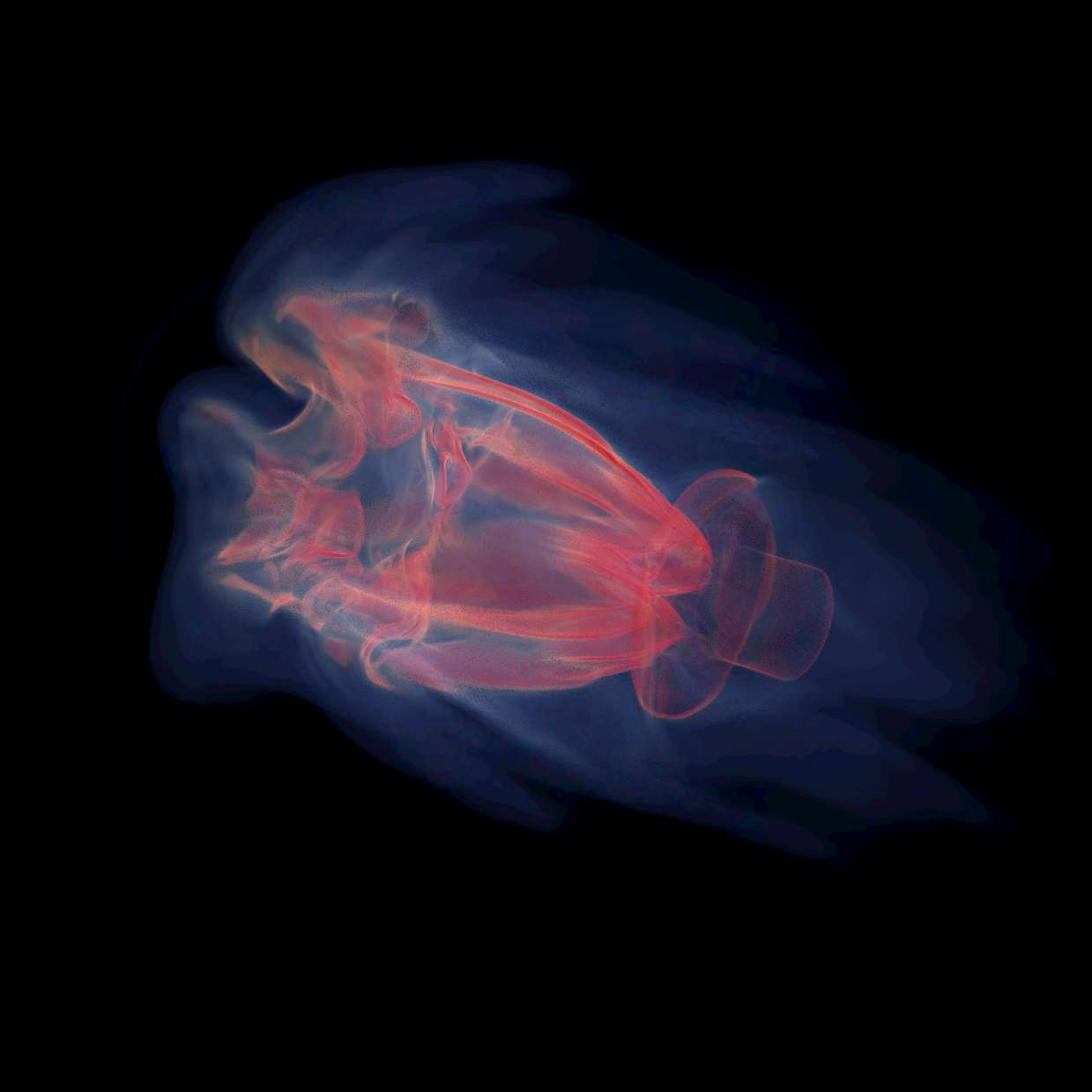}
         \caption{All 72 Distrib. Domains (15.6s/frame)}
         \vspace{-0.5em}
    \end{subfigure}
    \begin{subfigure}{0.32\linewidth}
         \centering
         \includegraphics[width=\linewidth]{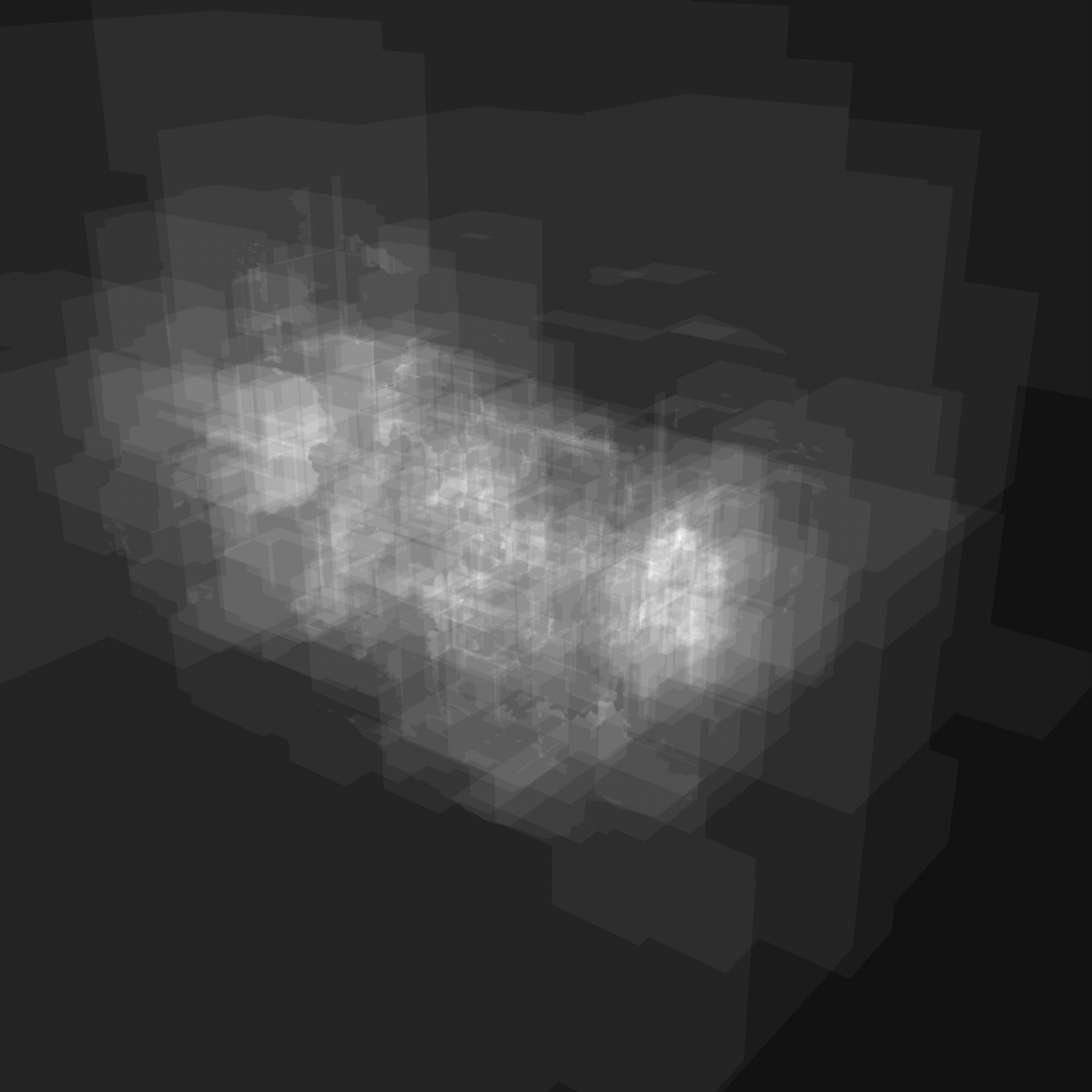}
         \caption{Segment Density Heatmap (max=28)}
         \vspace{-0.5em}
    \end{subfigure}
    \vspace{-0.5em}
    \caption{%
  	Large-scale moment-based order-independent (MBOIT) distributed transparency rendering with the FUN3D Mars Lander A/143M dataset, consisting of 72 subdomains and 798M elements. Figures (a) and (b) are rendered at 2560$\times$2560 using TACC Frontera Intel Xeon Platinum 8280 ("Cascade Lake") nodes with 192GB memory. (c) A heatmap of the per-pixel segment counts with a range of [0, 28]. The segment lists must be individually sorted and blended
   in sort-last compositing due to the overlapping boundaries of data on the ranks, resulting in large data transfers and bottlenecks.
   Our approach ensures a constant, fixed, and small amount of communication for compositing arbitrary data distributions. 
   }
   \label{fig:teaser}
}

\graphicspath{{figs/}{figures/}{pictures/}{images/}{./}} 

\usepackage{tabu}                      
\usepackage{booktabs}                  
\usepackage{lipsum}                    
\usepackage{mwe}                       

\usepackage{mathptmx}                  
\usepackage{amsfonts}
\usepackage{amsmath}
\usepackage{dblfloatfix}

\begin{document}


\firstsection{Introduction}

\maketitle



Volume visualization is a crucial part of the analysis pipeline, and is used by domain scientists to analyze their data in fields ranging from biology and medicine to engineering and geoscience.
Although continuing advances in data acquisition, simulation, and computation power provide ever more accurate
data this comes at the cost of single-node computation and storage space.
As scientists solve ever larger and more complex problems, it becomes necessary to render the data in situ
\new{because it can no longer be saved out frequently enough.}
High performance computing systems offer massive amounts of computing power for simulation and visualization;
however, scalable distributed volume rendering is non-trivial for simulations with non-convex data distributions.
\new{Such distributions are common in unstructured or octree AMR mesh simulations, e.g., using fill-reducing partitioners such
as ParMETIS~\cite{karypisParmetisParallelGraph1997} or p4est's~\cite{bursteddeP4estScalableAlgorithms2011}
Morton index partitioning.}


Although sort-last data-parallel rendering is a standard approach for distributed volume rendering,
it requires that the data partition on each node be convex and disjoint to ensure each node
\new{can produce a single depth-sortable partial image for compositing.}
\new{This restriction is typically not satisfied in distributed simulations on
unstructured~\cite{anderson2023fun3d,palaciosStanfordUniversityUnstructured2013,economonSU2OpenSourceSuite2016,offermansAdaptiveMeshRefinement2020,anderson2023fun3d}
or octree AMR~\cite{arndtDealIIFiniteElement2021,bursteddeP4estScalableAlgorithms2011} grids.
Elements in such simulations are distributed to optimize compute and networking costs, e.g., through
fill reducting orderings~\cite{karypisParmetisParallelGraph1997},
resulting in data partitions with jagged boundaries where the bounding box of each rank's partition
overlaps somewhat those of its neighbors (\Cref{fig:example_umesh_boundaries}).}
\new{The partial images from each rank can no longer be sorted since the jagged boundaries cause them to overlap in depth along the view axis.}
\new{Although sort-last compositing could be extended}
to produce and blend individual fragments for each nonoverlapping segment
of the volume on each rank, such an approach would be prohibitively expensive
in both bandwidth and computation costs.


We propose a novel, scalable compositing approach that works on arbitrary data distributions, does not require
sorting, and minimizes \new{network} communication to two \new{parallel} reductions. 
We achieve this by adapting an order-independent transparency technique, Moment-Based Order-Independent Transparency~\cite{munstermann2018moment},
to data-parallel compositing.
Our final compositing pipeline consists of two stages, a moments generation phase, where communication consists
of a single all-reduce add, and a final gather stage where the final image is produced through a reduce add to the display rank.
We describe our proposed Approximate Puzzlepiece Compositing (APC) pipeline and evaluate its performance on
the NASA FUN3D Mars Lander datasets~\cite{anderson2023fun3d}, the NASA Exajet dataset~\cite{Moran_2023},  
and a \new{worst-case} synthetic scaling test case on the TACC Frontera HPC system.
We compare our proposed method to sort-last compositing on segments as the state of the art and
demonstrate that APC is fast, applicable to any large-scale volumetric data,
and produces accurate images.
Our contributions are:
\begin{itemize}
    \item A highly scalable two-stage distributed volume rendering method that imposes no restrictions on the data distribution;
    \item Evaluation of the image quality of Moment-Based Order-Independent Transparency for compositing distributed volumetric data; and
    \item A thorough performance study on a \new{worst-case} synthetic test volume and two real-world large-scale datasets,
    the FUN3D and the Exajet, demonstrating our method's \new{minimal communication costs} and high scalability.
\end{itemize}

\section{Related Work}
\edit{We review related work in distributed volume rendering and order-independent rendering
of transparent objects.}
Sort-last compositing is a widely studied technique for distributed volume rendering (\Cref{sec:rel_work_compositing})
The problem of real-time rendering of complex transparent objects is frequently
encountered in real-time graphics and games, and has been the subject of extensive study (\Cref{sec:rel_work_oit}).
\edit{We further review order-independent transparency methods and their potential application to distributed
volume rendering in}~\Cref{sec:background}.

\subsection{Distributed Volume Rendering}
\label{sec:rel_work_compositing}
A common way to render large volumes is to parallelize rendering over a cluster of machines in image-, object-, or hybrid-order.
\new{Distributing the rendering workload allows accelerating rendering or rendering data sets that cannot fit on a single machine}.
\new{Each machine now has just a subpiece of the data, and independently produces a partial image of whole dataset.
The fundamental scaling challenge in distributed rendering is combining these partial images into a final single image of the
entire dataset}~\cite{hsu1993segmented,meuer01,ma94,zheng20,molnar94,yang01,wu18,brownlee12,grosset16,Usher_2019,peterka2009configurable}.

Based on the Porter and Duff over operation \cite{porter1984compositing}, Molnar et al. \cite{molnar94} summarized a theoretical model of distributed rendering based on where the sorting happens: \new{sort-first, sort-middle, and sort-last}.
Sort-last is a practical and scalable object-order approach \new{used for large volumes where the data is distributed and}
rendered fragments are exchanged between each node~\cite{yang01,wu18,brownlee12,grosset16,Usher_2019,hsu1993segmented,ma94,peterka2009configurable}.
\new{Hsu~\cite{hsu1993segmented} proposed segmented raycasting, where data is distributed and fragments
are sent to the node that owns a given pixel}.
Tree-based methods, such as binary swap~\cite{ma94} and Radix-k~\cite{peterka2009configurable}, introduce more
structured and scalable fragment exchange patterns to improve parallelism.
The IceT library~\cite{moreland2011icet} provides practical implementations of a number of sort-last algorithms and is widely used in the scientific visualization community.
OSPRay~\cite{wald16} provides a distributed rendering facility that can integrate with IceT~\cite{wu18} or leverage its own
scalable Distributed FrameBuffer~\cite{Usher_2019} to allow for more flexible data distributions.
\new{However, these techniques are all based on sort-last compositing and require that the partial image
produced on each node can be uniquely ordered in depth relative to other node's partial images so that
they can be composited to produce the final image. This requirement does not hold for the data distributions
we consider in this paper.}

\new{Layer and deeper fragment buffer approaches have been proposed to handle some level of depth order overlap.  The A-buffer algorithm enables unordered rendering by storing and sorting fragments afterwards \cite{carpenter1984buffer}.
Then, to avoid sorting an arbitrarily long list, the k-buffer allows merging extra pixel segments heuristically \cite{bavoil2007multi}. }
However, even though these layer-based approaches accumulate only a fixed number of layers, the compositing pipeline still needs to be executed in order.

\subsection{Order-Independent Transparency for Rendering}
\label{sec:rel_work_oit}
\begin{figure}[t]
     \centering
     \hspace{2em}
     \begin{subfigure}{0.15\textwidth}
         \centering
         \includegraphics[width=0.6\textwidth, angle=270]{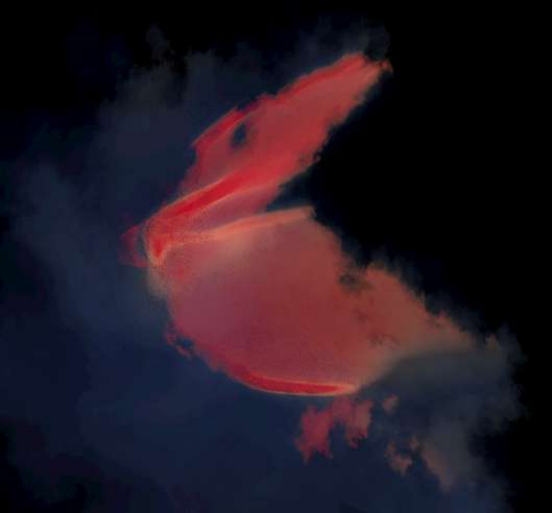}
         \caption{One Subdomain}
     \end{subfigure}
     \begin{subfigure}{0.25\textwidth}
         \centering
         \includegraphics[width=0.7\textwidth]{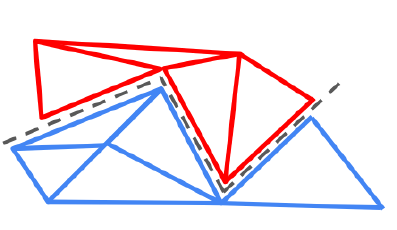}
         \caption{\label{fig:example_umesh_boundaries}%
         Example Boundaries}
     \end{subfigure}
     \hfill

    \vspace{-0.5em}
    \begin{subfigure}{0.42\textwidth}
         \centering
         \includegraphics[width=\textwidth]{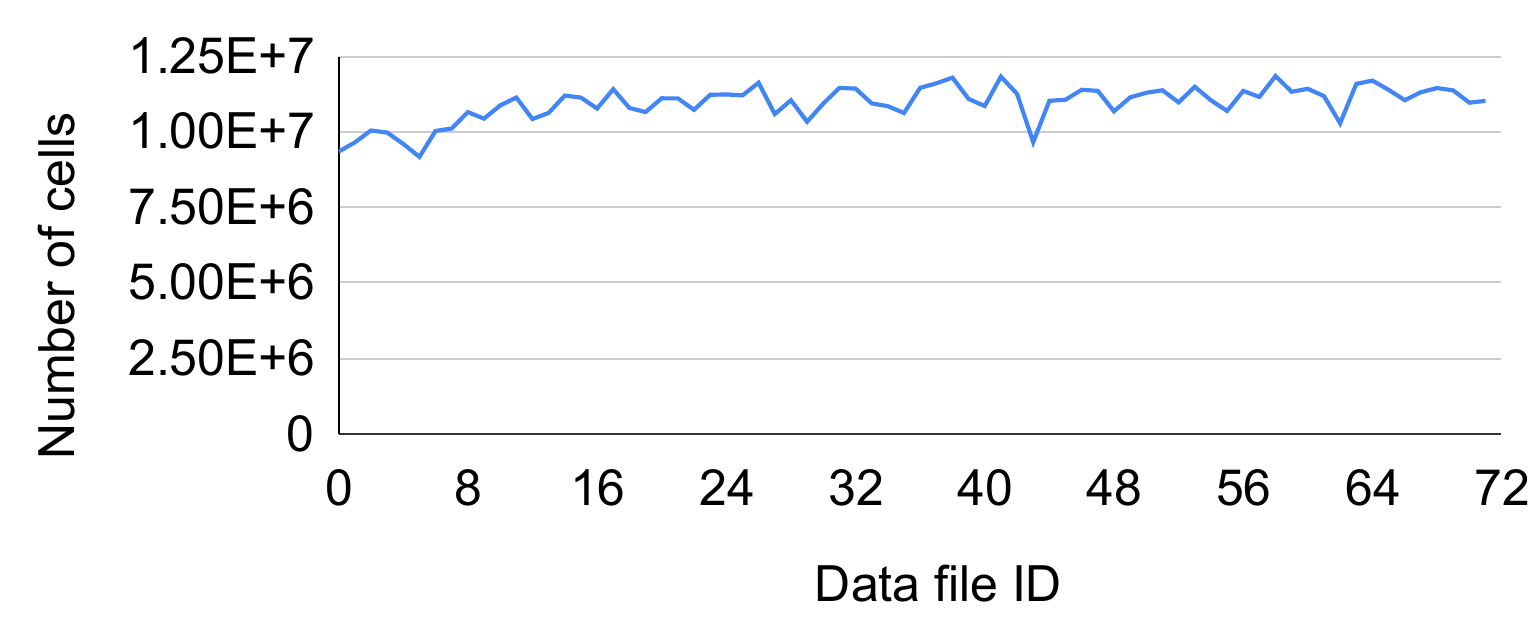}
         \vspace{-1.5em}
         \caption{Number of Cells for Each Subdomain.}
    \end{subfigure}
     
     \vspace{-1em}
     \caption{\label{fig:fun3d_sizes}%
     The FUN3D data. Note the uneven and jagged data boundary of an individual subdomain's data in (a).
     An illustration of the volume boundaries of two subdomains is shown in (b).}
     \vspace{-1.5em}
\end{figure}

Enabling order-independent transparency (OIT) the need to sort fragments to produce the final image, which is especially costly in modern highly parallel renderers. The key is to find a reasonable heuristic that balances approximation accuracy with any additional computational overhead. Developing such techniques has been the focus of a substantial body of work in real-time rendering.
\new{Early OIT work on depth peeling}~\cite{Everitt2001InteractiveOT} utilized multiple
rendering passes to render and peel away layers of surfaces in depth order without explicitly sorting geometry before rendering.
\new{Although depth peeling accurately resolves} the transparency of objects, its performance is highly dependent on scene
complexity, resulting in unpredictable compute and memory costs.

\edit{To reduce the impact of scene complexity}, OIT works heuristically merge or discard fragments~\cite{maule2013hybrid,enderton2010stochastic,munstermann2018moment,
meshkin2007sort,mcguire2017phenomenological,McGuire2013Transparency,salvi2010adaptive} to achieve fast, scene-independent OIT, at the cost of image quality.
\new{However, OIT methods are primarily targeted at real-time applications such as games, incorporating assumptions about
limited depth-complexity and smoothness that do not hold for visualization applications.
Scenes with high depth or color complexity, as can be common in visualization, break these assumptions and are especially challenging for such methods.}
Sub-sampling based approaches such as hybrid transparency~\cite{maule2013hybrid},
which tries to pick the $k$ most important colors, and stochastic transparency~\cite{enderton2010stochastic},
which stochastically discards fragments, can encounter missing surfaces.
Single-layer heuristic techniques,
such as phenomenological transparency~\cite{mcguire2017phenomenological}, sort-independent alpha blending~\cite{meshkin2007sort},
and weighted-blended OIT~\cite{McGuire2013Transparency} operate in a fixed memory budget
can produce incorrect occlusion and other visual artifacts.
Although these methods achieve fast and scene-independent OIT, the artifacts introduced can make them less suited to
visualization applications, where accuracy is more important compared to real-time applications such as games.

\begin{figure}[t]
     \centering
     \captionsetup[subfigure]{justification=centering}
     \begin{subfigure}[b]{0.22\textwidth}
         \centering
         \includegraphics[width=\textwidth]{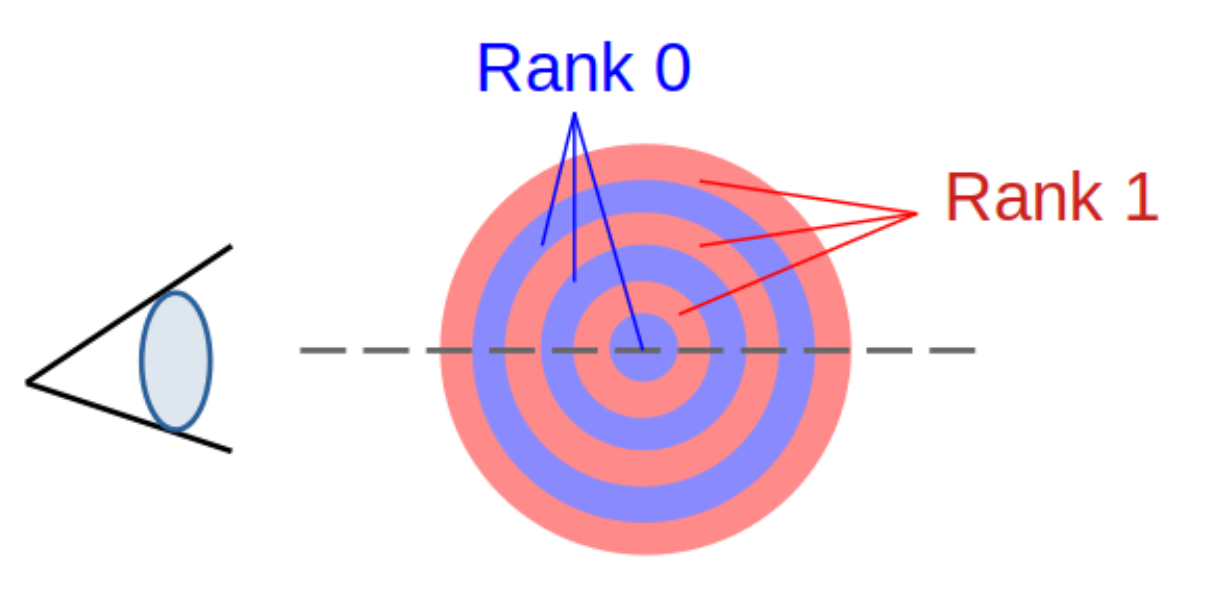}
         \vspace{-1.5em}
         \caption{Synthetic \\ sphere}
     \end{subfigure}
     \begin{subfigure}[b]{0.11\textwidth}
         \centering
         \includegraphics[width=\textwidth]{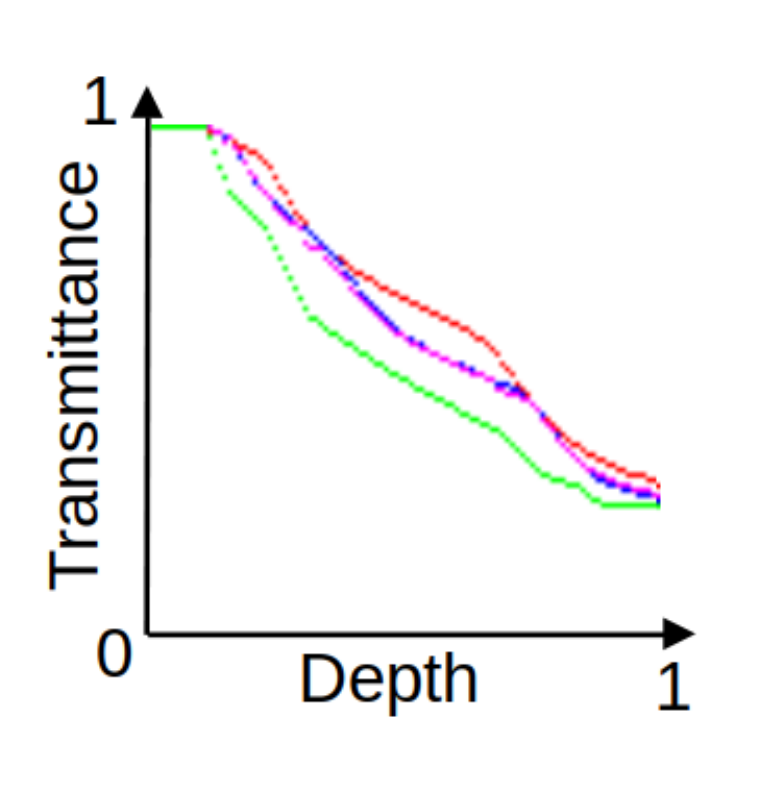}
         \vspace{-1.5em}
         \caption{Different moments}
     \end{subfigure}
     \begin{subfigure}[b]{0.11\textwidth}
         \centering
         \includegraphics[width=\textwidth]{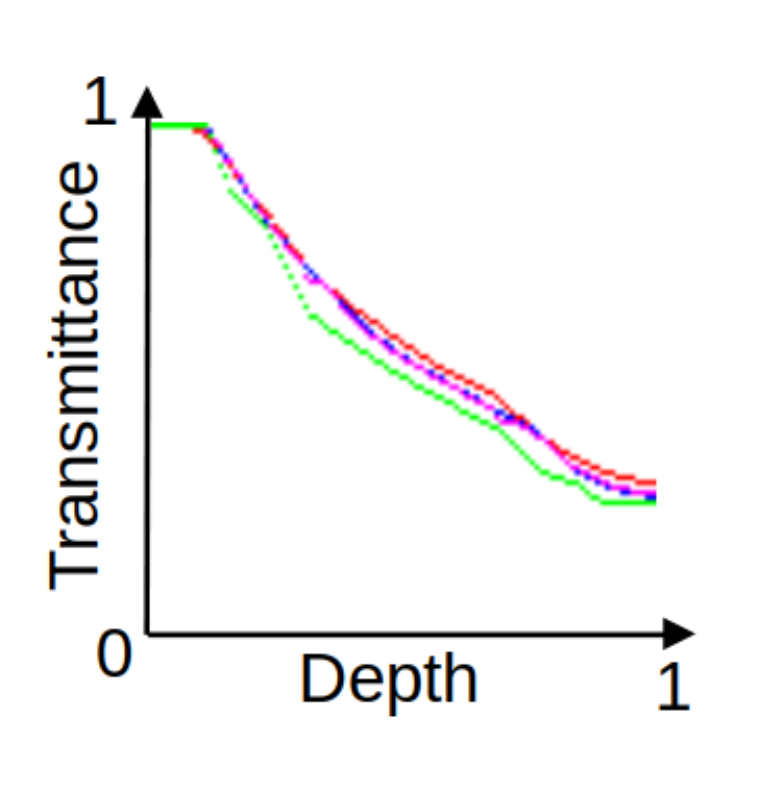}
         \vspace{-1.5em}
         \caption{\label{fig:sphere_mboit:bias}Moments with bias}
     \end{subfigure}
     
     \vspace{-.3em}
     \begin{subfigure}[b]{0.11\textwidth}
         \centering
         \includegraphics[width=\textwidth]{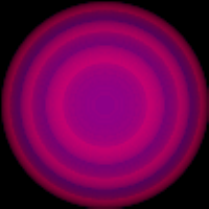}
         \vspace{-1.5em}
         \caption{MBOIT}
     \end{subfigure}
     \begin{subfigure}[b]{0.11\textwidth}
         \centering
         \includegraphics[width=\textwidth]{figures/sphere/output0_mboit_composited.pdf}
         \vspace{-1.5em}
         \caption{APC}
     \end{subfigure}
     \begin{subfigure}[b]{0.11\textwidth}
         \centering
         \includegraphics[width=\textwidth]{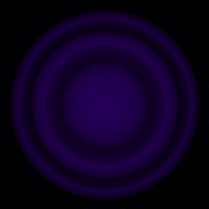}
         \vspace{-1.5em}
         \caption{APC Rank 0}
     \end{subfigure}
     \begin{subfigure}[b]{0.11\textwidth}
         \centering
         \includegraphics[width=\textwidth]{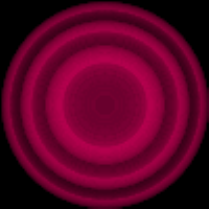}
         \vspace{-1.5em}
         \caption{APC Rank 1}
     \end{subfigure}

     \vspace{-.3em}
     \begin{subfigure}[b]{0.11\textwidth}
         \centering
         \includegraphics[width=\textwidth]{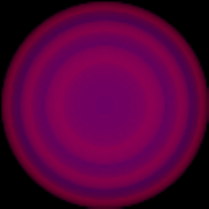}
         \vspace{-1.5em}
         \caption{Sort-last}
     \end{subfigure}
     \begin{subfigure}[b]{0.11\textwidth}
         \centering
         \includegraphics[width=\textwidth]{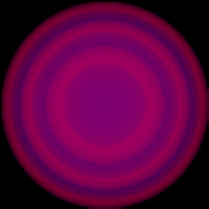}
         \vspace{-1.5em}
         \caption{APC w/ Bias}
     \end{subfigure}
     \begin{subfigure}[b]{0.11\textwidth}
         \centering
         \includegraphics[width=\textwidth]{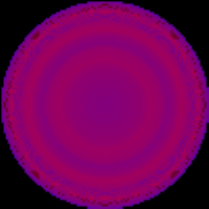}
         \vspace{-1.5em}
         \caption{HT}
     \end{subfigure}
     \begin{subfigure}[b]{0.11\textwidth}
         \centering
         \includegraphics[width=\textwidth]{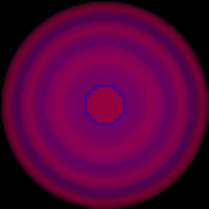}
         \vspace{-1.5em}
         \caption{WBOIT}
     \end{subfigure}
     \vspace{-1.5em}
     \caption{\label{fig:sphere_mboit}%
    \edit{A comparison of approximate OIT techniques on a synthetic red and blue concentric circles example.
    This configuration produces sharp changes in color, challenging approximate OIT methods.
     (b,c) compare different MBOIT configurations with sort-last (green line): 4 power moments (red),
     6 power moments (blue) and trigonometric moments (purple).
     We observe that APC achieves the same rendering quality as single-node MBOIT, providing a high-quality approximation.
     This is in contrast to the color artifacts of Hybrid Transparency (j) or occlusion errors from Weight-blended OIT (k).
     Furthermore, with 4 power moments and a bias (c), the APC image closely approximates the ground truth sort-last. 
     The respective image similarity measurements are \new {(h) vs (i):  SSIM=0.99, MSE=38.18, PSNR=32.34 (h) vs (j) SSIM=0.87, MSE=618.76, PSNR=20.24 (h) vs (k) SSIM=0.98, MSE=21.55, PSNR=34.82}}}
\end{figure}

In this work, we leverage Moment-Based Order-Independent-Transparency (MBOIT)~\cite{munstermann2018moment} as an off-the-shelf and efficient solution for OIT that is well suited to use in a distributed visualization environment due to its low data requirements and high image quality. MBOIT is based on moment-based shadow map approximation~\cite{peters2015moment,peters2017improved}, which offers compact, filterable, closed-form representation that is able to capture sparse signals accurately. Similar to OIT through Fourier opacity mapping, which views transmittance as a function of depth in logarithmic space to enable an additive accumulation, this approximation provides a more continuous and faithful representation of complex transparent scenes \cite{munstermann2018moment}. MBOIT requires little data to be transferred, does not require sorting or redistributing the mesh data, and provides high image quality, making it well suited to scalable rendering of unstructured and AMR datasets.
\new{\Cref{fig:sphere_mboit} compares rendering a synthetic volume using MBOIT with Hybrid Transparency and Weighted Blended OIT to
illustrate MBOIT's improved image quality.}

\begin{figure}
     \centering
     \vspace{-1.em}
     \captionsetup[subfigure]{justification=centering}
     \begin{subfigure}[]{0.23\textwidth}
         \centering
         \includegraphics[width=\textwidth]{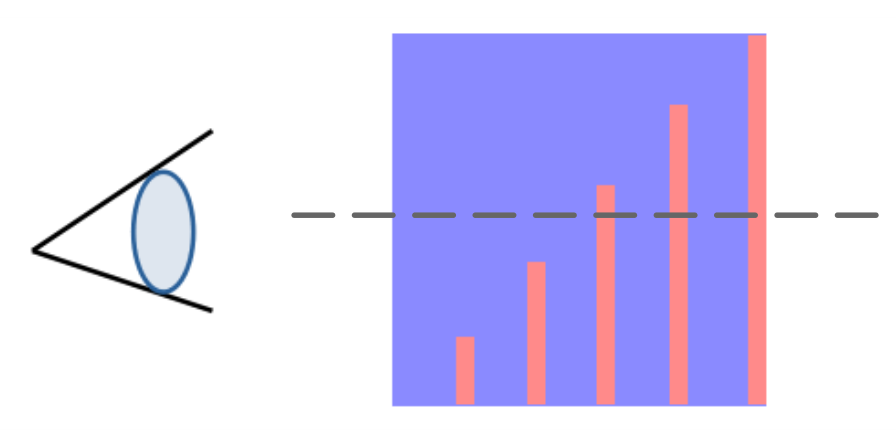}
         \vspace{-1.5em}
         \caption{Surface-Like Volume}
     \end{subfigure}
      \hspace{2mm}
    \begin{subfigure}[]{0.1\textwidth}
         \centering
         \includegraphics[width=\textwidth]{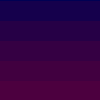}
         \vspace{-1.5em}
         \caption{Sort-last}
     \end{subfigure}
     \hspace{2mm}
     \begin{subfigure}[]{0.1\textwidth}
         \centering
         \includegraphics[width=\textwidth]{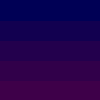}
         \vspace{-1.5em}
         \caption{APC}
     \end{subfigure}
     \vspace{-1.5em}
     \caption{\label{fig:apc_spikes_bad_case}%
     \new{A synthetic case where APC exhibits noticeable image quality loss. With thin, near-opaque red structures in the middle of a blue volume, we can see that MBOIT does not handle abrupt transmittance changes well such that (c) looks more blended, i.e., more purple, than (b).}}
     \vspace{-2em}
\end{figure}

\section{Background} \label{sec:background}

\new{Scalable sort-last compositing algorithms require that the data
be partitioned among the ranks in convex,  nonoverlapping pieces, but this requirement
is typically not satisfied by unstructured~\cite{anderson2023fun3d,palaciosStanfordUniversityUnstructured2013,economonSU2OpenSourceSuite2016,offermansAdaptiveMeshRefinement2020,anderson2023fun3d}
or octree AMR~\cite{arndtDealIIFiniteElement2021,bursteddeP4estScalableAlgorithms2011} simulations that use nonspatial
data distribution methods to accelerate the solver, e.g. ParMETIS for fill-reduction~\cite{karypisParmetisParallelGraph1997},
or Morton-order~\cite{bursteddeP4estScalableAlgorithms2011}.
Such simulations are common in computational fluid dynamics (CFD)~\cite{anderson2023fun3d,palaciosStanfordUniversityUnstructured2013,economonSU2OpenSourceSuite2016,offermansAdaptiveMeshRefinement2020},
medicine~\cite{neicEfficientComputationElectrograms2017}, and geoscience~\cite{arndtDealIIFiniteElement2021}}

\new{We evaluate our approach on two motivating CFD datasets that represent these mesh configurations. The FUN3D~\cite{anderson2023fun3d} Mars Lander uses
an unstructured mesh, whereas the Exajet~\cite{casalinoLatticeBoltzmannBased2016} uses a Cartesian AMR mesh.}
The FUN3D data were generated on Summit at Oak Ridge Leadership Computing Facility (OLCF) using a CFD code developed at the NASA Langley Research Center.
The mesh consists of a mix of tetrahedra, pyramid, and prism cells, and were
written in a total of 72 subdomains, each storing part of the mesh.
In our benchmarks we use the smaller version with 798M total elements that was run at Mach 2.4.
\new{Fun3D uses ParMETIS~\cite{karypisParmetisParallelGraph1997} for data decomposition, resulting in mesh cells being} distributed among ranks with nonuniform,
jagged boundaries that fit together much like puzzle pieces and served as the initial motivation for our work (\Cref{fig:fun3d_sizes}).
MBOIT is a high-accuracy two-pass OIT solution that operates on a moment-based representation of transparency. \new{To avoid collecting transparency by alpha blending along the depth with the Porter and Duff compositing operator, MBOIT looks into approximating the transmittance as a function of depth by two order-independent operations: one to construct the function, the other to recover the transmittance value through the generated function.}
\new{Defined by $[z, z^2, z^3, z^4]$ for a 4 moments implementation where $z$ represents the depth, the power moments serve as a collection of measures that record transmittance behaviors when traversing through transparent layers. Converted to logarithmic space operations, this representation can be additively constructed for global transmittance information, allowing for accurate, order-independent approximation reconstruction of the transmittance value given any depth.}
\new{Therefore, the first rendering pass accumulates per-pixel power moments that encode powers of transparency needed to generate the function, and the second rendering pass uses the moments to solve for transmittance value along depth to produce final colors that can also be additively blended.}
Two data summation phases are required in image space, and thus the scene complexity does not become the bottleneck of the rendering process.

\edit{We summarize the algorithm in \Cref{alg:getMboit} and the two key components in \Cref{alg:GeneratePowerMoments} and \Cref{alg:getTranmittance}. For additional details on the MBOIT computation, we refer to the paper by M\"unstermann et al.~\cite{munstermann2018moment} and a more detailed description in the preceding paper~\cite{peters2017improved}.}

\edit{Compared to prior OIT methods,} MBOIT has been shown to perform well for different signal frequencies, in that it is both truly order-independent throughout the computation, and \edit{well approximates reconstruction} of the object occlusions in their true depth order. The key steps of moments construction and moments-based transmittance reconstruction, namely the summation loops with \textit{GeneratePowerMoments} and \textit{getTransmittance}, can be executed in an arbitrary order, allowing for an order-independent rendering pipeline. This characteristic sets the foundation of our scalable distributed compositing method.

\begin{algorithm}[t]
\caption{The \textit{RenderMBOIT} algorithm that returns a pixel color rendered with Moment-Based Order-Independent Transparency}
\label{alg:getMboit}
\small
\begin{algorithmic}[1]
\Function{RenderMBOIT}{volume \textit{v}, bias $\beta$}
        \State $col \gets (0, 0, 0, 0)$, $\textbf{b} \gets (0, 0, ... 0)$ \label{alg:moments}
        \ForEach{sample $s$ in \textit{v}} \label{alg:stage1}
            \State \Call{GeneratePowerMoments}{$\textbf{b}$, $s.depth, s.transmittance$}
        \EndFor
        \ForEach{sample $s$ in \textit{v}} \label{alg:stage2}
            \State $col \gets col $ $+$ \Call{getTransmittance}{$\textbf{b}$, $s.depth$, $\beta$} $\cdot$ $s.col$
        \EndFor
        \State \Return $col$
\EndFunction
\end{algorithmic}
\end{algorithm}

\begin{algorithm}[t]
\caption{The \textit{GeneratePowerMoments} function that computes the moments at a given sample point}
\label{alg:GeneratePowerMoments}
\small
\begin{algorithmic}[1]
\Function{GeneratePowerMoments}{moments $\textbf{b}$, depth \textit{d}, transmittance $t$}
        \State $d \gets logDepthWarp(d)$ \Comment{rescale logged depth value to [-1, 1]}
        \State $absorbance \gets -log(transmittance)$ \Comment{get logged absorbance}
        \State $absorbance \gets$ \Call{min}{$absorbance$,  $ABSORBANCE\_MAX\_VALUE$}
        \For{moment $b_i$ in \textbf{b}} 
            \State $b_i \gets b_i$ $+$ \Call{pow}{$s.depth, i$} $\cdot$  $absorbance$ \Comment{store moments by powers of depth}
        \EndFor
\EndFunction
\end{algorithmic}
\end{algorithm}
\begin{algorithm}[t!]
\caption{The \textit{getTransmittance} function that reconstructs the final transmittance at a given depth by the moments}
\label{alg:getTranmittance}
\small
\begin{algorithmic}[1]
\Function{getTransmittance}{moments $\textbf{b}$, depth \textit{d}, transmittance $t$}
        \State $m \gets len(\textbf{b})$ \Comment{get number of moments}
        \State $d \gets logDepthWarp(d)$ \Comment{rescale logged depth value to [-1, 1]}
        \State $\textbf{b}_{\textbf{tmp}}$ $\gets (b_1, ... b_m)$ 
        \State $\textbf{b}_{\textbf{tmp}} \gets$ \Call{mix}{$\textbf{b}_{\textbf{tmp}}$,  $\beta.bias\_vector$} \Comment{bias input data to avoid artifacts}
        \State $\textbf{q} \gets$ $\Call{SolveMat}{\textbf{b}_{\textbf{tmp}}, d}$ \Comment{compute Cholesky factorization of the Hankel matrix}
        \State $\textbf{z} \gets$ $\Call{SolvePowerEquation}{\textbf{q}, d}$ \Comment{get roots of the power equation}
        \State $weights_0 \gets \beta.overstimation$ \Comment{adjust weight factors by overestimation}
        \For{$i < m$} 
        \State $weights_i \gets$ $(z_i < z_0)?$ $ 1.0$ : $0.0$
        \EndFor
        \State $\textbf{p}\gets $ $\Call{solvePolynomial}{ \textbf{weights}, \textbf{z}}$ \Comment{solve for final absorbance vector}
        \State $absorbance \gets$ $\textbf{p} \cdot vector(1.0, \textbf{b}_{\textbf{tmp}})$ \Comment{compute absorbance value}
        \State \Return \Call{clamp}{$exp(-b_0 \cdot absorbance)$} \Comment{return transmittance in the original depth range}
\EndFunction
\end{algorithmic}
\end{algorithm}

\section{Method}

\begin{figure}
    \centering
    \includegraphics[width=0.35\textwidth]{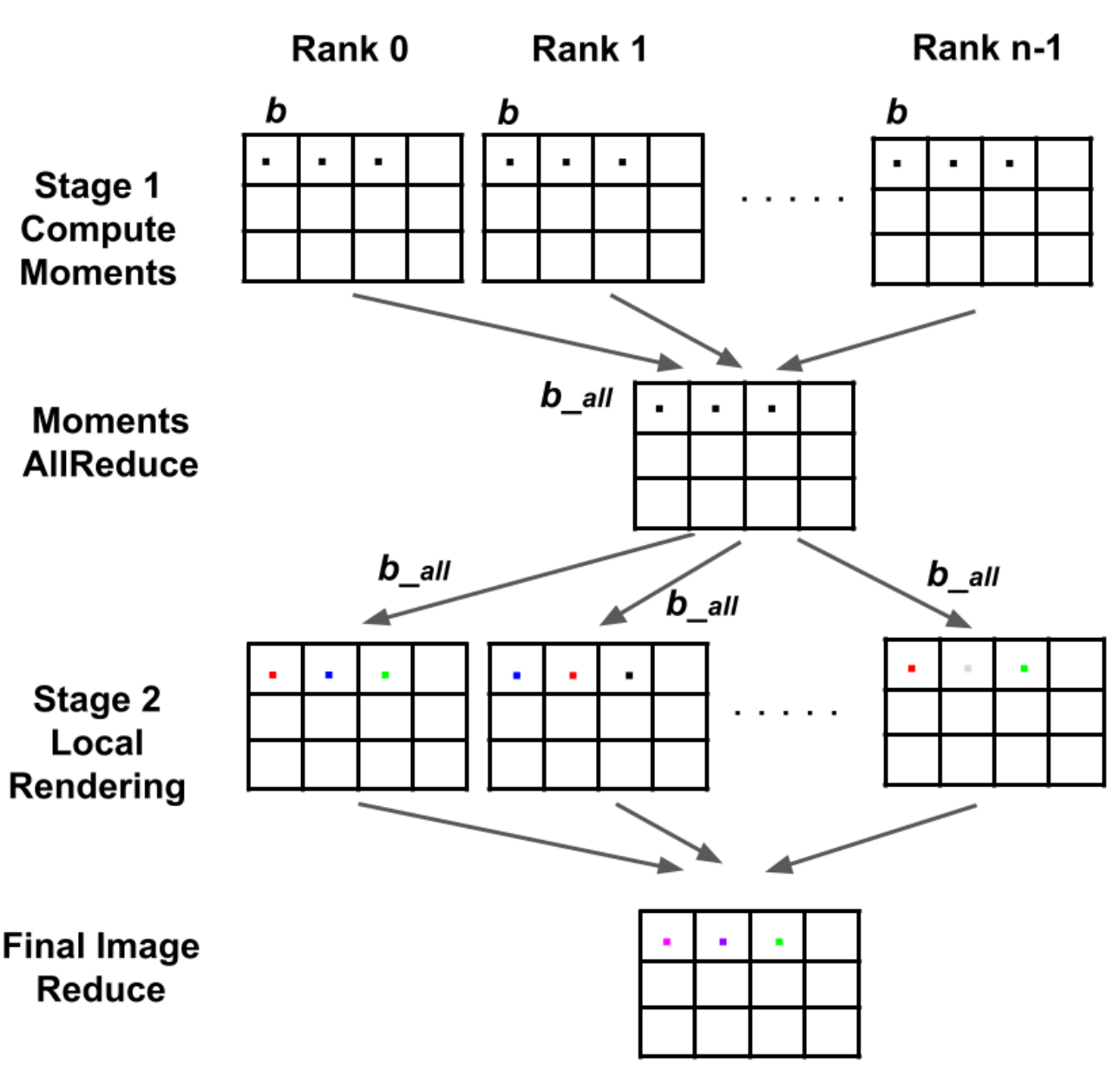}
    \vspace{-0.5em}
    \caption{\label{fig:pipeline_illustration}%
    An illustration of the APC pipeline.
    Stage one computes a local moments vector per pixel on each rank, adding them up in the moments AllReduce step to form the global moments.
    The global moments are used in stage two on each rank to approximate transmittance when rendering their local volume
    to produce final subimages.
    Finally, all subimages are added using a Reduce onto the display rank.} 
    \vspace{-0.5em}
\end{figure}

Our rendering pipeline extends the single-node MBOIT pipline to a distributed computing environment to
enable highly scalable approximate order-independent compositing.
Our Approximate Puzzlepiece Compositing pipeline consists of two main stages: stage one renders the local moments
on each rank and computes global moments through an \texttt{MPI\_Allreduce} add (\Cref{sec:method_stage_1});
stage two uses the global moments to approximate transmittance on each rank to render
local partial images (\Cref{sec:method_stage_1}).
The partial images are then combined to form the final image through an \texttt{MPI\_Reduce} add.
The advantage of our proposed method for distributed rendering is that each rank can
independently produce its part of the moments and final image, with global results produced
through single optimized MPI operations (AllReduce and Reduce). 
An illustration of our pipeline is shown in \Cref{fig:pipeline_illustration}.

\begin{figure}
     \centering
     \captionsetup{justification=centering}
     \begin{subfigure}{0.23\columnwidth}
         \centering
         \includegraphics[width=\textwidth]{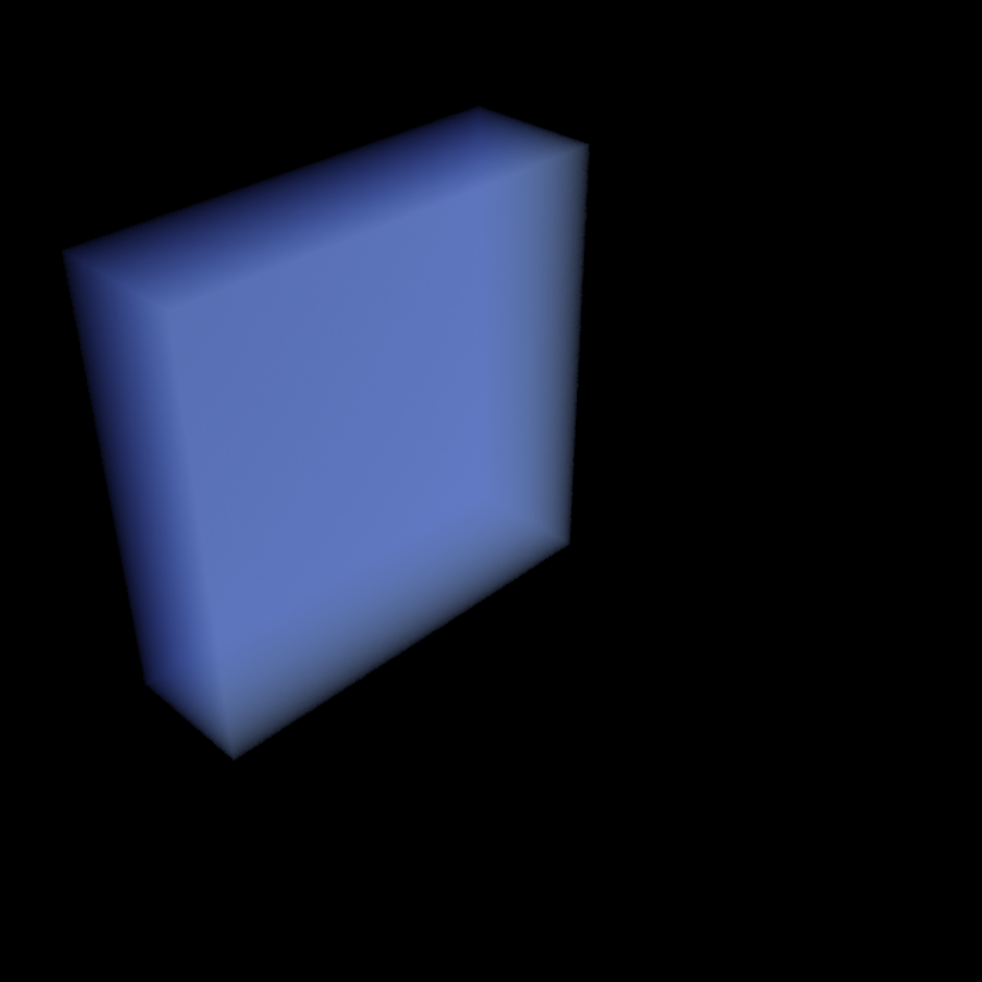}
         \caption{Local Object Rank 0}
         \label{fig:alpha blend rank 0}
     \end{subfigure}
     \begin{subfigure}{0.23\columnwidth}
         \centering
         \includegraphics[width=\textwidth]{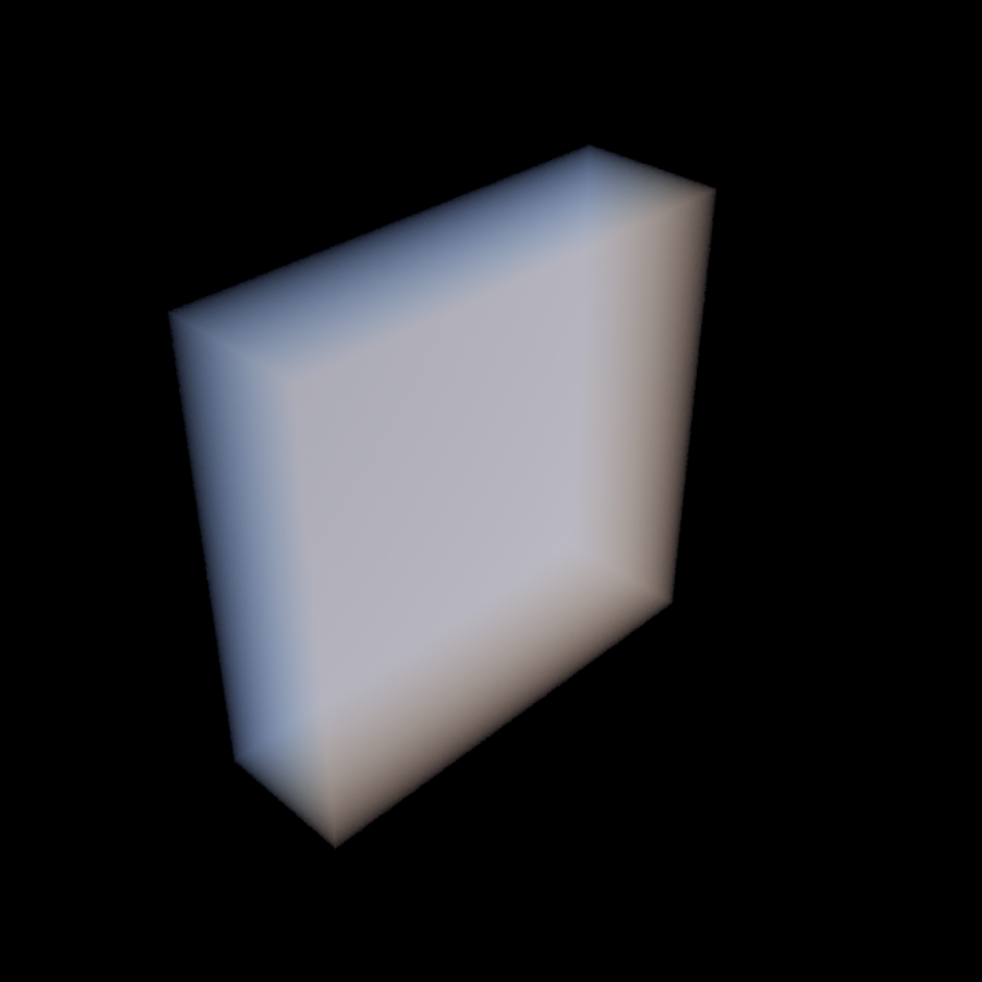}
         \caption{Local Object Rank 1}
         \label{fig:alpha blend rank 1}
     \end{subfigure}
     \begin{subfigure}{0.23\columnwidth}
         \centering
         \includegraphics[width=\textwidth]{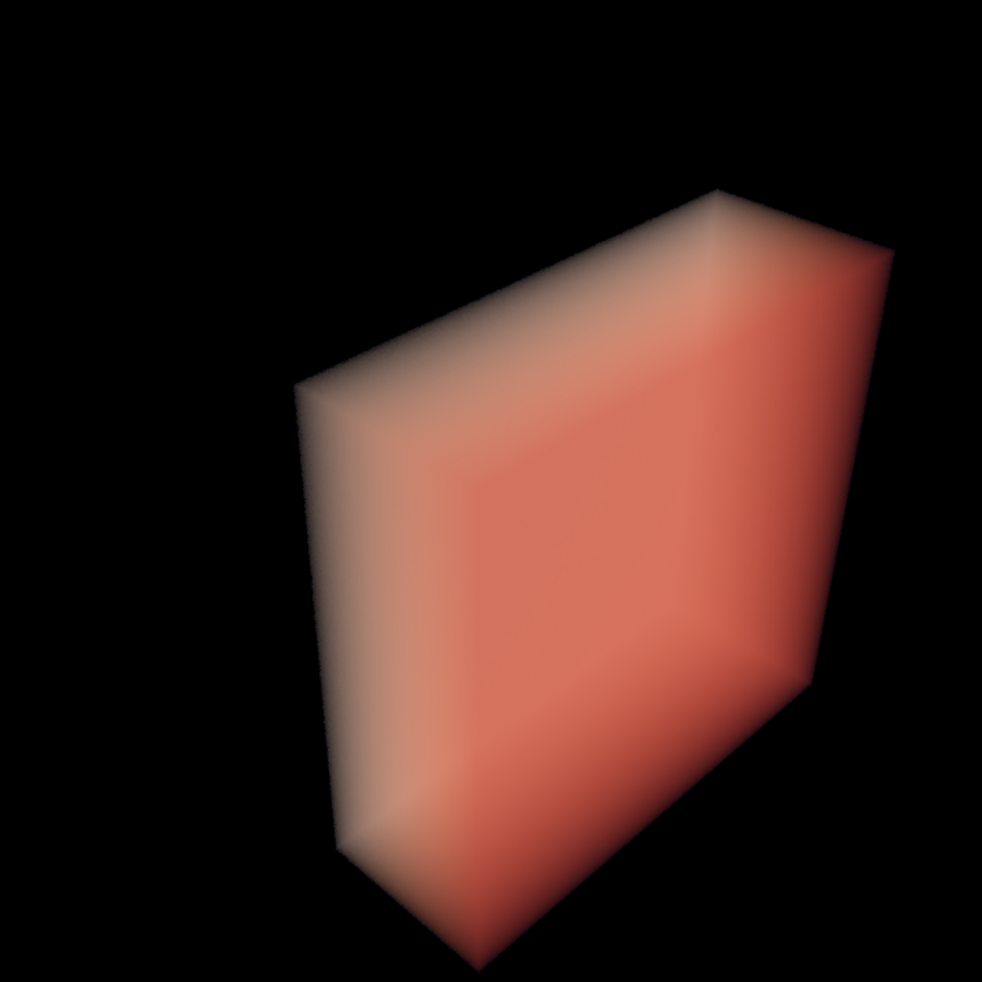}
         \caption{Local object Rank 2}
         \label{fig:alpha blend rank 2}
     \end{subfigure}
     \begin{subfigure}{0.23\columnwidth}
         \centering
         \includegraphics[width=\textwidth]{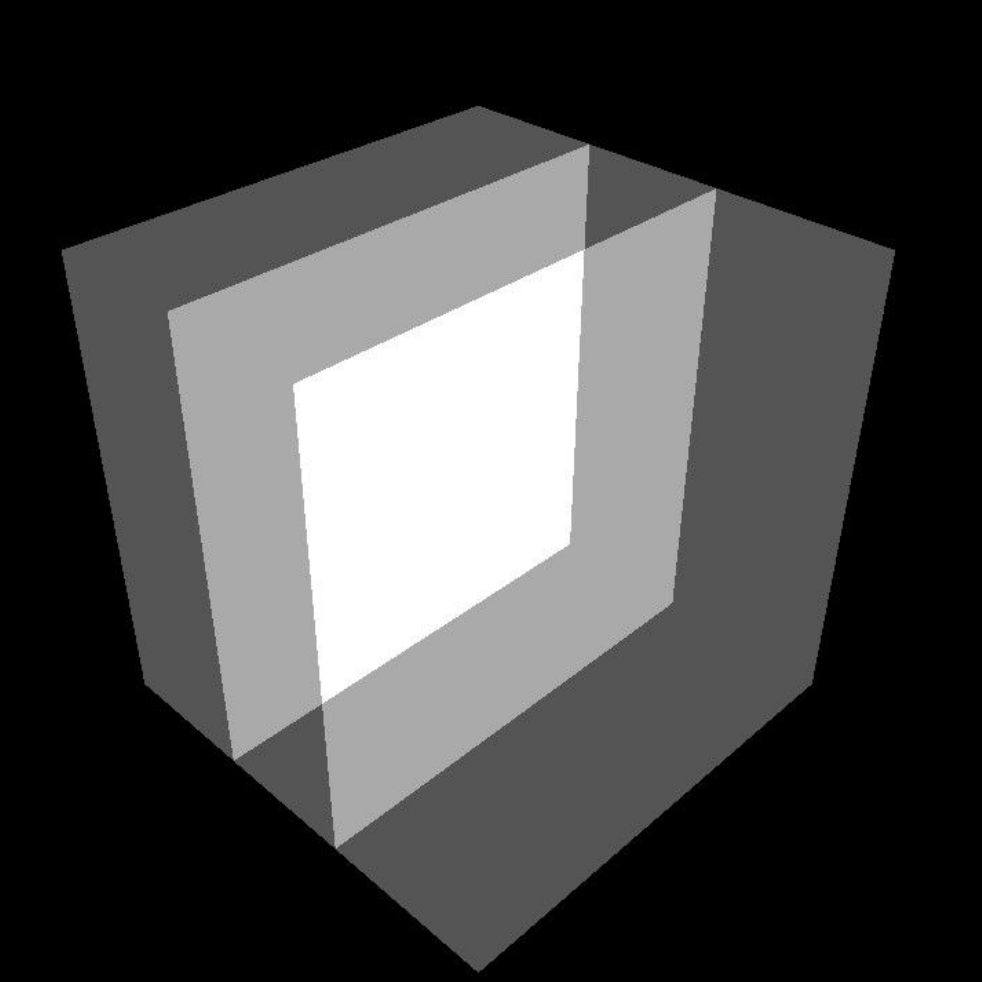}
         \caption{Segment Density}
     \end{subfigure}

     \vspace{-.3em}
    \begin{subfigure}{0.23\columnwidth}
         \centering
         \includegraphics[width=\textwidth]{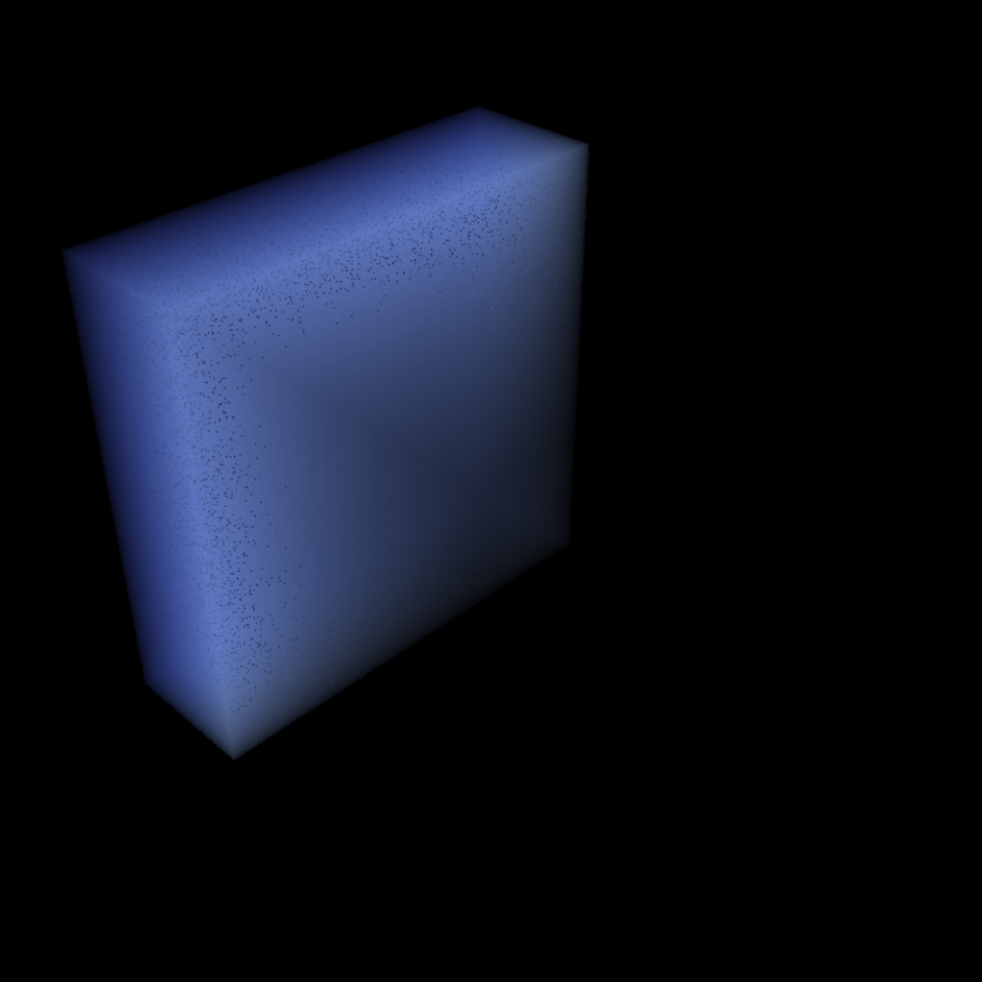}
         \caption{ APC \\ Rank 0}
         \label{fig:mboit rank 0}
     \end{subfigure}
     \begin{subfigure}{0.23\columnwidth}
         \centering
         \includegraphics[width=\textwidth]{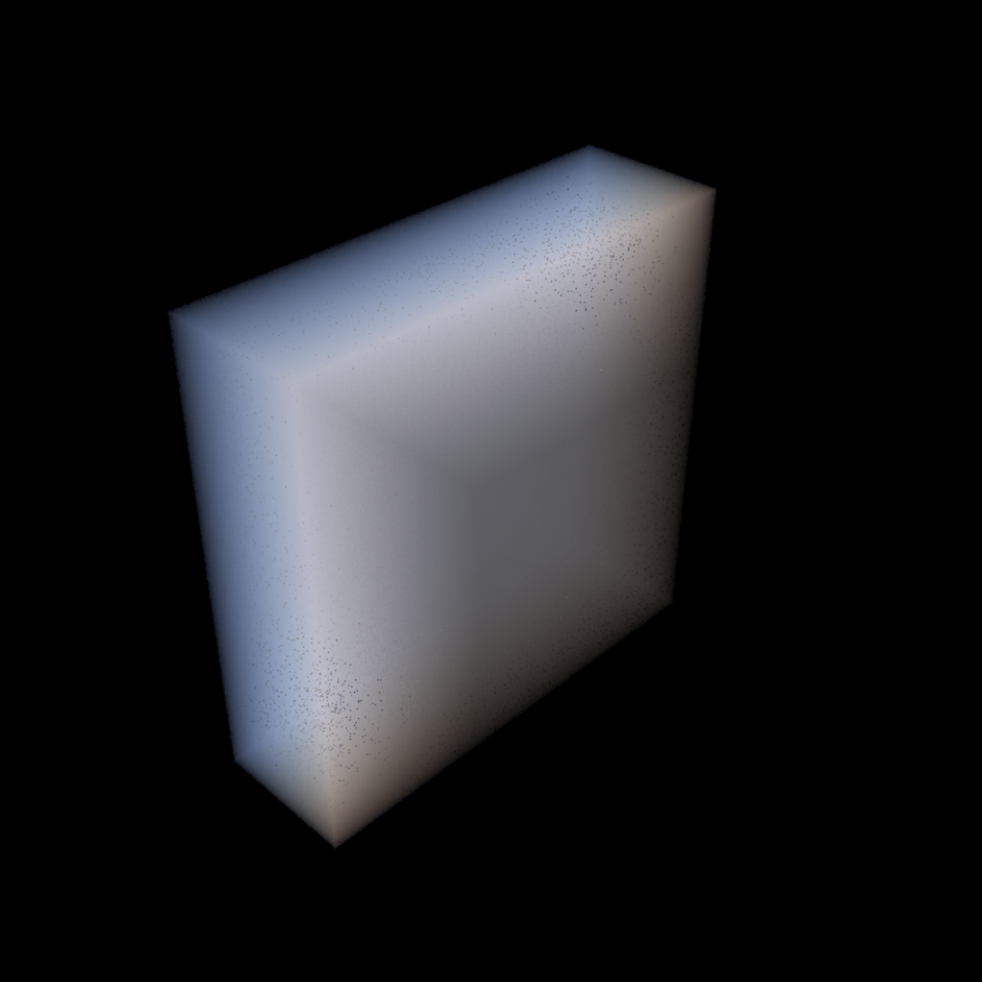}
         \caption{APC \\ Rank 1}
         \label{fig:mboit rank 1}
     \end{subfigure}
     \begin{subfigure}{0.23\columnwidth}
         \centering
         \includegraphics[width=\textwidth]{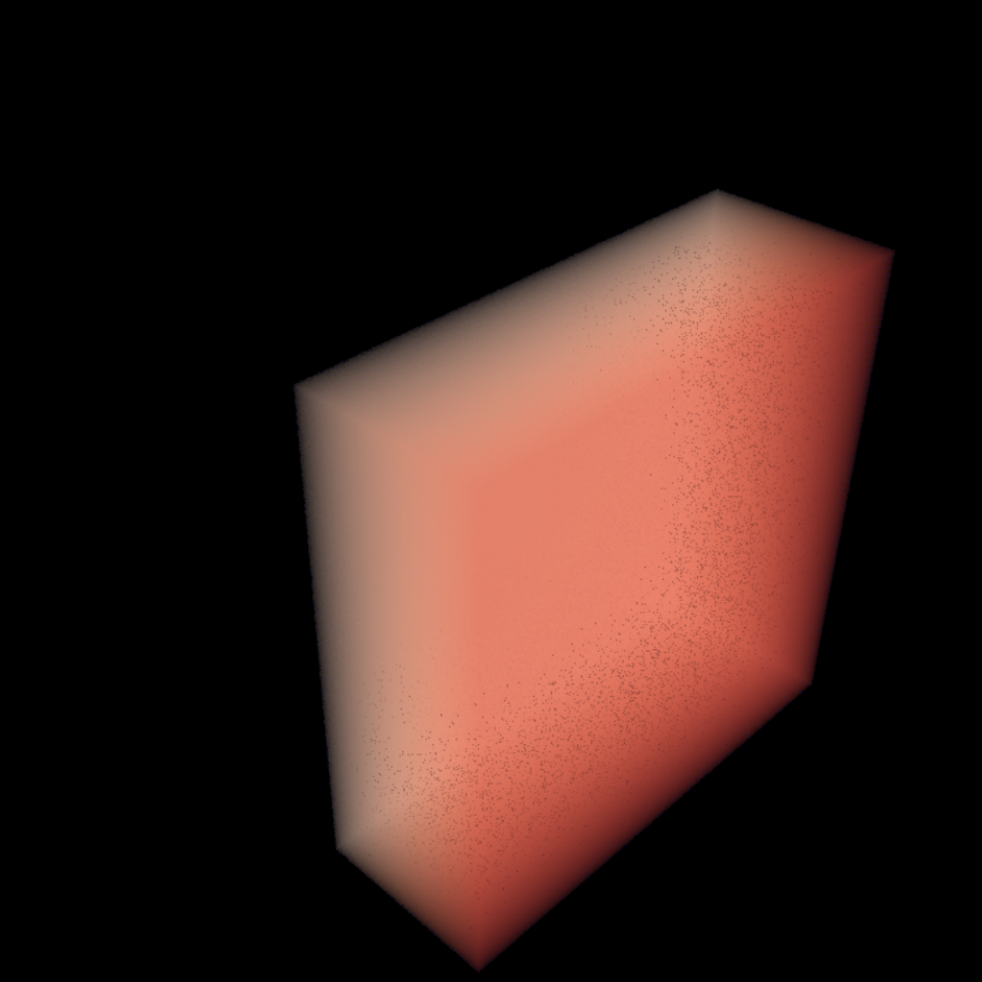}
         \caption{APC \\ Rank 2}
         \label{fig:mboit rank 2}
     \end{subfigure}
     \begin{subfigure}{0.23\columnwidth}
         \centering
         \includegraphics[width=\textwidth]{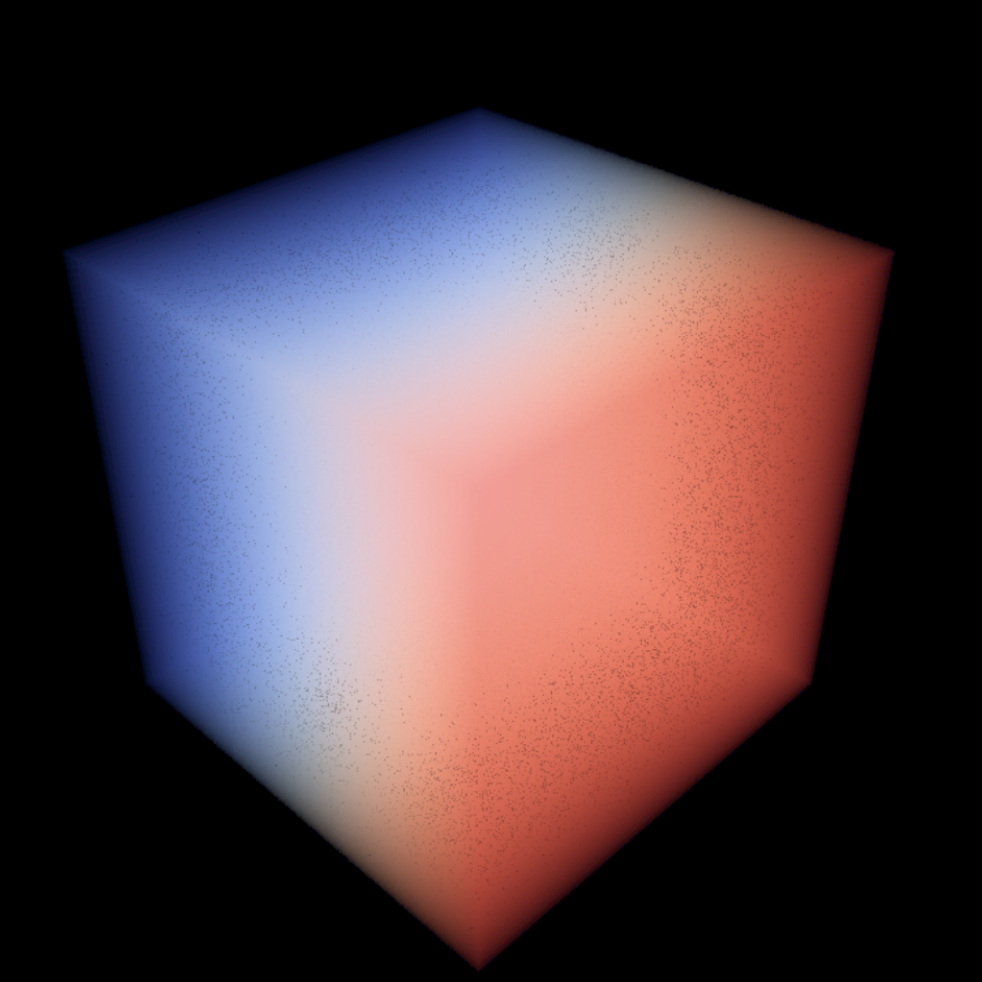}
         \caption{APC \\ by (e) - (g)}
         \label{fig:image sum add}
     \end{subfigure}
     \vspace{-1.2em}
     \captionsetup{justification=raggedright}
     \caption{An example image produced from distributed rendering of a sandwiched volume over three ranks.
     Each rank has two interleaved slices of a cube.
     The occluded part has its opacity corrected with MBOIT's approximation, i.e., each rank here has its local objects rendered with the "correct alpha" as it would appear in the final image, which can thus be produced by a simple image space summation.}
     \label{fig:sandwich}
     \vspace{-1.5em}
\end{figure}

\subsection{Stage One: Computing Local and Global Moments}\label{sec:method_stage_1}
The first step is the same as the first stage  in~\Cref{sec:background}.
We must produce a local moments image on each rank that we can combine to produce
a global moments image to use for representing absorbance.
Each rank traces rays through its local data; however, instead of rendering out final colors as in a
typical rendering pass, we compute the moment for the pixel by summing up moments along the ray.
These moments are written out to an image to produce a local moments image on each rank.
After all ranks have computed their local moments image, we compute the global moments
image by performing an \texttt{MPI\_Allreduce} add on the local moments images.
As moments in MBOIT are computed through addition, which is commutative, it makes no difference whether
the addition occurs locally on a node or globally over MPI.
Thus, the moments computed in our distributed pipeline match those produced by single-node rendering.

\subsection{Stage Two: Using Moments to Approximate Transmittance}
\label{sec:method_stage_2}

After the Allreduce add, each rank has the same global moments image, and we can now approximate
the global transmittance along each ray as we traverse the volume.
In stage 2, each rank produces a local partial image of its data by tracing rays through its local data again,
using the global moments image to approximate transmittance.
Our rendering algorithm is similar to standard ray-tracing; however, instead of getting the standard RGBA at the voxel position, we drop alpha when blending the sample into the final color and replace the transmittance value with what we get from the MBOIT function.
By combining the sampled color and the approximated transparency, we are able to blend the pixel segments in any order.

When all ranks have completed rendering their data using the approximate transmittance data, we have the local images ready with the correct global transmittance.
As discussed in~\Cref{sec:background}, each local image is completely order-independent because the reconstruction function calculates the transmittance with correct occlusion. Therefore, a \texttt{MPI\_Reduce} add in image space will produce the result with the correct object ordering.

A three-rank distributed rendering using this pipeline is shown in \Cref{fig:sandwich}. \Cref{fig:alpha blend rank 0,fig:alpha blend rank 1,fig:alpha blend rank 2} are the subvolumes each local renderer owns, rendered in traditional alpha blending. Each rank consists of two interleaved slices that are occluded by either other rank's subvolume or partially by themselves.
\Cref{fig:mboit rank 0,fig:mboit rank 1,fig:mboit rank 2} show the local rendering of our method after stage 2, where the global transmittance has been approximated to render each rank's local data with the global transmittance.
At the final image reduce step in our method, all that needs to be done is to add all the local images together,
whereas alpha blending requires sorting for each pixel to determine a correct sequence of operations to produce the final color.

\subsection{Implementation}\label{sec:method_impl}
We implement our method within OSPRay~\cite{wald16} by modifying its distributed rendering framework~\cite{Usher_2019} to take advantage of OSPRay's
high-fidelity ray-tracing engine for fast volume rendering on modern CPU architectures.
To boost performance, we group pixels into tiles for locality, exchange only nonempty tiles, and utilize ISPC~\cite{pharr2012ispc} to leverage SIMD hardware to accelerate moment computation and transmittance estimation.

MBOIT can also be customized by using different sets of moments, which require slightly different mathematical operations for \textit{GeneratePowerMoments} and \textit{GetTranmittance}, enabling finer adjustments of local rendering computation, memory overheads, and numeric precision. However, the pipeline we describe here remains the same regardless of particular moments of choice. \Cref{fig:sphere_mboit} shows a sample sphere volume and comparisons between different configurations. The distributed computation with APC produces a faithful image of the single-node computation quality due to the order-independent nature of MBOIT. Compared to popular alternatives, the MBOIT images provide a more faithful depth perception, presenting a smooth transition for low-frequency volume intervals while preserving high-frequency occlusion details, \new{whereas the method still suffers from inaccurate blending when facing surface-like, extremely thin structures in volume rendering as in \Cref{fig:apc_spikes_bad_case}.} A closer look at image-quality evaluation with real-world datasets will be presented in ~\Cref{img_qual}.

As shown in the transmittance curves, the MBOIT method is able to provide an accurate and smooth approximation to the sort-last techniques but tends to overestimate the current transmittance, resulting in slightly brighter pixels in higher-density regions. Different from mainstream compositing techniques, the MBOIT method is not guaranteed to be energy conserving, which is compensated for by additional renormalization operations and bias vectors, as described in section 3 in the original paper~\cite{munstermann2018moment}. Thus, a constant bias parameter $\beta$ (as shown in \Cref{alg:getMboit}) is set in the implementation to offset the overestimation, allowing more convincing rendering outcomes. Under this adjusted configuration, the choice of moments functions does not introduce significant image difference as seen in \Cref{fig:sphere_mboit:bias}. Therefore, we adopt the most compact representation, i.e., the 4 power moments computation, for maximum performance.

\section{Evaluation}

\begin{figure}
     \centering
     \begin{subfigure}{0.4\textwidth}
         \centering
         \includegraphics[width=\textwidth]{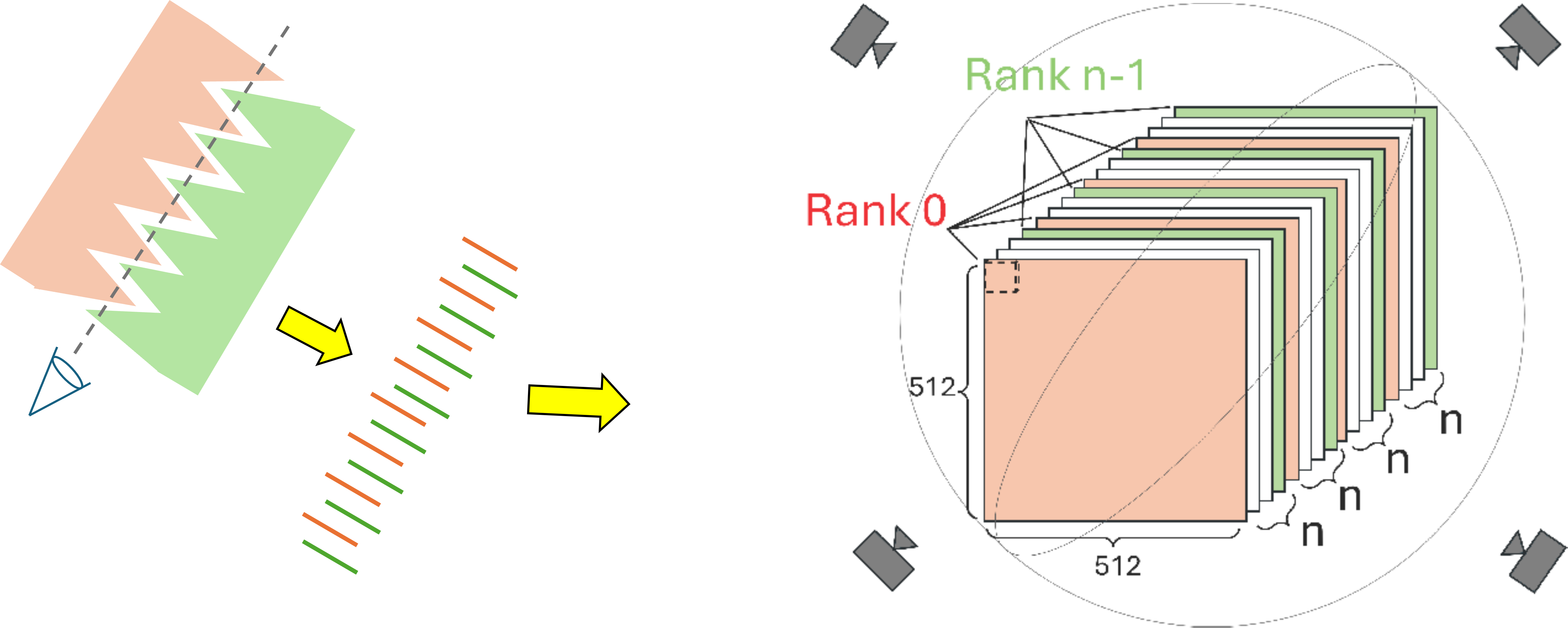}
     \end{subfigure}
     \vspace{-1em}
     \caption{\label{fig:syn_illustration_plot}%
     \new{Illustration of the weak-scaling stress-test volume, which represents a worst-case ``gear teeth'' boundary.}
     Each rank owns four $512 \times 512$ hexahedra layers, interleaved with $n-1$ layers from other ranks in between.
     \new{Traditional sort-last rendering would require producing and blending individual color fragments for each
     slice to produce a correct image.}
     Images are captured around a sphere camera orbit in our benchmarks.}
    \vspace{-0.5em}
\end{figure}

\begin{table}
\centering
\begin{tabular}{@{}lrrr@{}} 
\toprule
    Dataset & \makecell{Num of \\ Subdomains} & \makecell{Num of Cells} & \makecell{Cell Type(s)}\\  [0.5ex] 
    \midrule
    \vspace{0.1em}
    Synthetic & 64  &  8388608 & \makecell{hexahedra \tabularnewline  }\\
    \vspace{0.1em}
    FUN3D     & 72  & 788841511 & \makecell{tetrahedra, \\pyramids, prisms}\\ 
    Exajet    & 128 & 656444884 & \makecell{hexahedra \tabularnewline  }\\
    \bottomrule
    \end{tabular}
    \vspace{-0.5em}
    \caption{\label{tab:data}All unstructured mesh datasets used in the evaluation, with total number of subdomains and total cell numbers and cell types. For the performance experiments, each subdomain is assigned to one MPI rank run by a single compute node. 
    }
    \vspace{-1em}
\end{table}
We evaluate our method's image quality and rendering performance through distributed rendering
benchmarks performed on the TACC Frontera HPC system, \new{with 56 cores and 192~GB of memory on each Intel Xeon Platinum 8280 compute node.} We perform
runs on up to 128 ranks with one rank per compute node. For the MBOIT renderer implementation, we use 4 power moments with an overestimation bias $\beta=0.3$. All images are rendered at a 2560$\times$2560 resolution. 

We note that aggregating the entire dataset to a single node during in situ rendering is sometimes not possible due to the memory restriction on the compute nodes, or the heavy data transfer that is order-of-magnitude slower than the rendering. To compare our work to the state-of-the-art sort-last distributed rendering algorithm, we implemented a segment layer-based distributed rendering method, which alpha blends the fragments on every continuous sample interval into a single segment. The local render will thus render into order-based layers, leading to less total data transfer traffic and a reduced sorting workload compared to working with all fragments as in the full individual segment pipeline.

We present all used datasets and the corresponding settings in ~\Cref{datasets}. The image quality and performance results are shown in ~\Cref{img_qual} and ~\Cref{eval_runtime}. 
In addition, we perform an algorithmic level comparison of communication scaling against standard sort-last compositing
methods adapted to support these jigsaw puzzlepiece data boundaries in~\cref{sec:eval_algorithmic_analysis}.

\subsection{Evaluation Datasets} \label{datasets}
\new{The synthetic volume dataset is created based on the worst-case scenario where the camera looks through a zigzagging boundary between different ranks' distributions. This situation
produces an extreme case of
segment overlap, where each rank generates four local volume slices of $512 \times 512 \times 2$, which are separated
by $n-1$ slices from other ranks in between.}
As seen in ~\Cref{fig:syn_illustration_plot}, the result is a scene where each rank's local data segments overlap all other
ranks in-depth, posing a severe challenge to traditional sort-last compositors.
\new{Without an order-independent transparency method, a traditional sort-last compositor will need to exchange and sort $4n$ segments (i.e., RGBA color fragments) per pixel to produce a correct image. Note that although the memory requirement can be arbitrarily large for a real-world dataset to store all rendered fragments, we deliberately construct this synthetic where the number of per-pixel fragments grows linearly with the number of nodes for the purpose of scaling pattern comparisons.}

\begin{figure}[t]
    \centering
    \begin{minipage}{.5\linewidth}
        \begin{subfigure}[t]{.95\linewidth}
            \includegraphics[width=\textwidth]{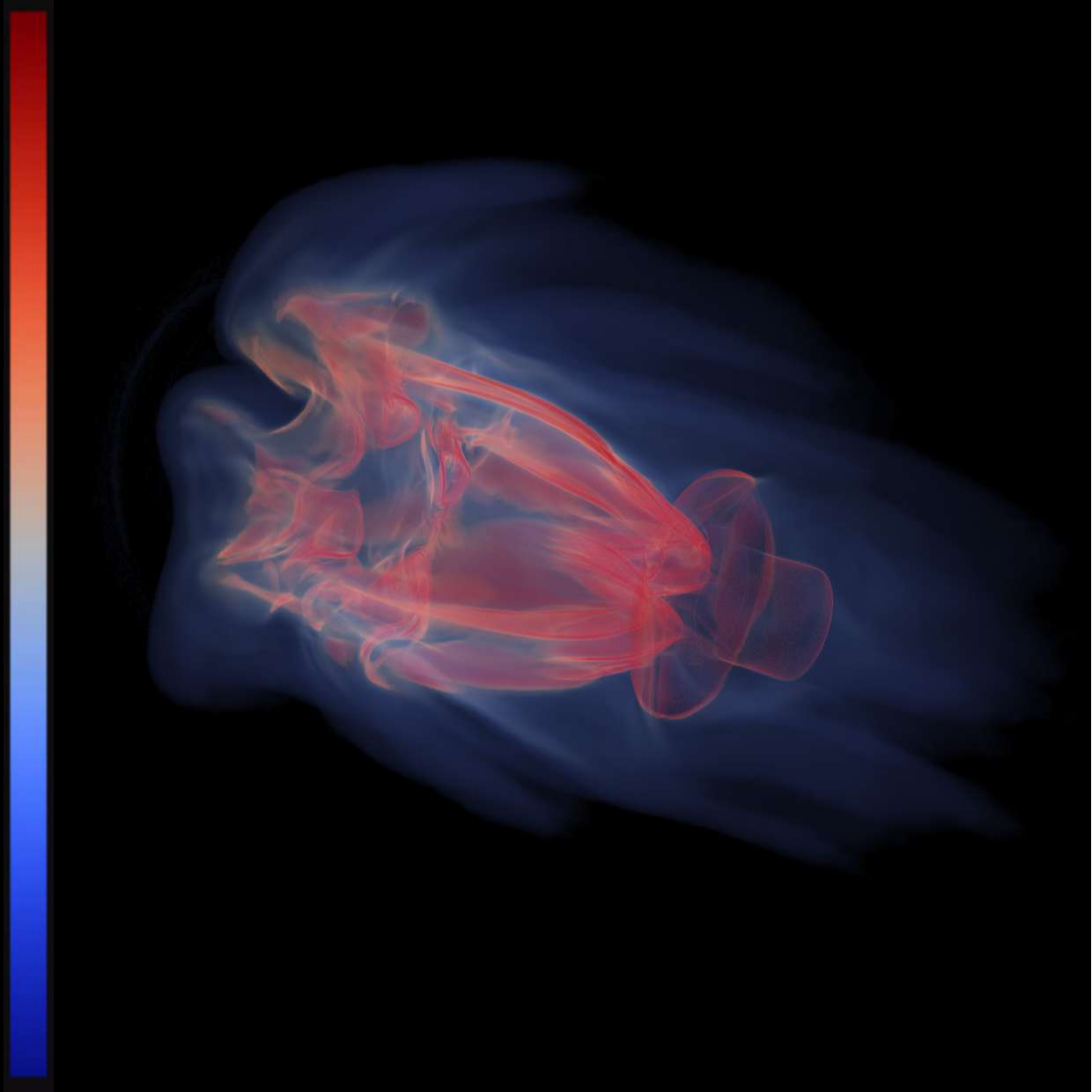}
            \vspace{-1em}
            \caption{The FUN3D Mars Lander}
        \end{subfigure}
    \end{minipage}
    \begin{minipage}{.45\linewidth}
        \begin{subfigure}[t]{.45\linewidth}
            \includegraphics[width=\textwidth]{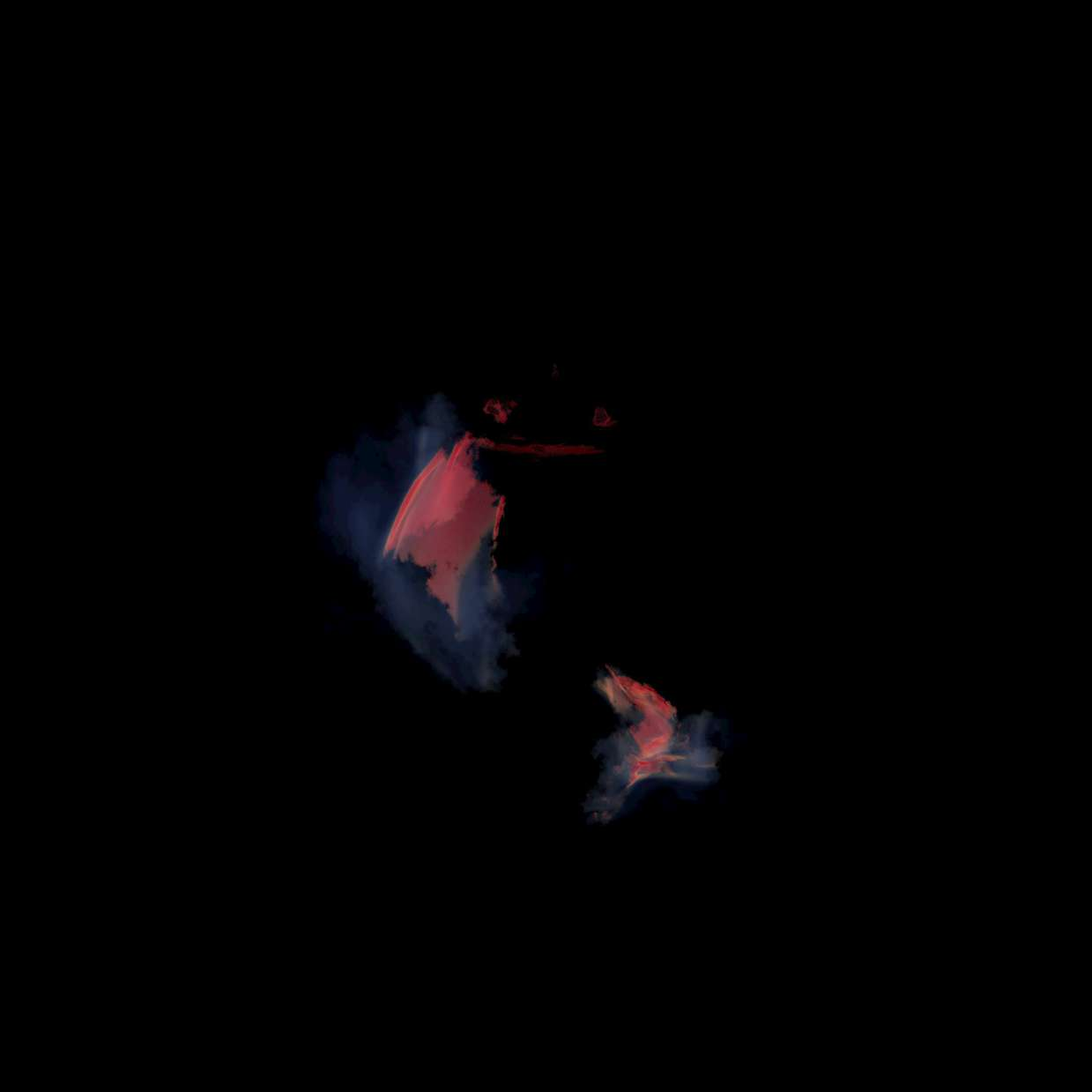}
            \vspace{-1em}
            \caption{n=4}
            \vspace{.25em}
        \end{subfigure} 
        \begin{subfigure}[t]{.45\linewidth}
            \includegraphics[width=\textwidth]{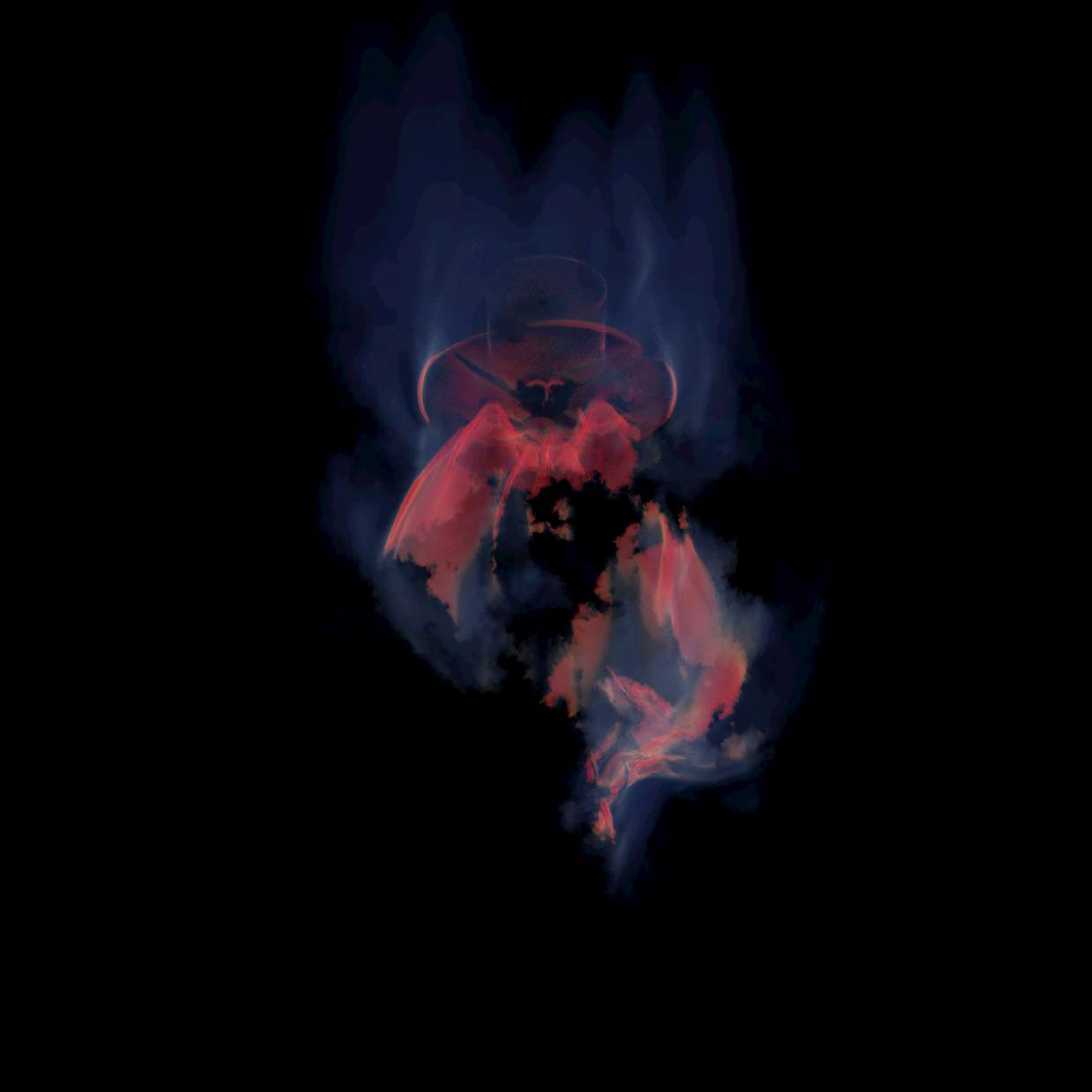}
            \vspace{-1em}
            \caption{n=16}
            \vspace{.25em}
        \end{subfigure}
        
        \begin{subfigure}[b]{.45\linewidth}
            \includegraphics[width=\textwidth]{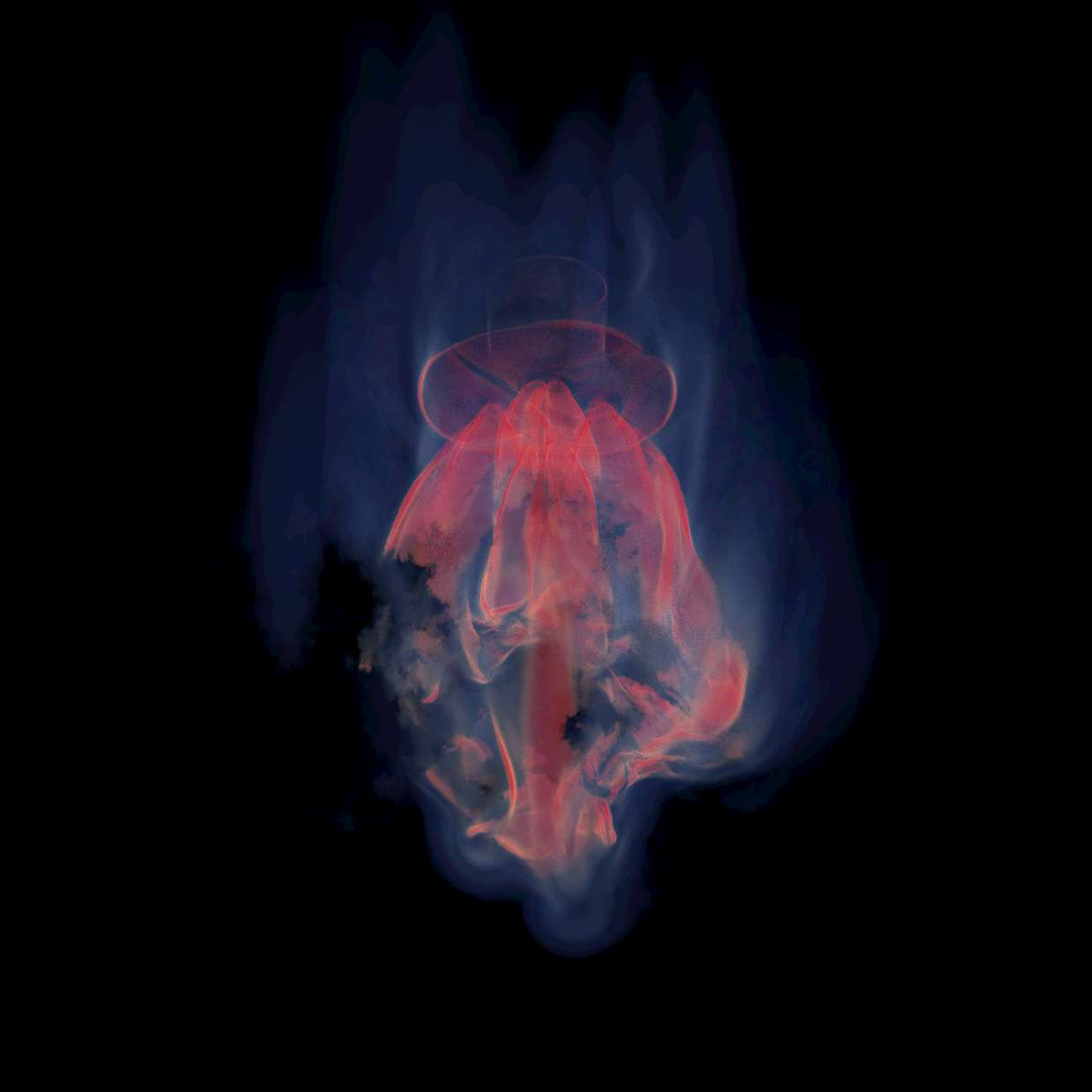}
            \vspace{-1em}
            \caption{n=32}
        \end{subfigure}
        \begin{subfigure}[b]{.45\linewidth}
            \includegraphics[width=\textwidth]{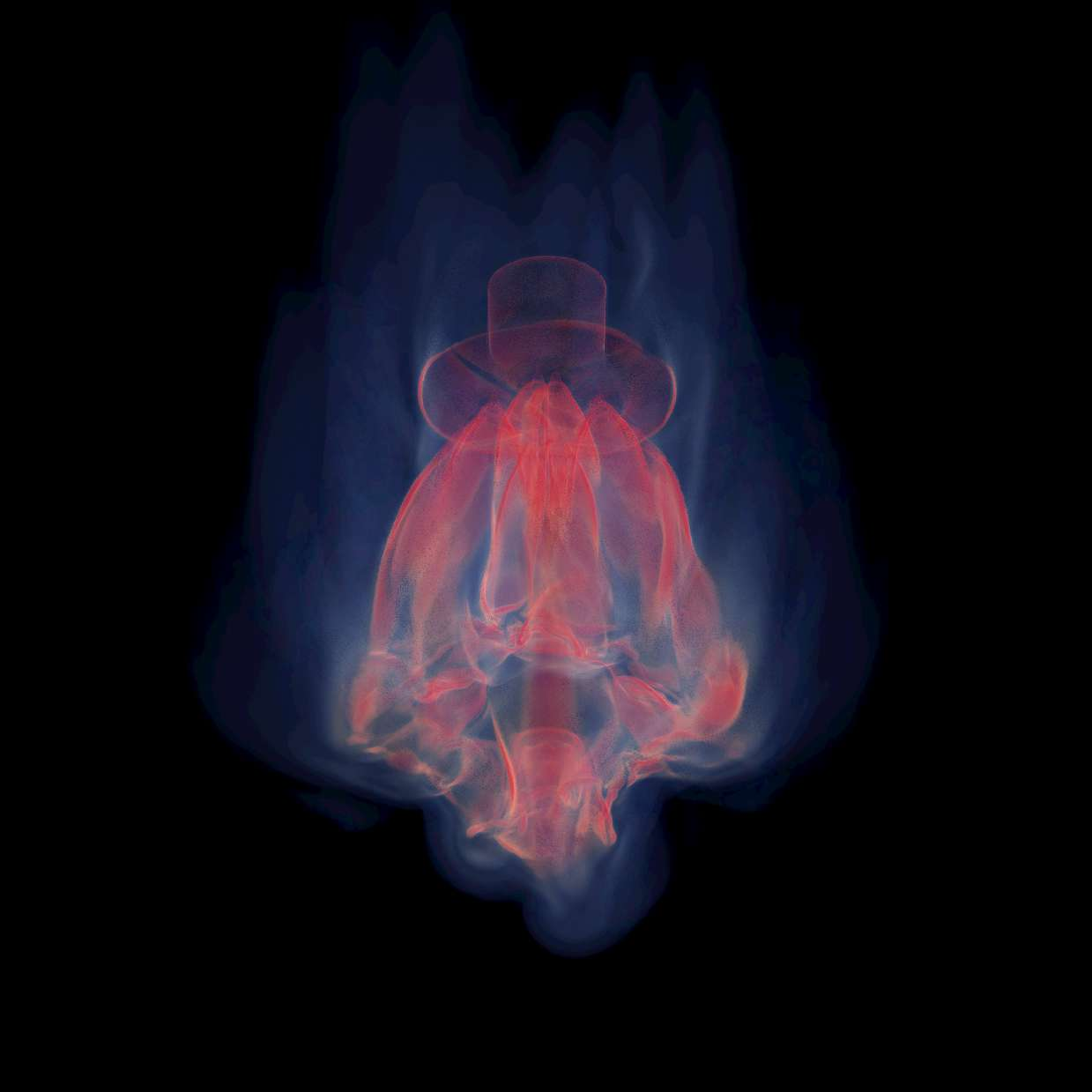}
            \vspace{-1em}
            \caption{n=72}
        \end{subfigure}         
    \end{minipage}
    \vspace{-1em}
    \caption{The FUN3D from partial domain to full domain. Resulting from a CFD simulation, the subdomains are spatially sparse and consist of nonuniform boundaries, as mentioned before in \Cref{fig:fun3d_sizes}. (b-e) The dataset rendered with 4, 16, 32, and 72 (all) subdomains combined. Even though the number of cells in each subvolume stays roughly the same, the shapes and spatial locations can be unpredictable.}
    \label{fig:fun3d_inprogress}
    \vspace{-1em}
\end{figure}

\begin{figure}
    \centering
    \begin{minipage}{.62\linewidth}
        \begin{subfigure}[t]{.95\linewidth}
            \includegraphics[width=\textwidth]{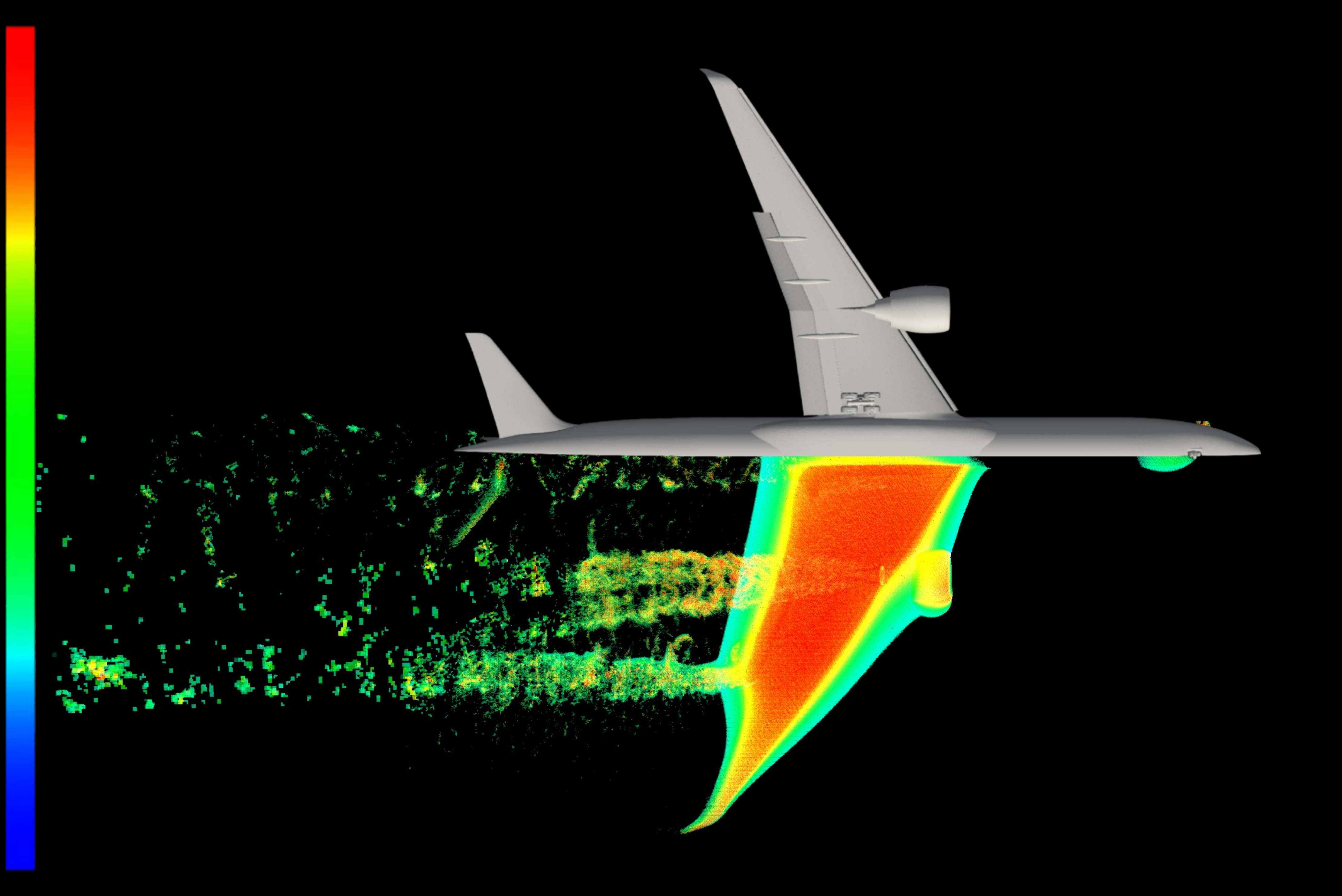}
            \caption{The Exajet }
        \end{subfigure}
    \end{minipage}
    \hspace{-0.5em}
    \begin{minipage}{.38\linewidth}
        \begin{subfigure}[t]{.45\linewidth}
            \includegraphics[width=\textwidth]{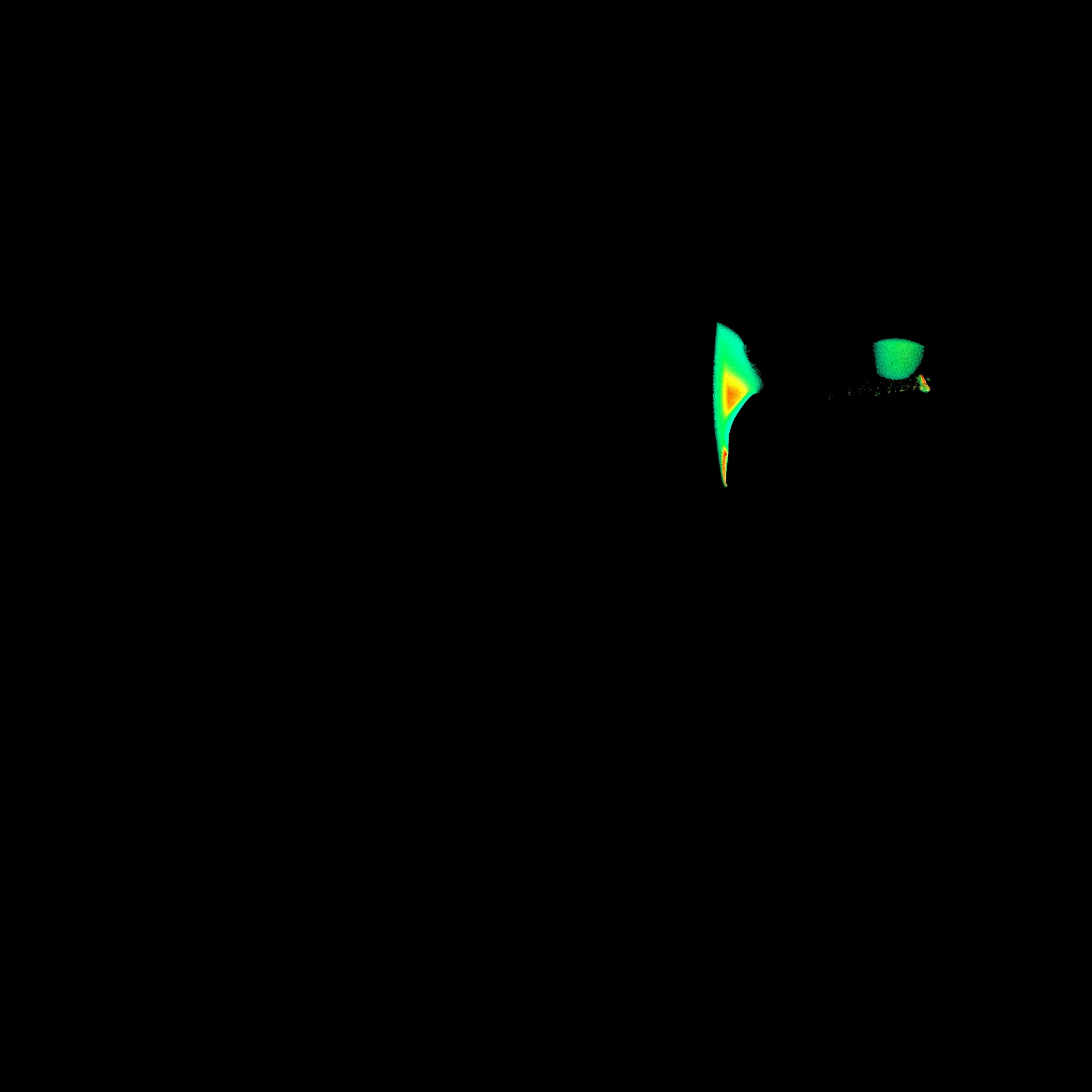}
            \vspace{-1em}
            \caption{n=16}
            \vspace{.25em}
        \end{subfigure} 
        \begin{subfigure}[t]{.45\linewidth}
            \includegraphics[width=\textwidth]{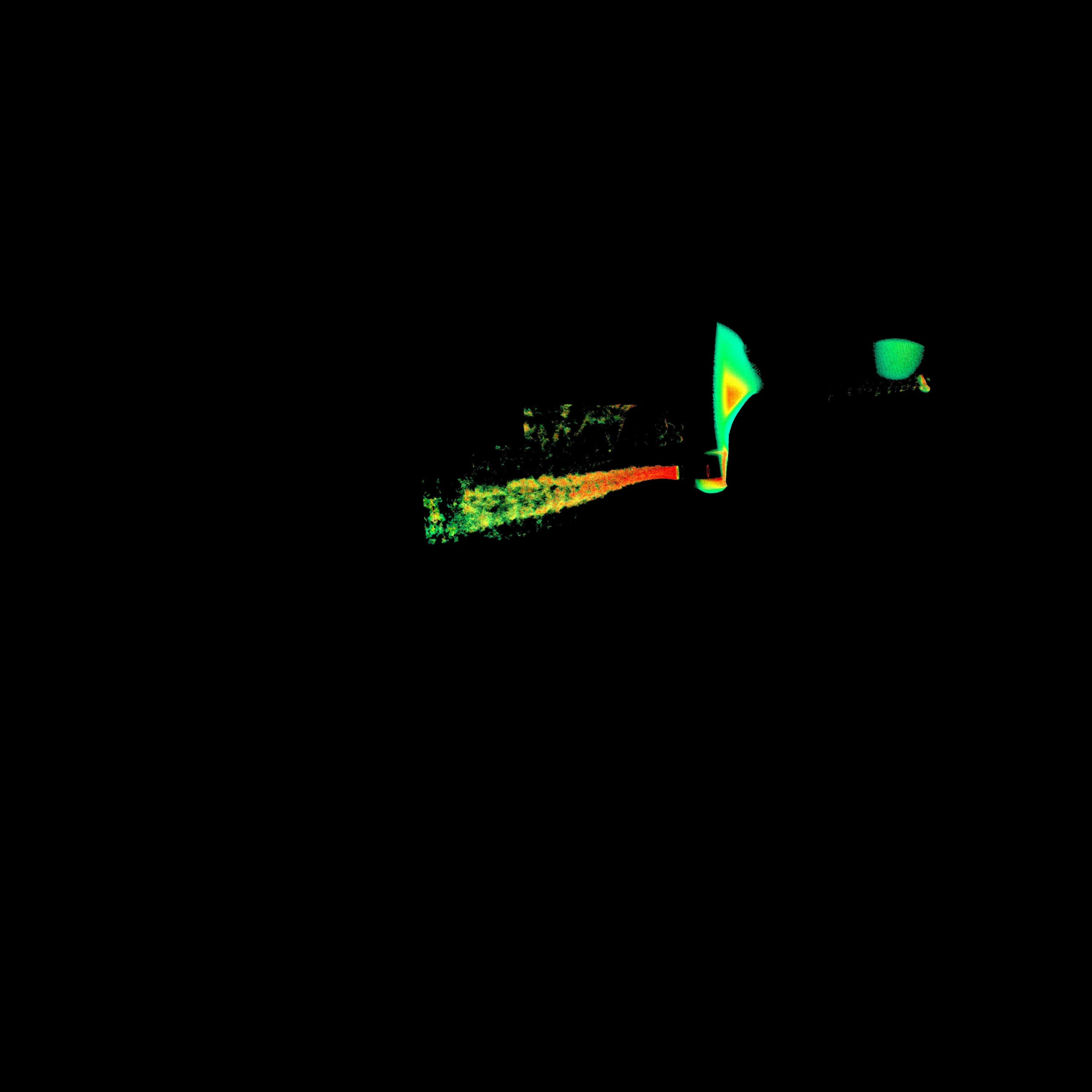}
            \vspace{-1em}
            \caption{n=64}
            \label{fig:jetn64}
            \vspace{.25em}
        \end{subfigure}
        
        \begin{subfigure}[b]{.45\linewidth}
            \includegraphics[width=\textwidth]{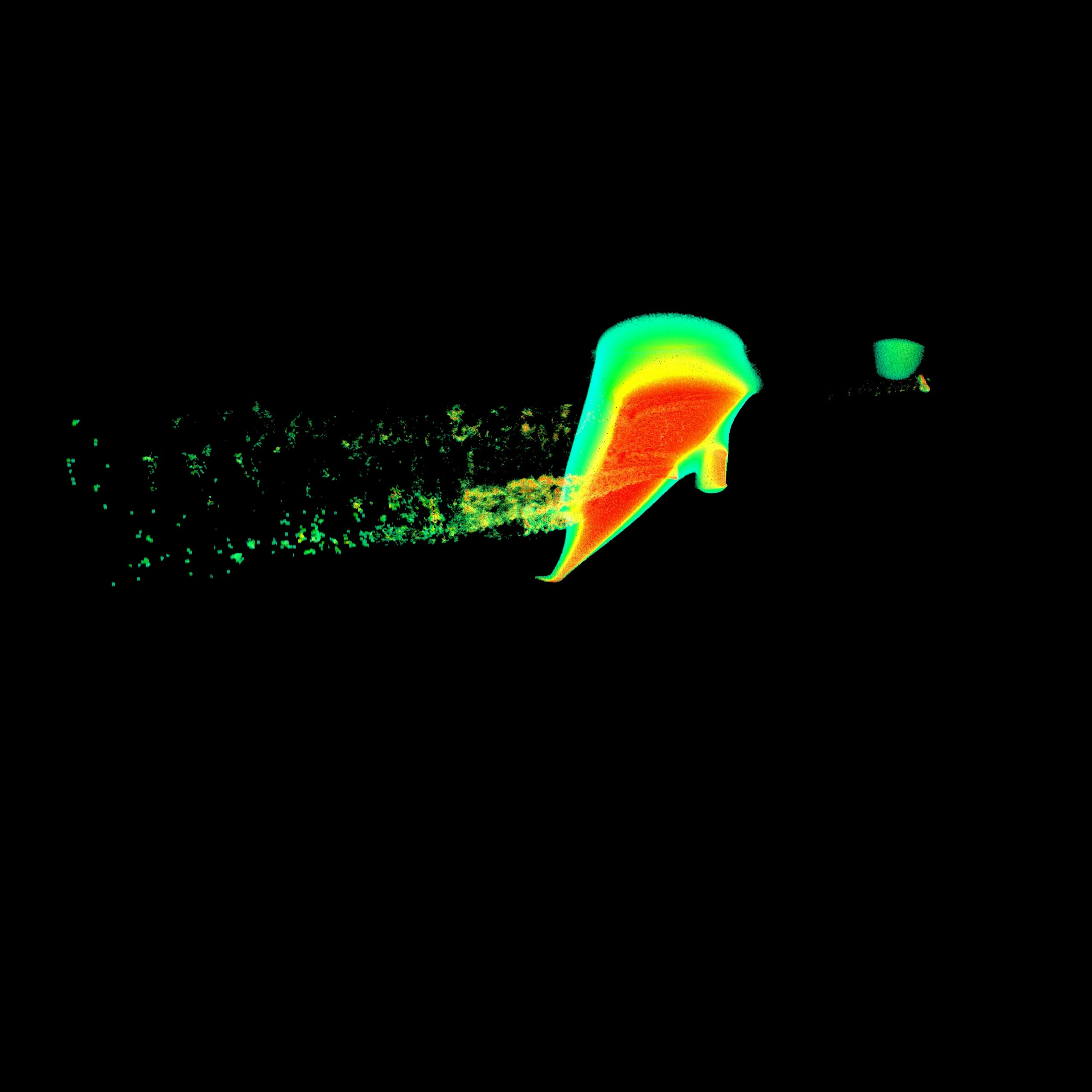}
            \vspace{-1em}
            \caption{n=96}
        \end{subfigure}
        \begin{subfigure}[b]{.45\linewidth}
            \includegraphics[width=\textwidth]{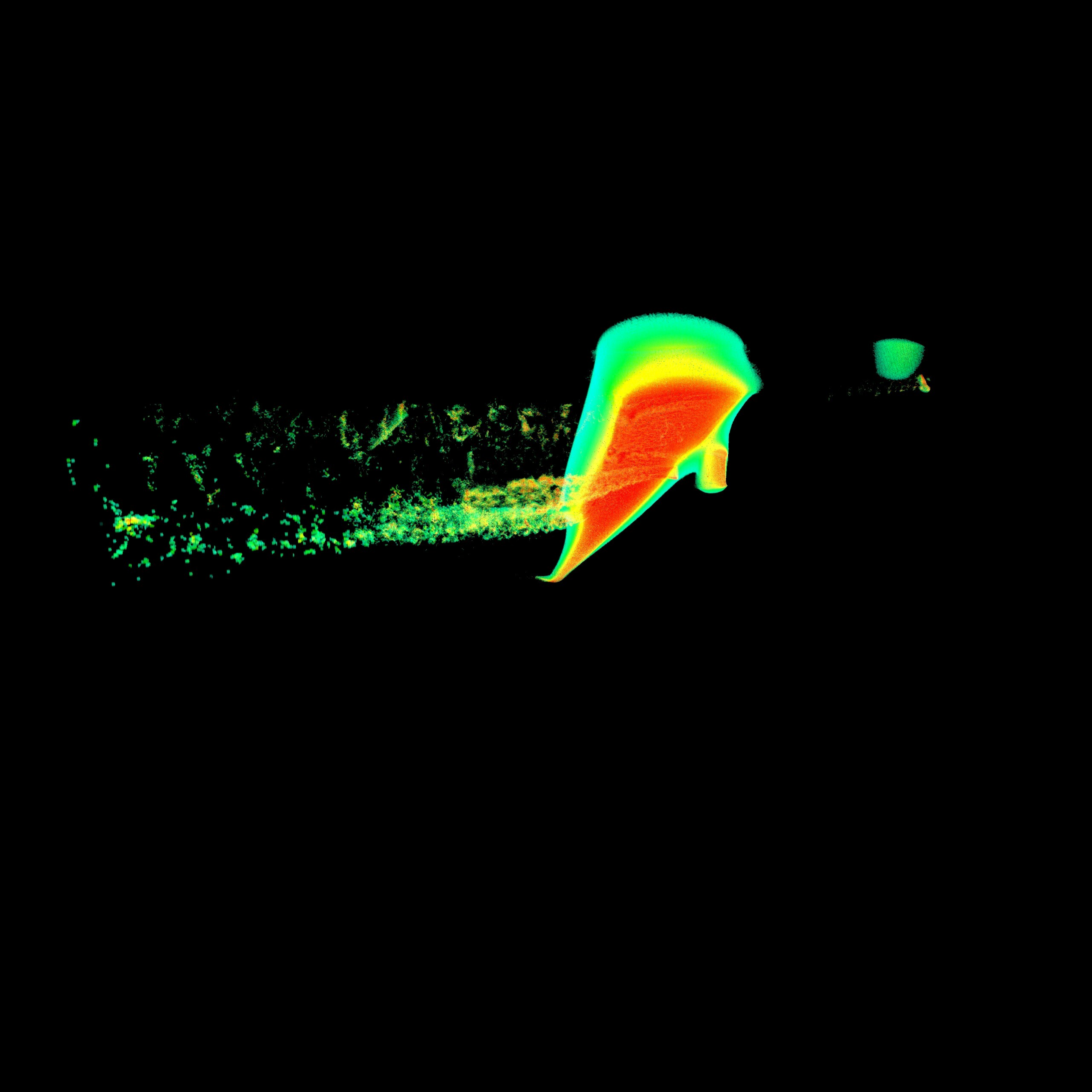}
            \vspace{-1em}
            \caption{n=128}
        \end{subfigure}         
    \end{minipage}
    \vspace{-1.5em}
    \caption{\label{fig:exajet_inprogress}%
    The Exajet. (a) A bottom-up view of the semi-span Exajet data with the half-plane model on the symmetric side of the axis. (b-e) show Exajet rendered with 16, 64, 96, and 128 (all) subdomains. The data are distributed by splitting the simulation into equal numbers of cells on each rank.
    The Exajet has a more irregular spatial distribution and complex boundaries than the FUN3D.}
    \vspace{-1em}
\end{figure}

We then examine the efficiency of our method on the driving dataset, the FUN3D Mars Lander. This dataset comprises spatially dispersed subvolumes from the simulation, resulting in an uneven distribution across nodes and unpredictable boundaries. Our performance analysis involves loading one subvolume per node up to all 72 subdomains.
The performance evaluation is conducted by varying the number of computational ranks. \Cref{fig:fun3d_inprogress} displays the sets of FUN3D subvolumes loaded at different rank counts for the benchmark.

Furthermore, to demonstrate the generality of our method, we include another real-world unstructured CDF dataset, NASA's Exajet, which describes a half-span model of a large civilian transport aircraft consisting of 656 million cell-centered cubic hexahedra. Different from the FUN3D dataset distribution, the Exajet mesh is generated as a single file outputted by the simulation solver. The original Exajet dataset comprises two components: the geometry file detailing the positions of all hexahedra and the value file encoding the velocity magnitude per cell.
To create the distributed version of this dataset, we segment the global files into 128 subfiles, each containing an equal number of elements in the original sequence.  As displayed in \Cref{fig:exajet_inprogress}, the resulting boundaries reflect a blend of the native simulation layouts, such as the clustering of turbulence structures from airplane engines, and artificial boundaries introduced by enforcing a consistent element count per rank. This nonuniform distribution intensifies the imbalance in segment overlap across compute nodes.

\begin{figure*}[]
    \vspace{2em}
     \centering
     \begin{subfigure}{0.24\textwidth}
         \centering
         \includegraphics[width=\textwidth]{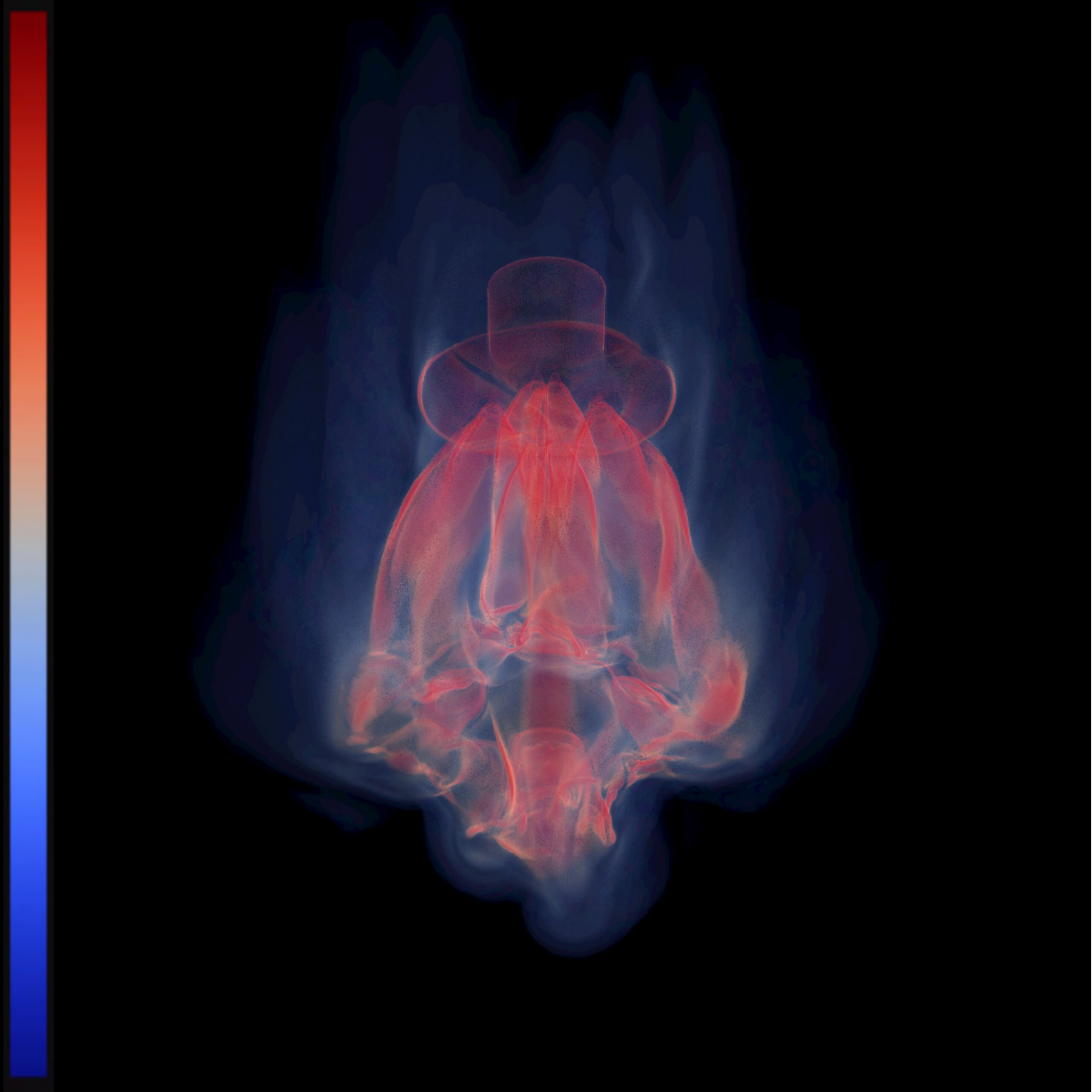}
         \caption{\label{fig:FUN3D_cmp_apc72}72 Nodes APC}
     \end{subfigure}
    \begin{subfigure}{0.24\textwidth}
         \centering
         \includegraphics[width=\textwidth]{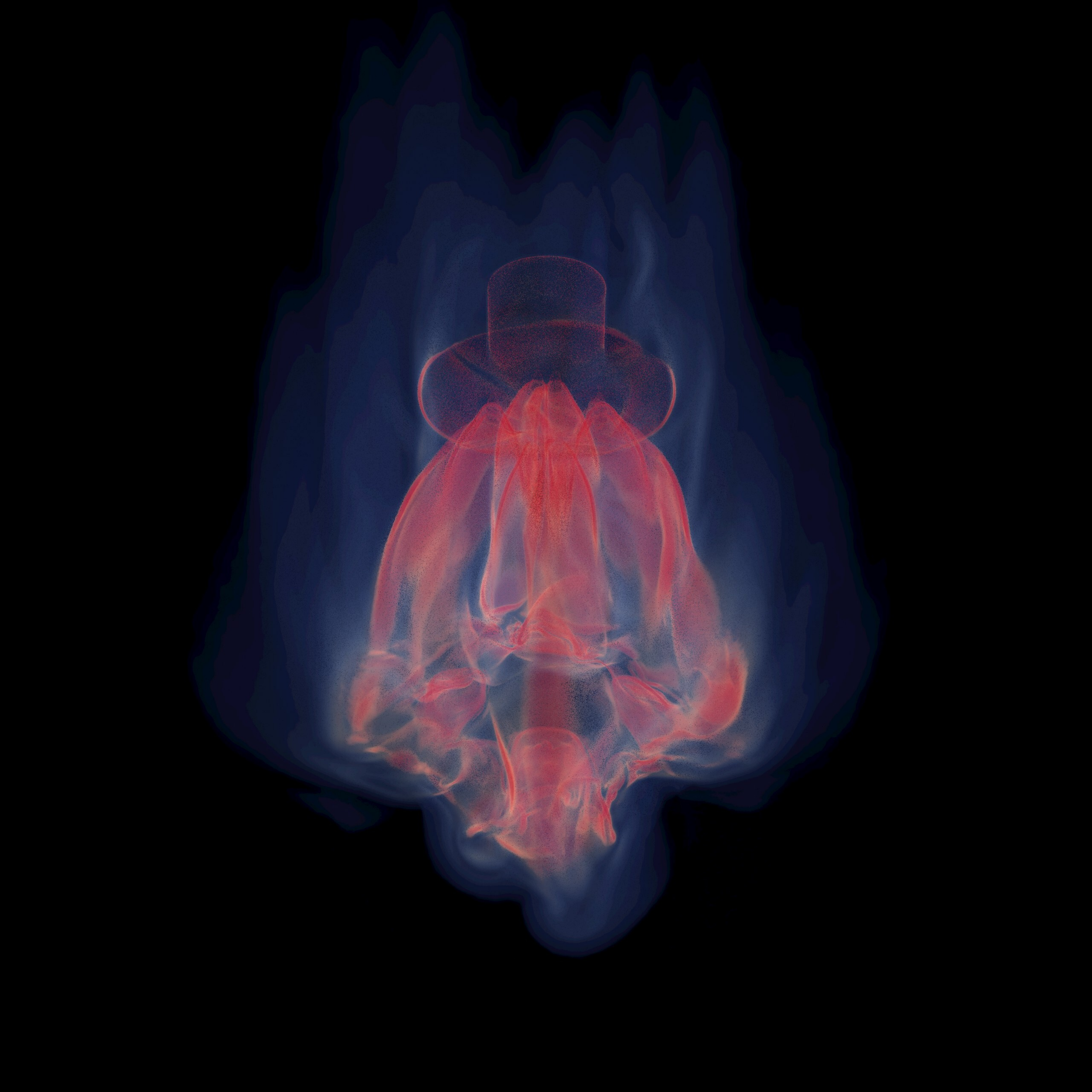}
         \caption{\label{fig:FUN3D_cmp_apc_all}Single-node MBOIT}
    \end{subfigure}
    \begin{subfigure}{0.24\textwidth}
         \centering
         \includegraphics[width=\textwidth]{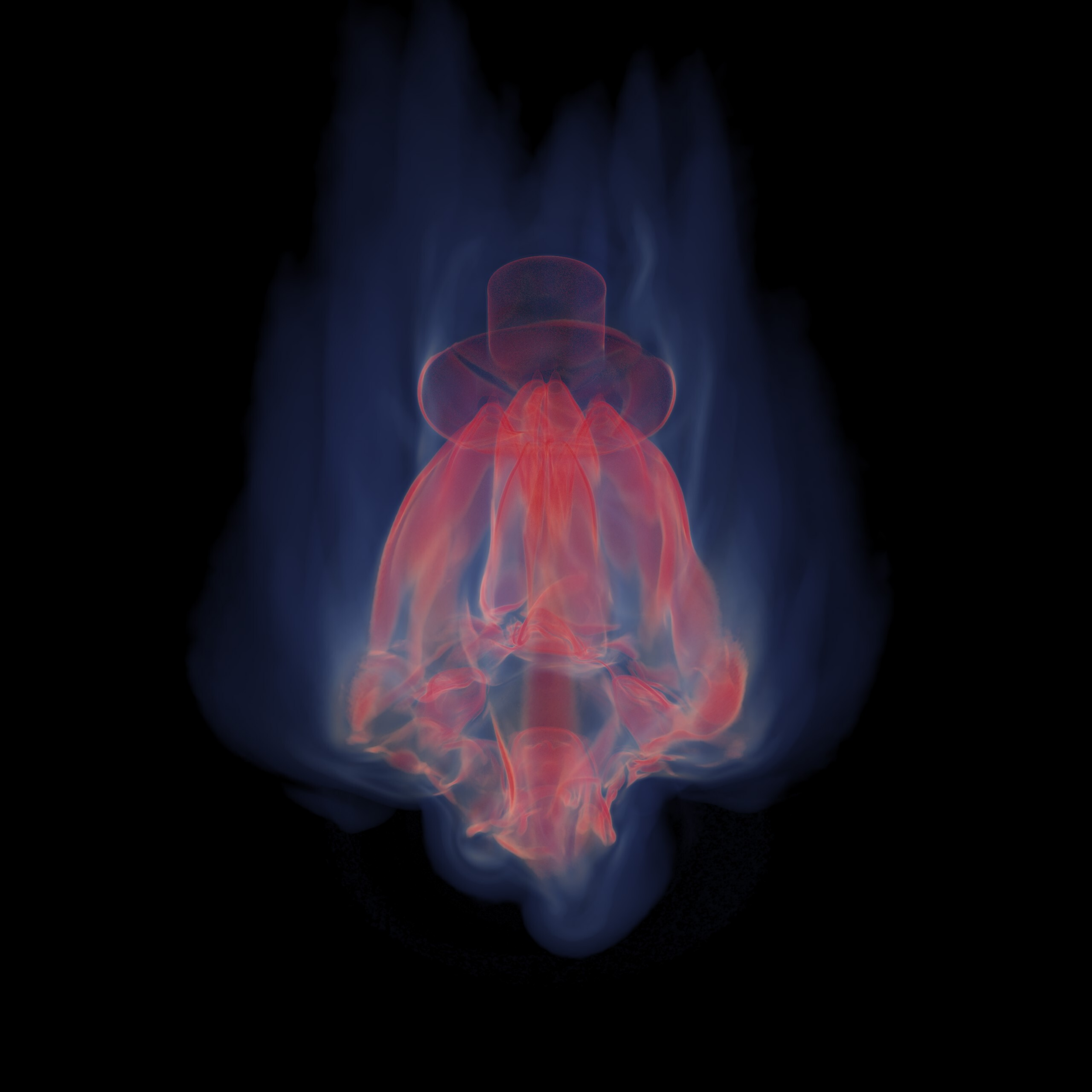}
         \caption{Single-node sort-last}
    \end{subfigure}
    \begin{subfigure}{0.24\textwidth}
         \centering
         \includegraphics[width=\textwidth]{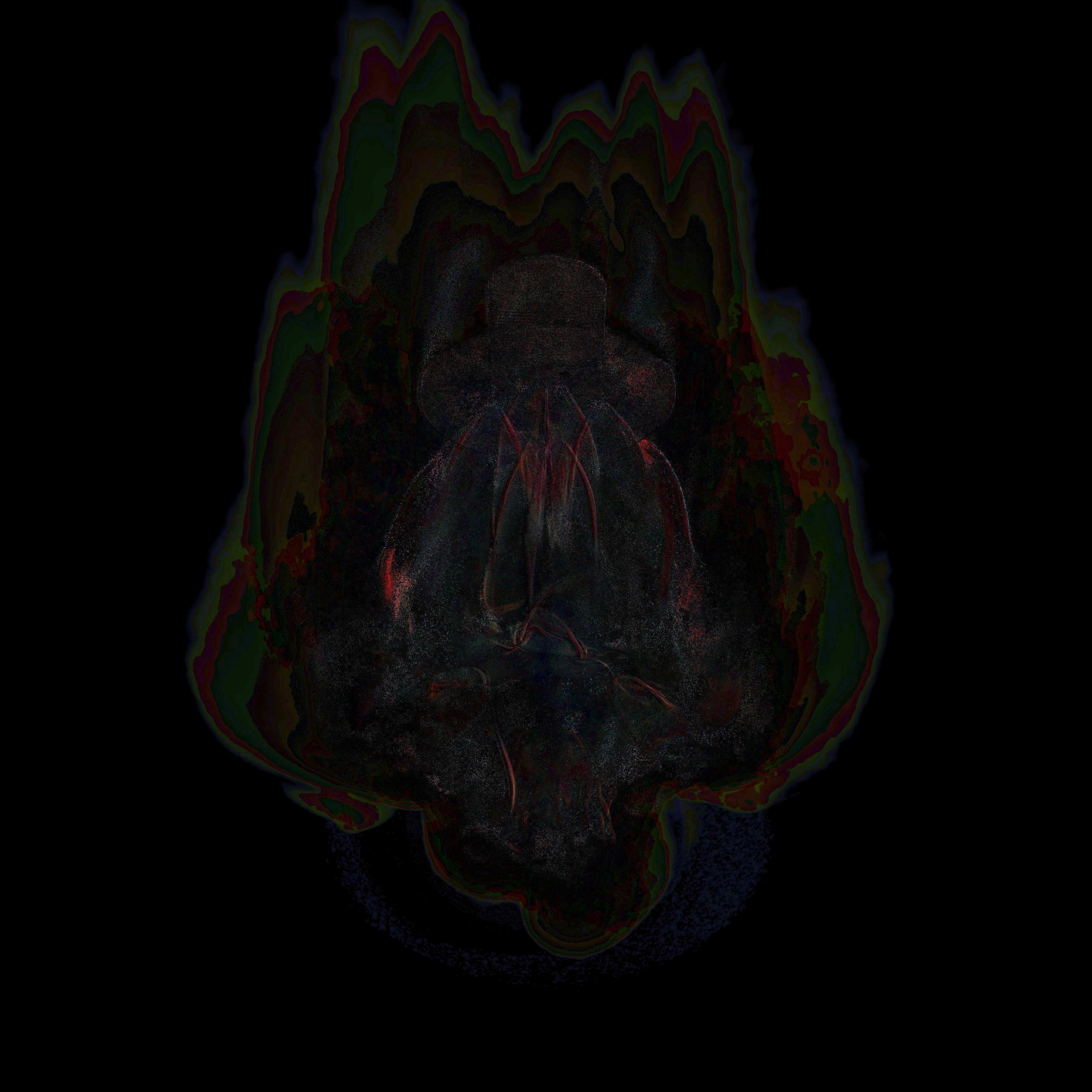}
         \caption{Diff (a) and (c), brightness $\times 3$}
    \end{subfigure}
    
    \begin{subfigure}{0.24\textwidth}
         \centering
         \includegraphics[width=\textwidth]{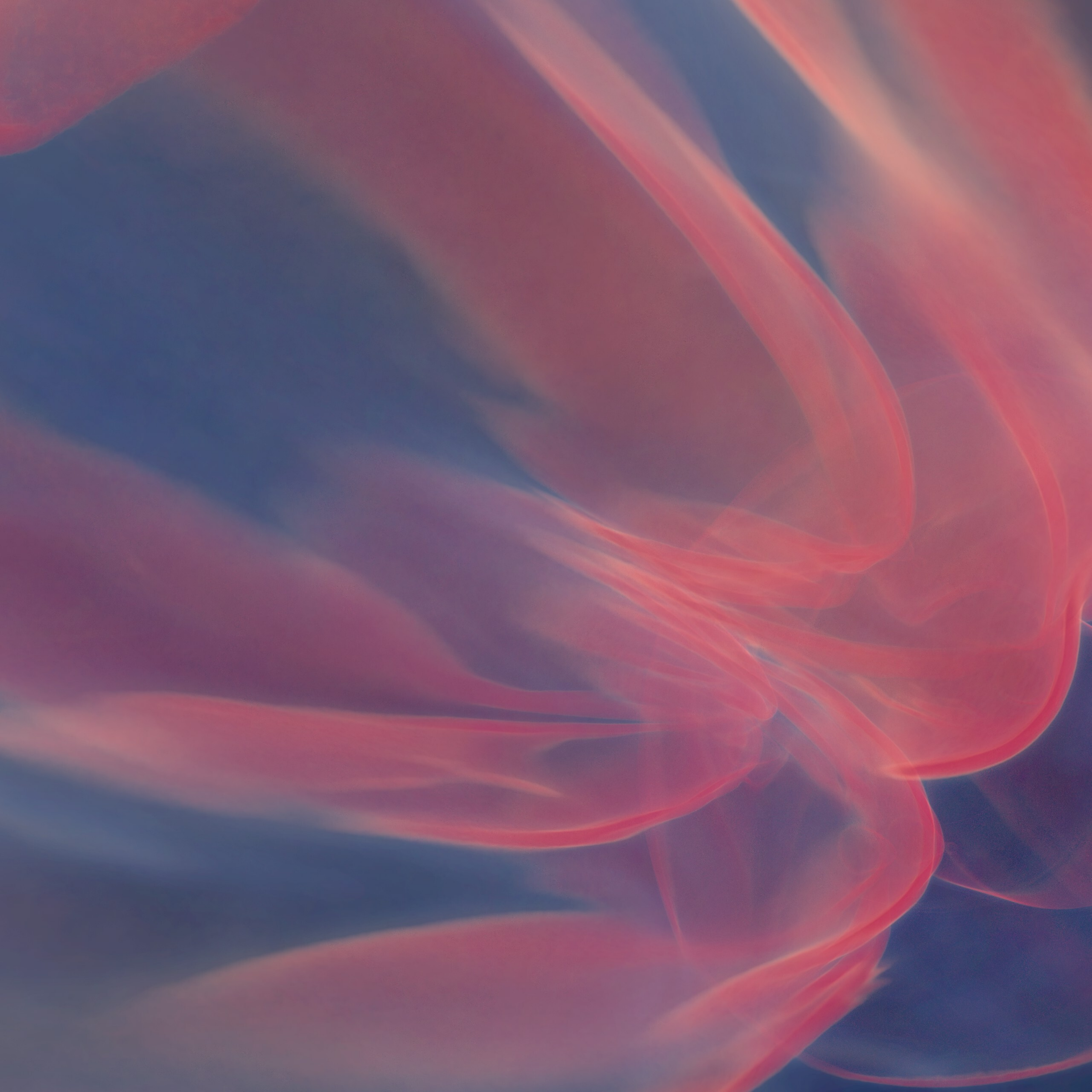}
         \caption{\label{fig:FUN3D_cmp20_apc72}72 Nodes APC}
    \end{subfigure}
    \begin{subfigure}{0.24\textwidth}
         \centering
         \includegraphics[width=\textwidth]{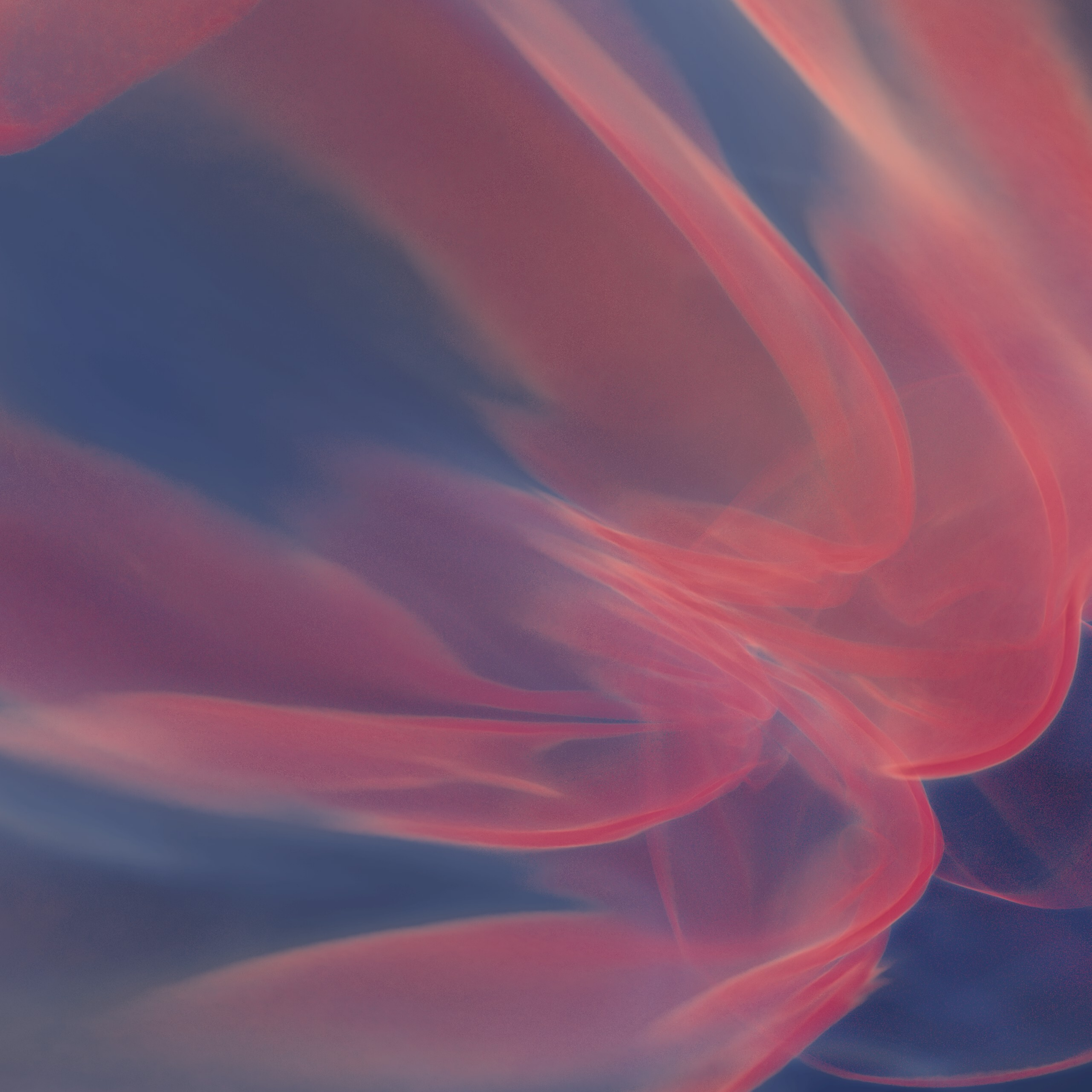}
         \caption{\label{fig:FUN3D_cmp20_apc_all}Single-node MBOIT}
    \end{subfigure}
    \begin{subfigure}{0.24\textwidth}
         \centering
         \includegraphics[width=\textwidth]{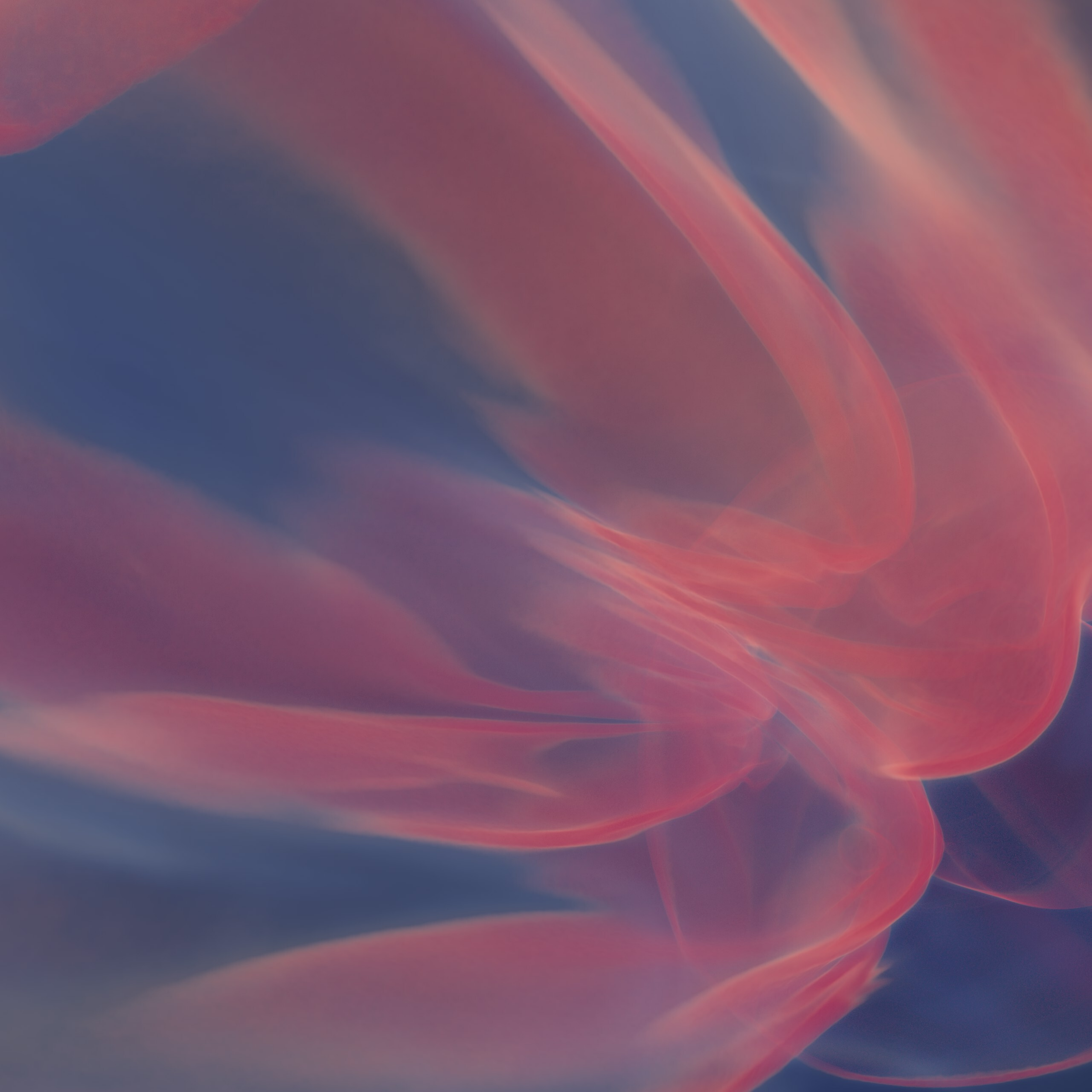}
         \caption{Single-node sort-last}
    \end{subfigure}
    \begin{subfigure}{0.24\textwidth}
         \centering
         \includegraphics[width=\textwidth]{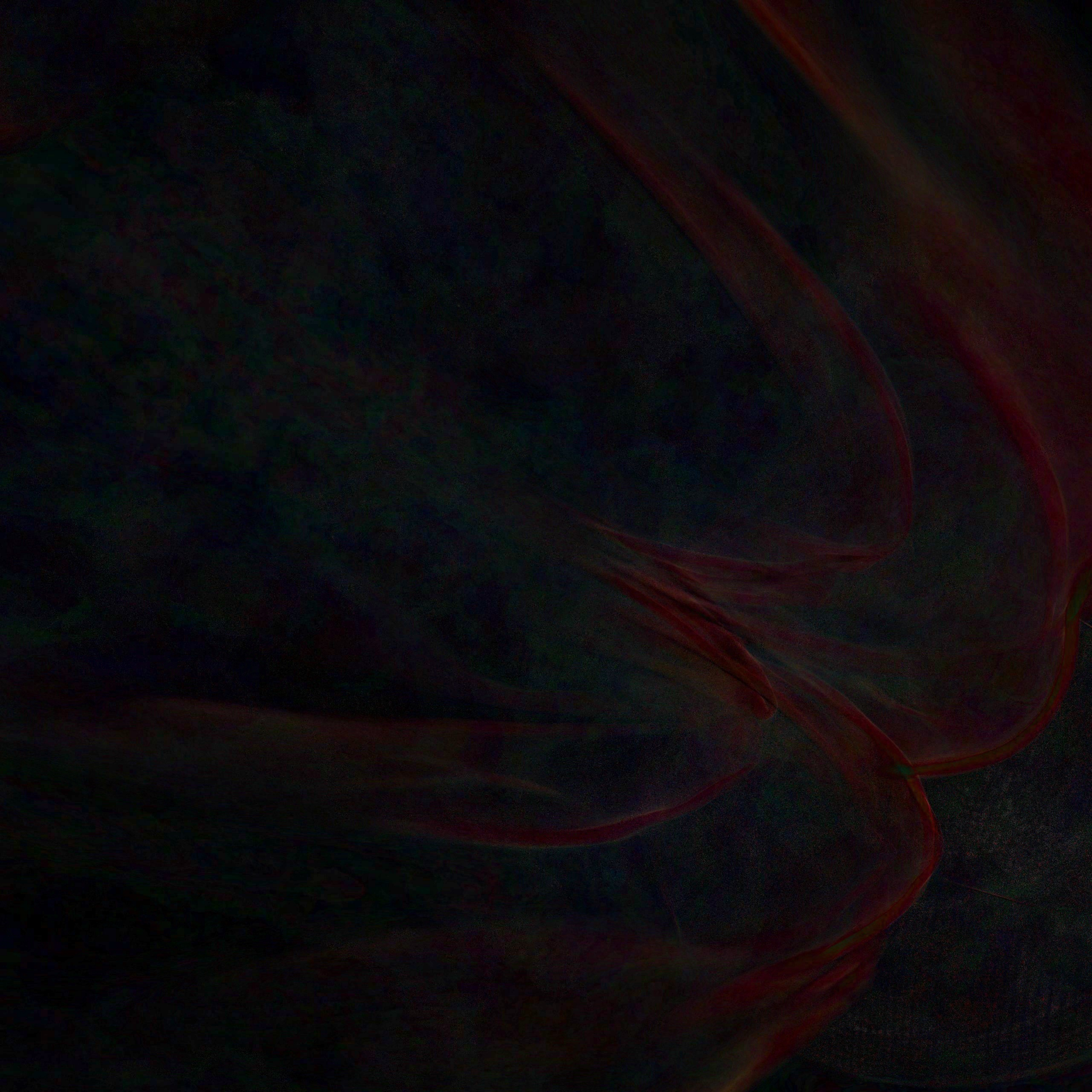}
         \caption{Diff (e) and (g), brightness $\times 3$}
    \end{subfigure}

    \centering
    \begin{subfigure}{0.24\textwidth}
         \centering
         \includegraphics[width=\textwidth]{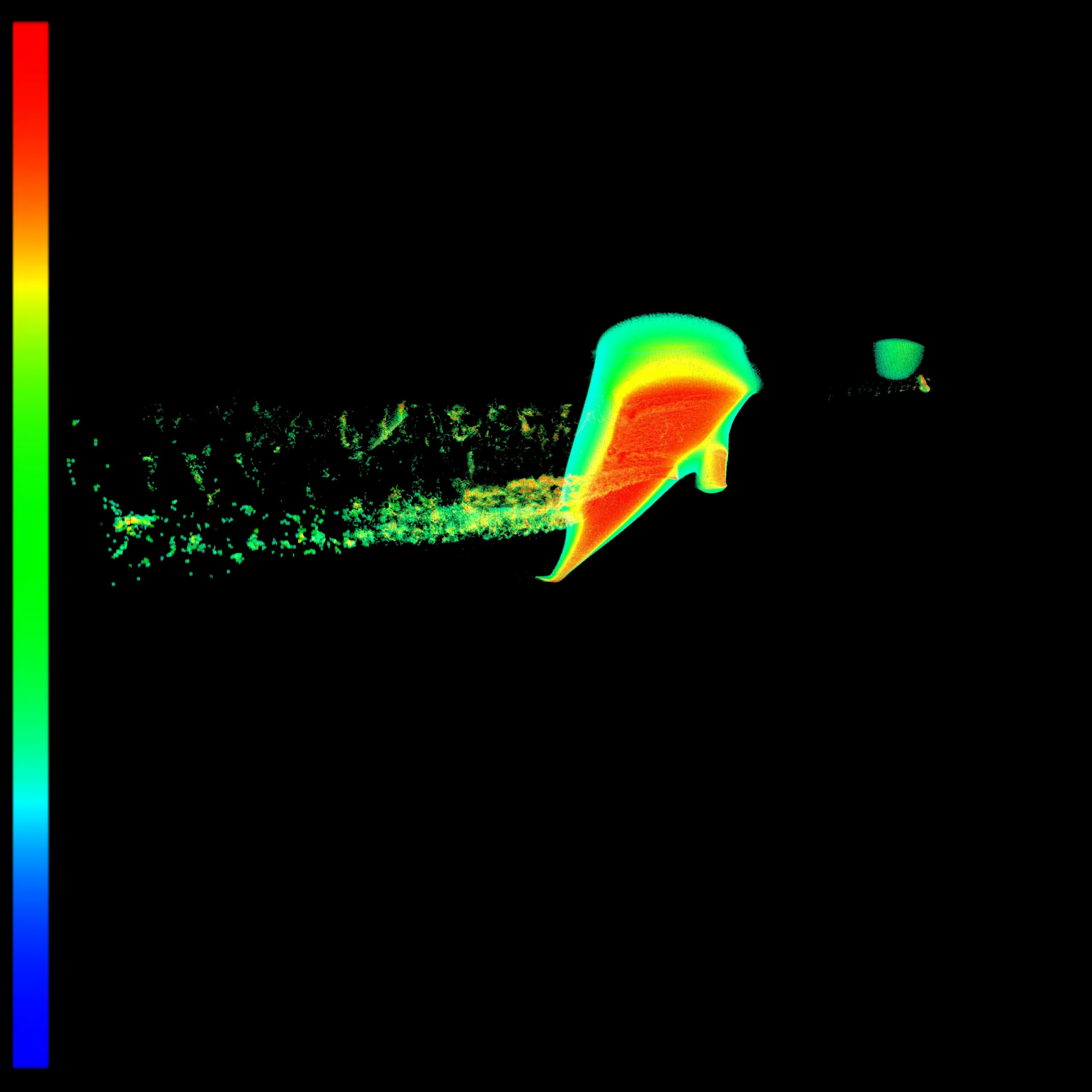}
         \caption{\label{fig:jet_cmp_apc72}128 Nodes APC}
    \end{subfigure}
    \begin{subfigure}{0.24\textwidth}
         \centering
         \includegraphics[width=\textwidth]{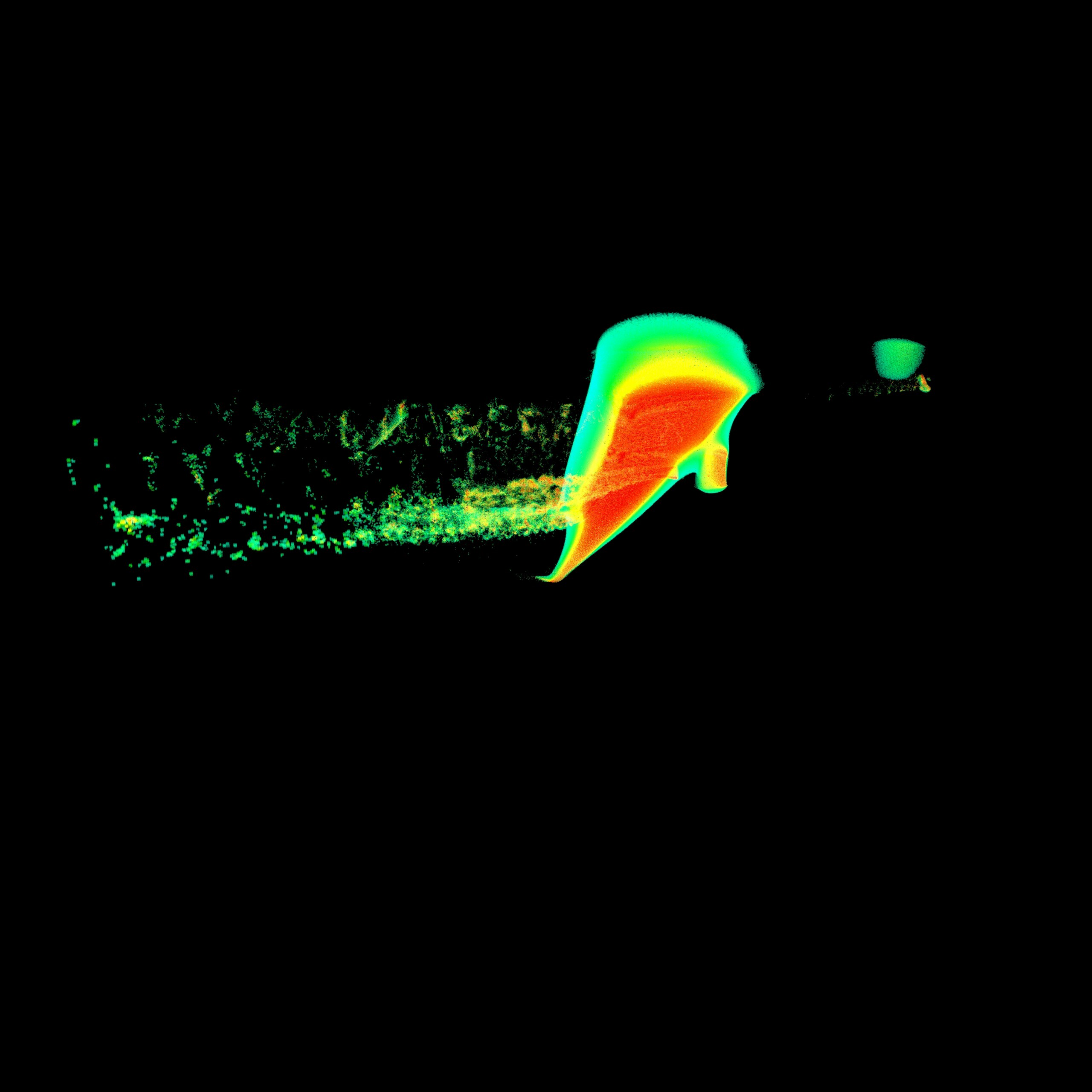}
         \caption{\label{fig:jet_cmp_apc_all}Single-node MBOIT}
    \end{subfigure}
    \begin{subfigure}{0.24\textwidth}
         \centering
         \includegraphics[width=\textwidth]{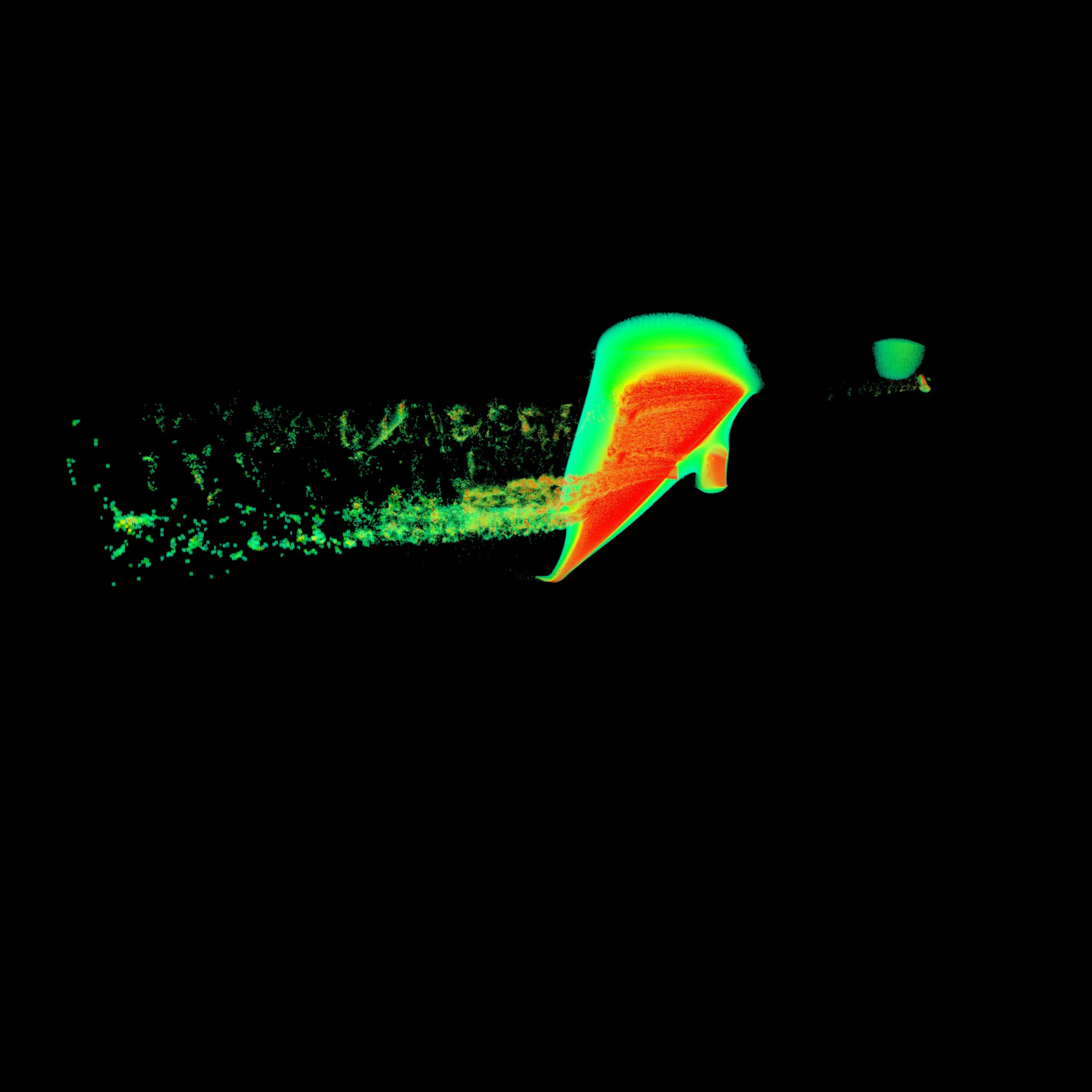}
         \caption{Single-node sort-last}
    \end{subfigure}
    \begin{subfigure}{0.24\textwidth}
         \centering
         \includegraphics[width=\textwidth]{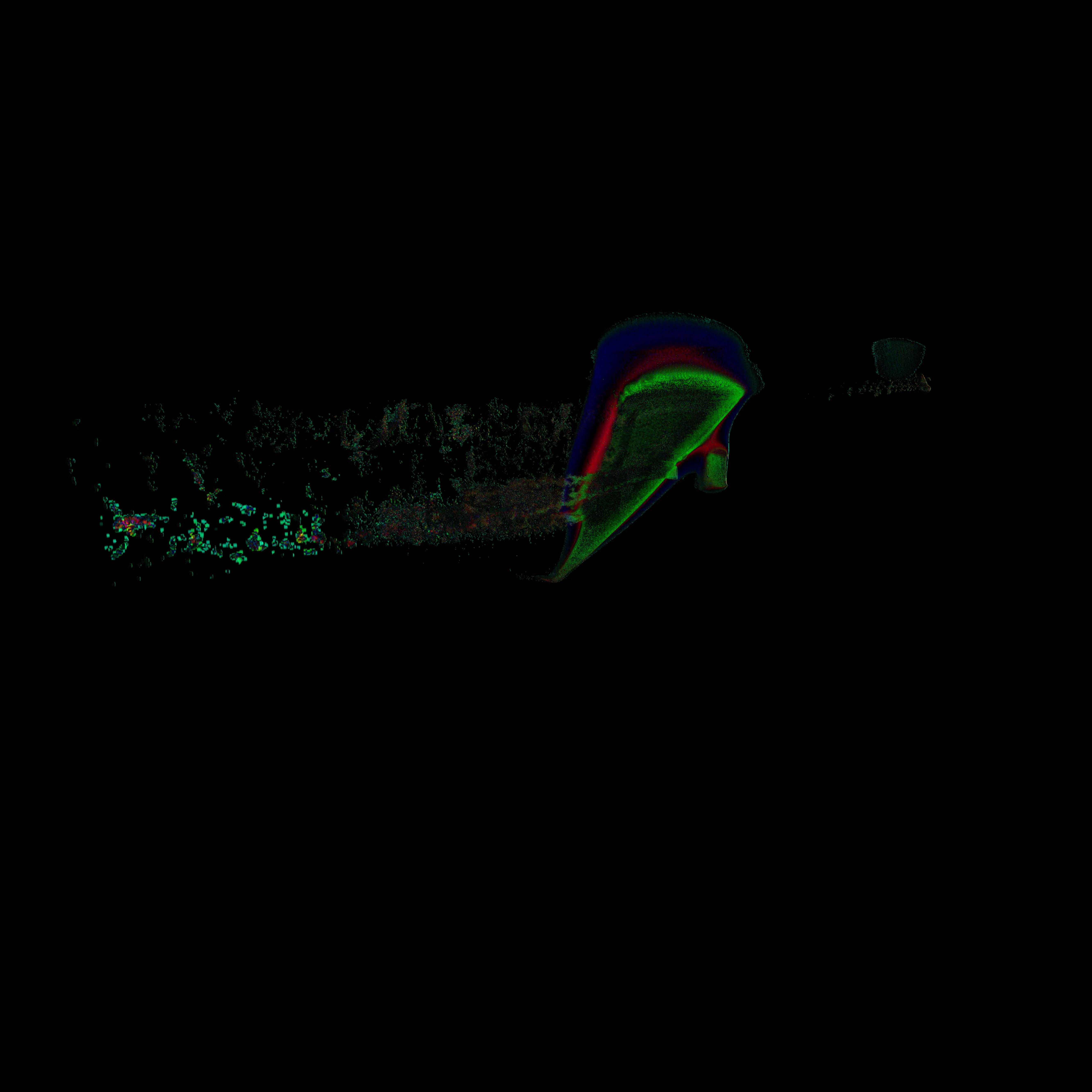}
         \caption{Diff (i) and (k)}
    \end{subfigure}
    
    \centering
    \begin{subfigure}{0.24\textwidth}
         \centering
         \includegraphics[width=\textwidth]{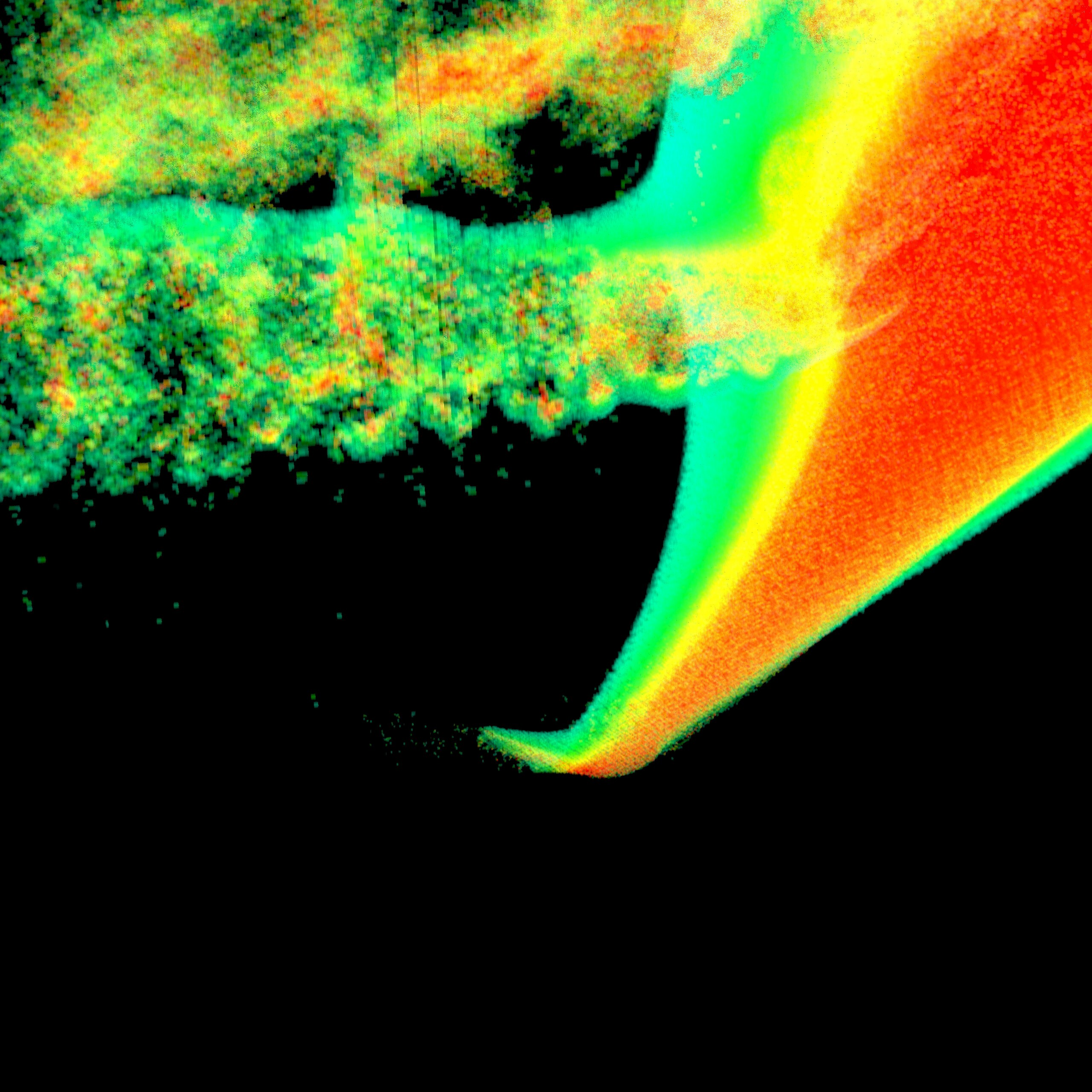}
         \caption{\label{fig:jet_near_cmp_apc72}128 Nodes APC}
    \end{subfigure}
    \begin{subfigure}{0.24\textwidth}
         \centering
         \includegraphics[width=\textwidth]{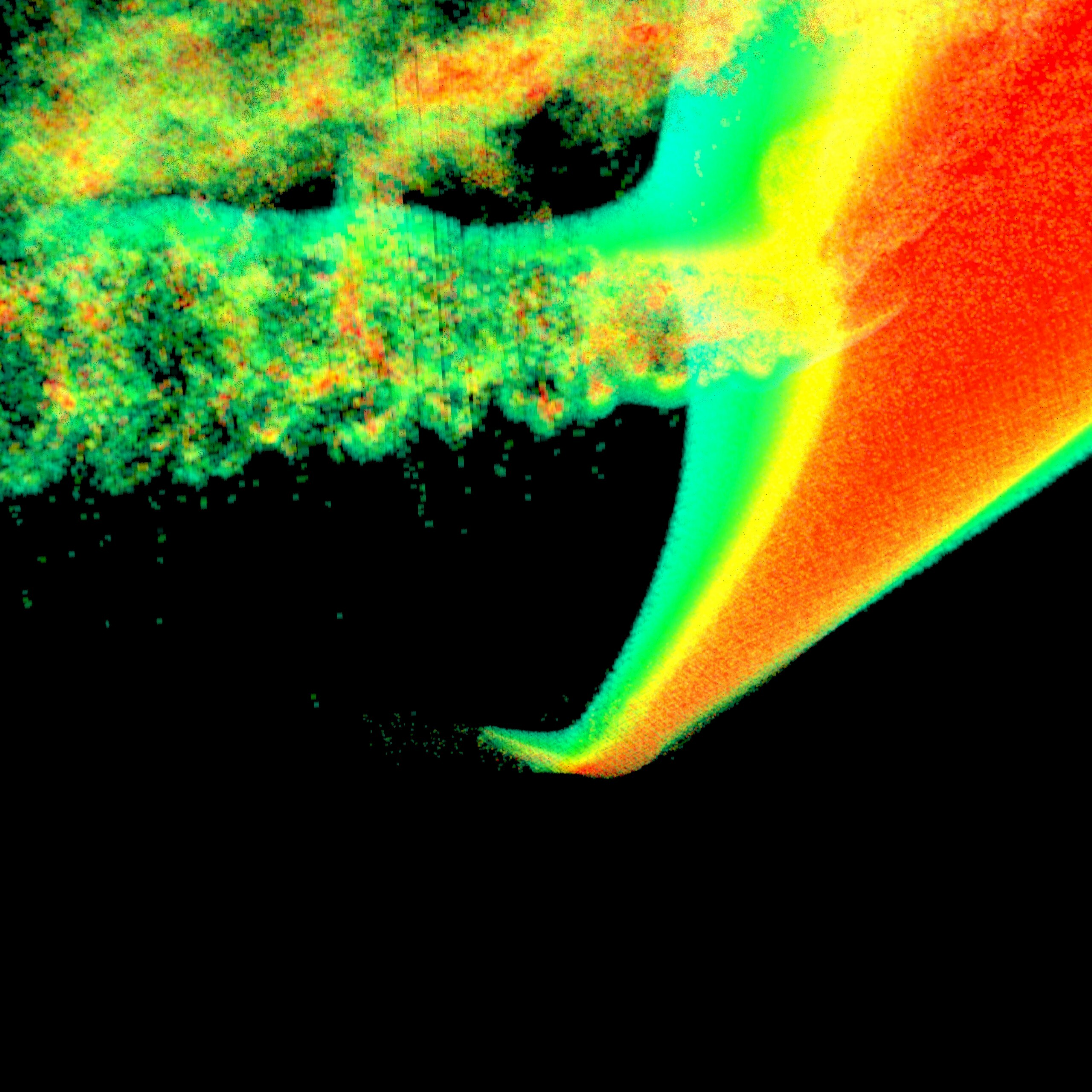}
         \caption{\label{fig:jet_near_cmp_apc_all}Single-node MBOIT}
    \end{subfigure}
    \begin{subfigure}{0.24\textwidth}
         \centering
         \includegraphics[width=\textwidth]{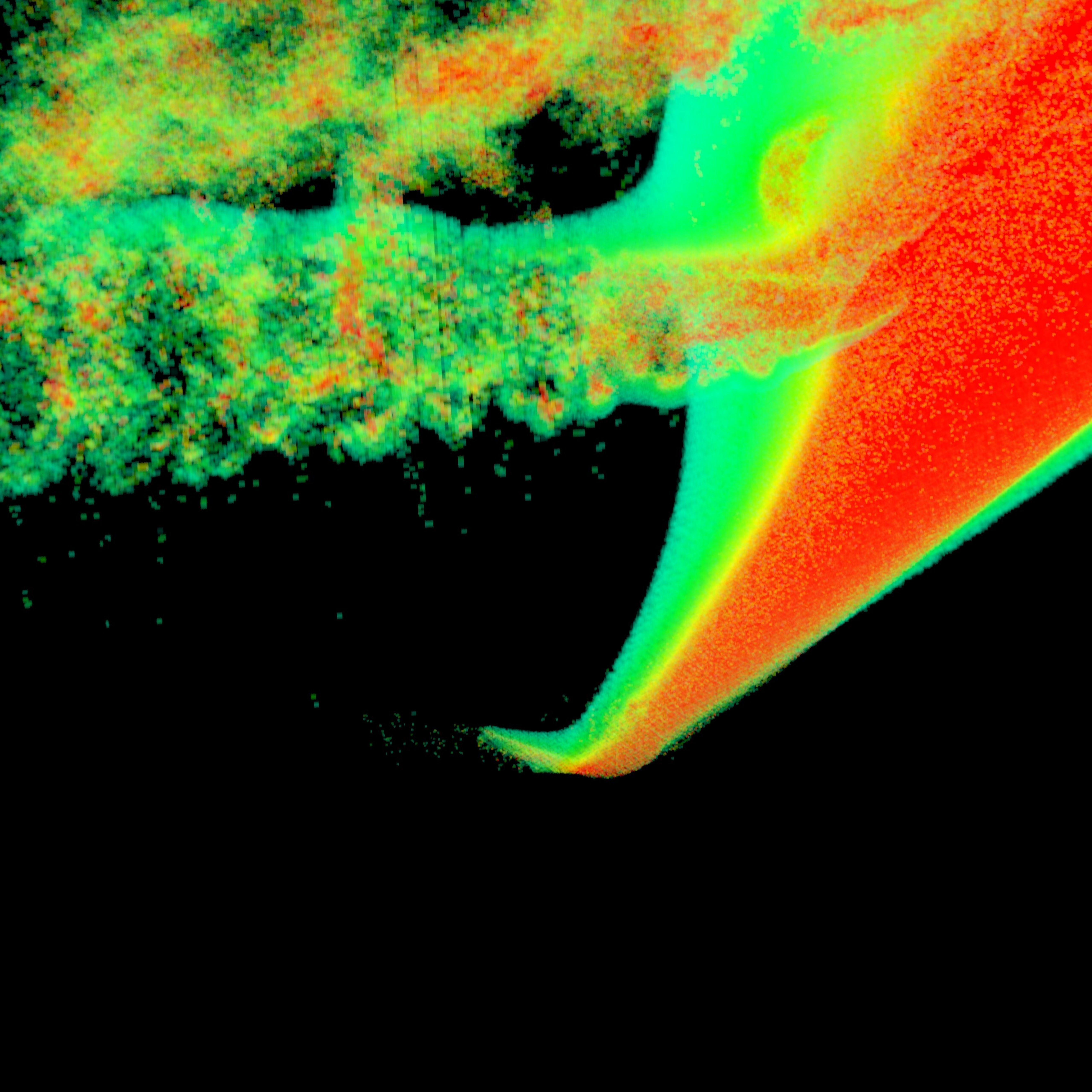}
         \caption{Single-node sort-last}
    \end{subfigure}
    \begin{subfigure}{0.24\textwidth}
         \centering
         \includegraphics[width=\textwidth]{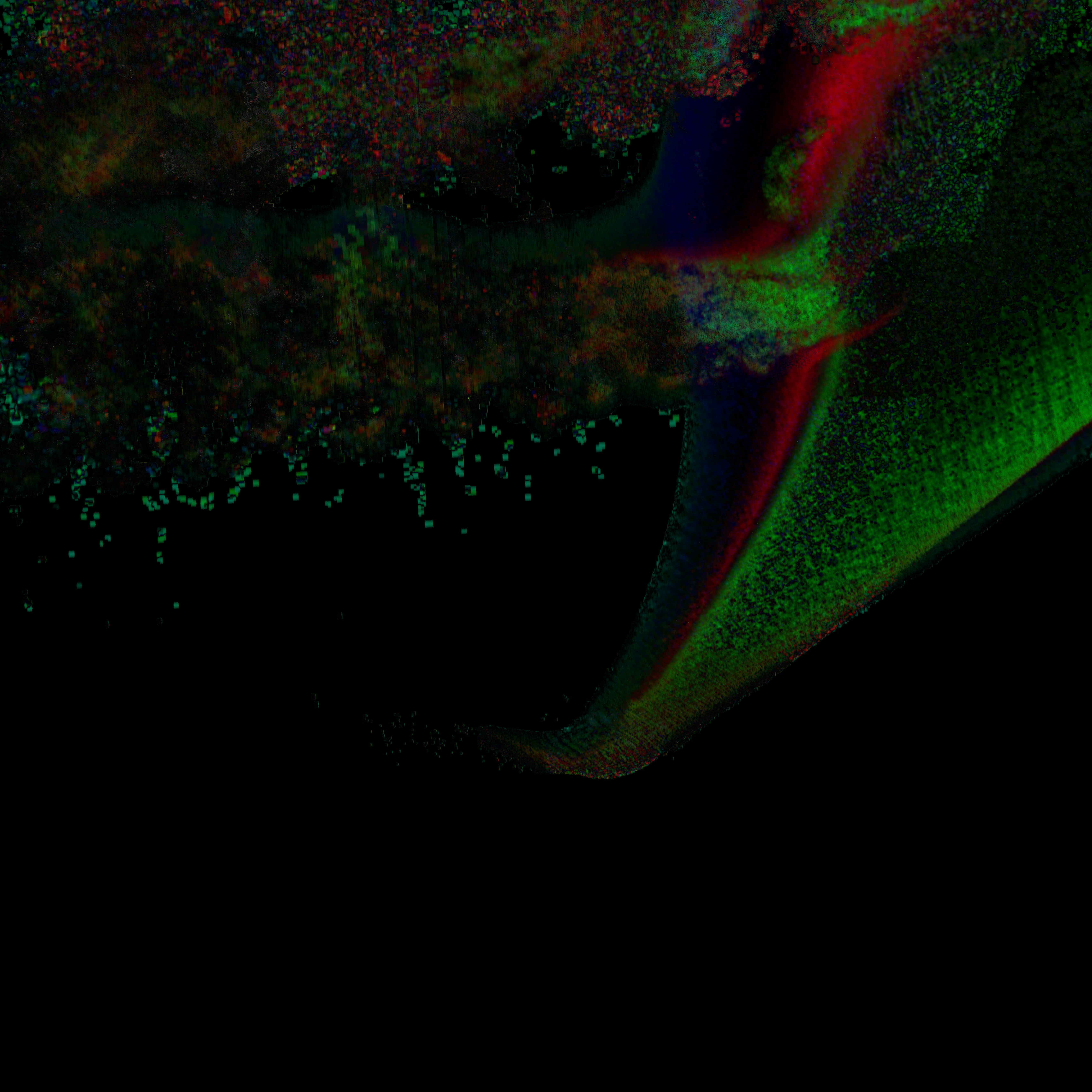}
         \caption{Diff (m) and (o)}
    \end{subfigure}
    
     \vspace{-1em}
     \caption{ \edit{Image comparison with real-world datasets. The first two rows show an overview and a closeup of the FUN3D dataset with distributed APC, single node MBOIT rendering, single node sort-last rendering and the difference image with brightness scaled by 3 for visibility. The following two rows show corresponding images for the Exajet. The FUN3D dataset is rendered with a cold-warm transfer function for its smooth overall structure. The Exajet dataset is rendered with a more opaque rainbow transfer function for a distinct view of the scattered turbulence elements.
     The respective image similarity measurements for the renderings
     \new{(a, e, i, m) vs. ground truth images (c, g, k, o) are
     (a) SSIM=0.88, MSE=19.86, PSNR=35.18 (e) SSIM=0.96, MSE=28.5, PSNR=33.6045 (i) SSIM=0.97, MSE=63.4532, PSNR=30.1402 (m) SSIM=0.82, MSE=315.405, PSNR=23.1761.}
     We can see that by faithfully representing the MBOIT rendering results, APC enables smooth volume rendering while preserving precise high-frequency details, and the rendered images exhibit a close resemblance to sort-last results with the color differences more noticeable at higher-density regions.}
     } 
     \label{fig:cmp_image_quality}
\end{figure*}

\subsection{Image Quality}\label{img_qual}
To evaluate the rendering quality of APC, we perform an image quality comparison with the single-node MBOIT and single-node sort-last results. As the entire dataset will not fit into a regular Frontera compute node, for the single node rendering each dataset is aggregated to run on one large memory Intel Xeon Platinum 8280M node with 2.1TB of Optane memory. We compare the images with the two real-world datasets in \Cref{fig:cmp_image_quality}. 

Due to its more uniform element distribution, the FUN3D dataset is rendered in a smoother and more transparent manner to showcase the internal composition. In contrast, the Exajet has more significant cell size variation, becoming extremely dense near the fuselage and wing, but sparse further from the object. Thus, this dataset is rendered in a more opaque setting for higher visibility of the smaller features. Note that we are able to capture the thin structures close to the landing gear (\Cref{fig:FUN3D_cmp_apc72}) and the air turbulence at the bottom of the wings (\Cref{fig:jet_cmp_apc72}) with correct object occlusions.

\edit{The APC rendered images showcase the capabilities of our method in ensuring smooth volume rendering while capturing finer grain details within AMR meshes, producing high-quality transmittance approximation similar to sort-last's results. We further note that both APC and single-node MBOIT produce identical images for both datasets, demonstrating that APC does not introduce artifacts to the MBOIT computation and that it is able to provide high-fidelity details in various rendering settings.} 

\edit{The inherent image difference between the MBOIT method and sort-last is visually more prominent at higher density, transparent regions like the Exajet wings areas, where a slight overestimation in the transmittance curve accumulates faster by including more samples before termination. As a result, a near-opaque pixel rendered from a tight cluster of transparent elements tends to appear brighter in the final image. Nevertheless, the approximation results still resemble the sort-last images in that the overall color difference is small, and they succeed in preserving important occlusion clues to avoid inaccurate depth perception. Furthermore, since some accuracy loss is expected from using an approximation method, an overall minor overestimation configuration is preferable in practice as an underestimated transmittance curve could lead to structural changes such as surface loss.}

\subsection{Compositing Performance} \label{eval_runtime}

As we are primarily interested in supporting in situ rendering of large-scale datasets,
we focus our performance evaluation on weak-scaling benchmarks for the synthetic sandwich stress test case (\Cref{fig:sandwich,fig:syn_illustration_plot}), and the incremental subdomain loading benchmarks for the FUN3D dataset (\Cref{fig:fun3d_inprogress}) and the Exajet dataset (\Cref{fig:exajet_inprogress}).

For the synthetic data, we record performance over a four-camera position orbit (illustrated on the synthetic data in~\Cref{fig:syn_illustration_plot}); on
the FUN3D data, we measure performance over a five-camera position orbit performed at three distances
from the dataset; and we use a three-camera position orbit on the Exajet dataset.
We further note that the performance of traditional sort-last compositors is affected by the camera position 
as it dictates the projected area of each rank's local data and the sorting. The compositing steps in our pipeline
are viewpoint oblivious since only two reductions over the image are needed.

\begin{figure}[ht!]
    \centering  
    \begin{subfigure}{0.45\textwidth}
    \centering 
    \includegraphics[width=\textwidth]{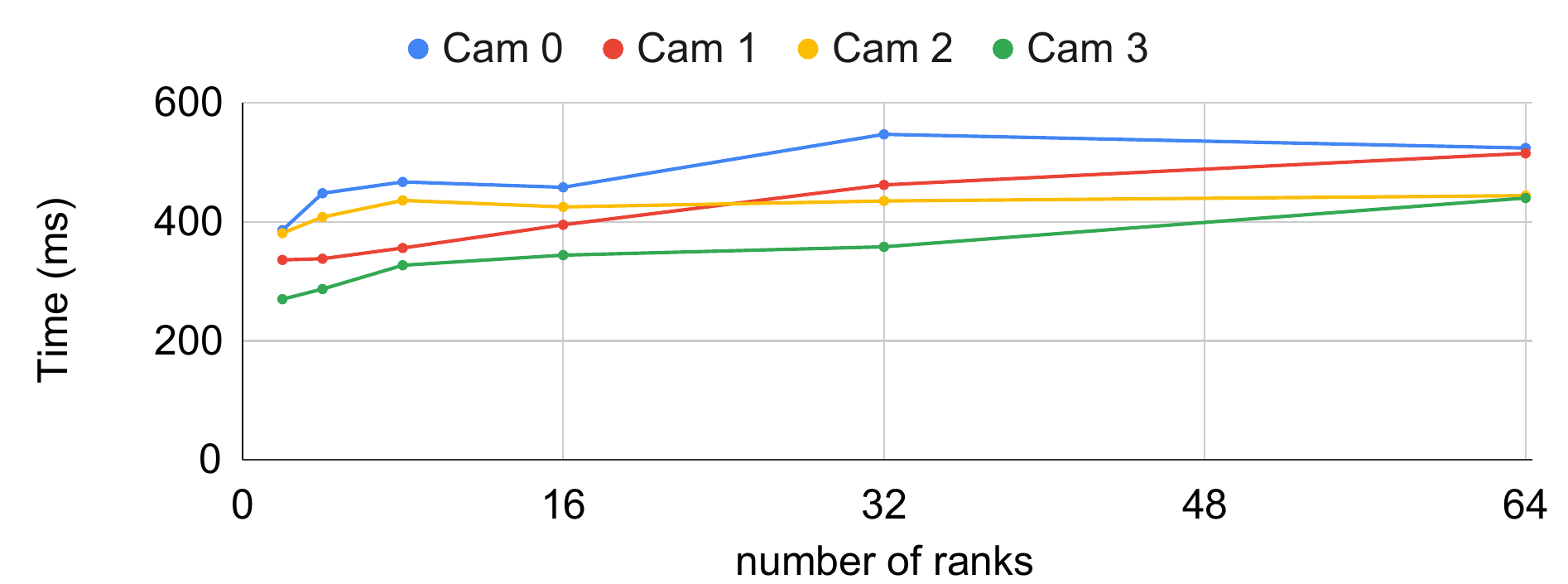}
    \caption{Different camera angles}
    \vspace{-.5em}
    \end{subfigure}
    \begin{subfigure}{0.45\textwidth}
         \centering
         \includegraphics[width=\textwidth]{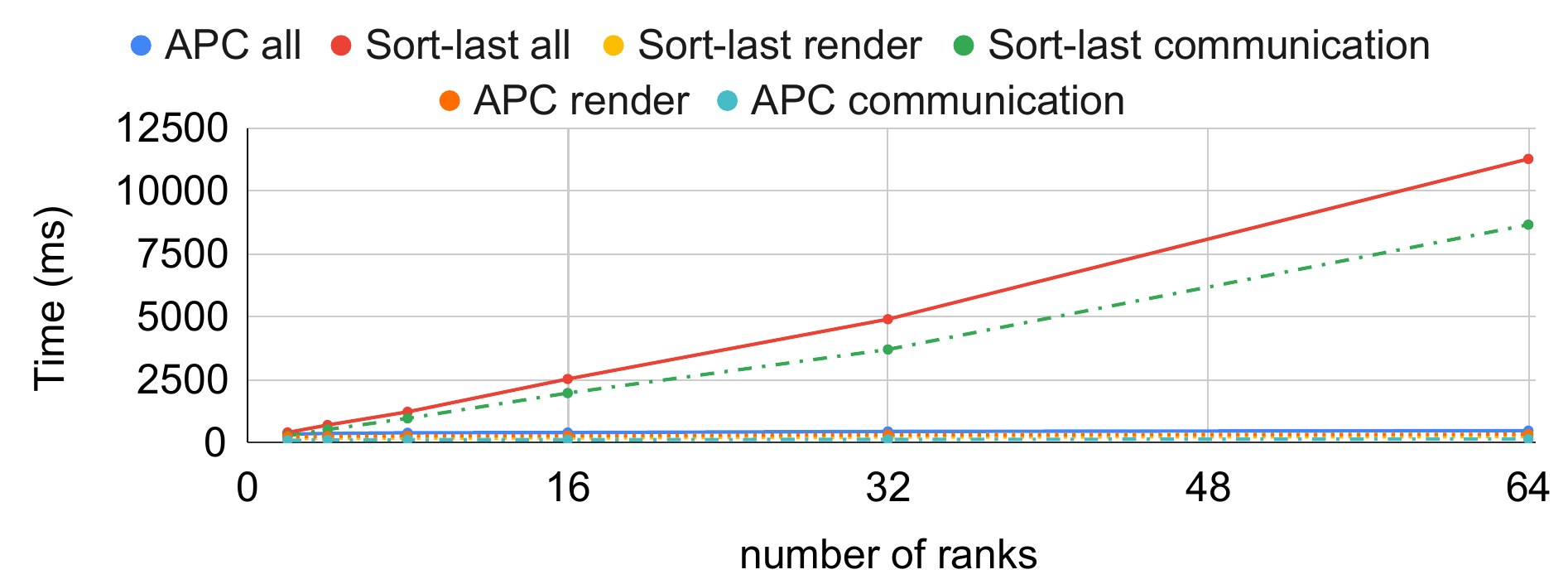}
         \vspace{-1.2em}
         \caption{\label{fig:syn_result_all}All stages}
         \vspace{-.5em}
     \end{subfigure}
    \begin{subfigure}{0.45\textwidth}
         \centering
         \includegraphics[width=\textwidth]{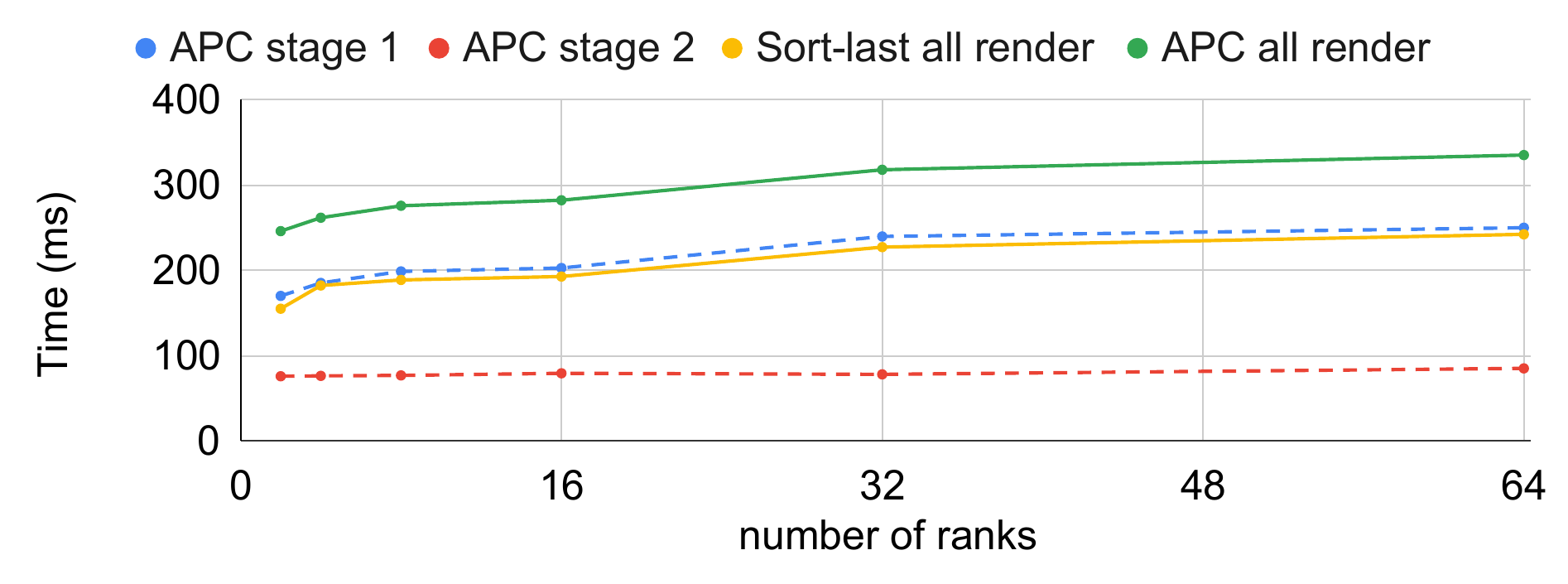}
         \vspace{-1.5em}
         \caption{\label{fig:syn_result_render}Rendering stages}
         \vspace{-.5em}
     \end{subfigure}
    \begin{subfigure}{0.45\textwidth}
         \centering
         \includegraphics[width=\textwidth]{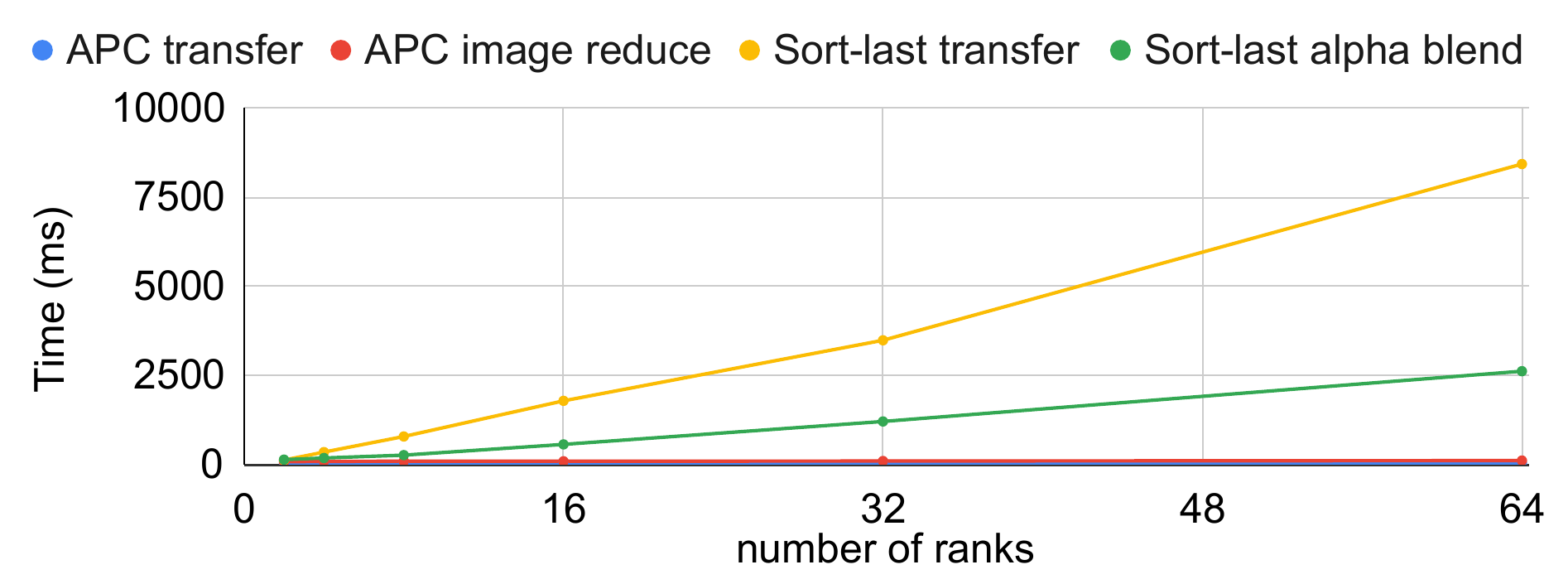}
         \vspace{-1.5em}
         \caption{\label{fig:syn_result_communication}%
         Compositing stages}
     \end{subfigure}
         \vspace{-1em}
    \caption{\label{fig:syn_result}%
    Synthetic example. (a) APC results remain consistent over different views. (b) The overall APC outperforms
    and scales better than sort-last.
    (c) Both the sort-last and APC scale well in local rendering, with APC's second pass doubling the local rendering runtime, as expected.
    (d) The compositing step (moments transfer and image sum for APC, and segment transfer and alpha-blend for sort-last) is the dominating cost in sort-last, whereas APC's compositing cost remains small and near constant. } 
    \vspace{-2em}
\end{figure}

\Cref{fig:syn_result} displays overall rendering performance results on the synthetic dataset benchmark on up to 64 ranks.
We find that APC closely follows the ideal weak-scaling trend of a flat trend line with both the rendering and the compositing, displaying a clear performance advantage over the sort-last method.
When breaking down performance to inspect the stages of our proposed Approximate Puzzlepiece Compositing algorithm (\Cref{fig:syn_result}), we notice that the data rendering stages are similar for both methods since data-parallel rendering is trivially parallel (\Cref{fig:syn_result_render}), allowing for good local scaling. With the same fixed step size sampling, the slight variation of the performance curves is caused by respective data structure overheads, which are more prominent on the chart in this experiment due to a smaller local rendering load.
We also break out compositing alone in~\Cref{fig:syn_result_communication}, which consists of data communication and final image blending. For APC this part includes the moments transfer and the final image reduction, and for sort-last, it includes segment layers transfer and alpha blend with sorting.
APC shows a near-constant scaling and outperforms the sort-last algorithm, which is quickly dominated by communication cost.

\begin{figure}
    \centering
     \begin{subfigure}{0.45\textwidth}
         \centering
         \includegraphics[width=\textwidth]{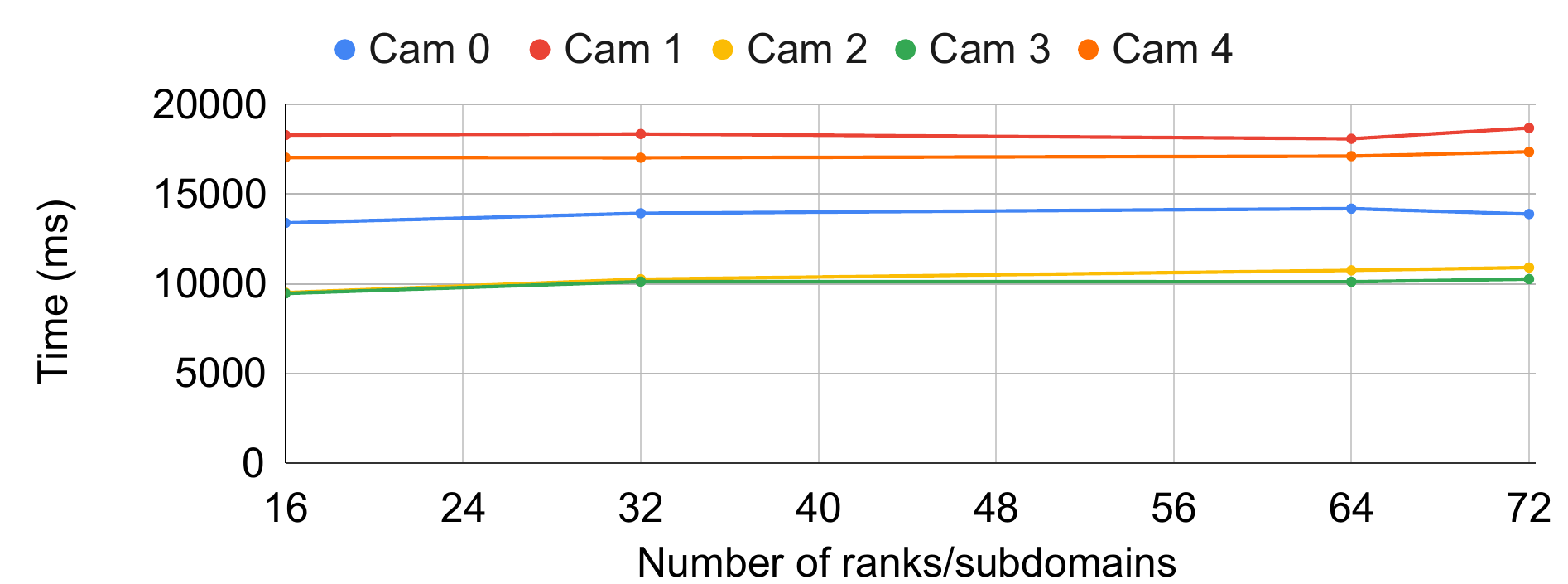}
         \caption{\label{fig:FUN3D_scaling_cam_dist}%
         Average time for each distance over all cameras.}
     \end{subfigure}
    \begin{subfigure}{0.45\textwidth}
         \centering
         \includegraphics[width=\textwidth]{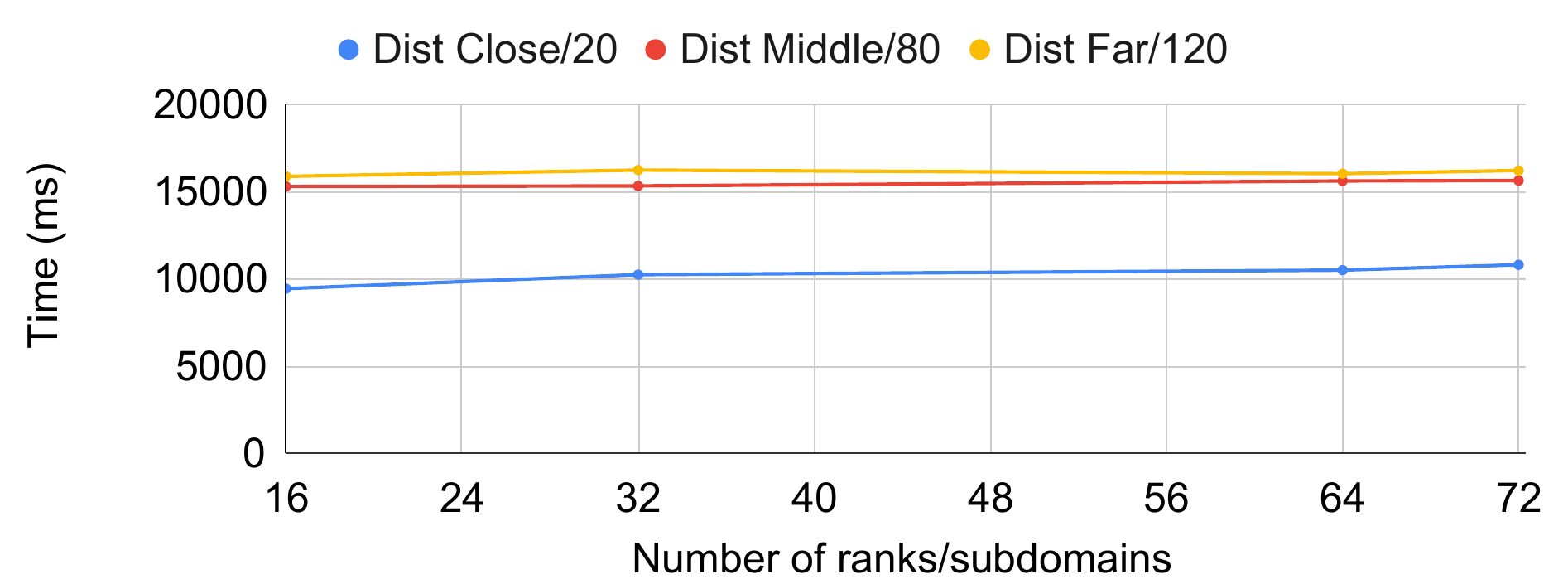}
         \caption{\label{fig:FUN3D_scaling_cam_orbit}%
         Average time for each camera over all distances.}
     \end{subfigure}
     \vspace{-1em}
     \caption{\label{fig:FUN3D_scaling_cam}FUN3D rendering performance on various camera settings. (a) The results from different camera angles. (b) Results from different camera distances. We see that our method performs consistently to different viewing configurations.} 
     \vspace{-0.5em}
\end{figure}

To match the in situ analysis process of the distributed dataset derived from the simulation, where each rank handles its own local data, the FUN3D experiments are conducted by loading one subvolume per rank and run up to 72 ranks. 
The benchmarks on FUN3D are run at three orbit radii, with five camera positions in each orbit.
This nonuniform distribution with jagged boundaries results in local ranks' data bounds overlapping, posing a challenge to
standard sort-last compositing methods.

\begin{figure}[ht!]
    \centering
    \vspace{-1em}
    \begin{subfigure}{0.45\textwidth}
         \centering
         \includegraphics[width=\textwidth]{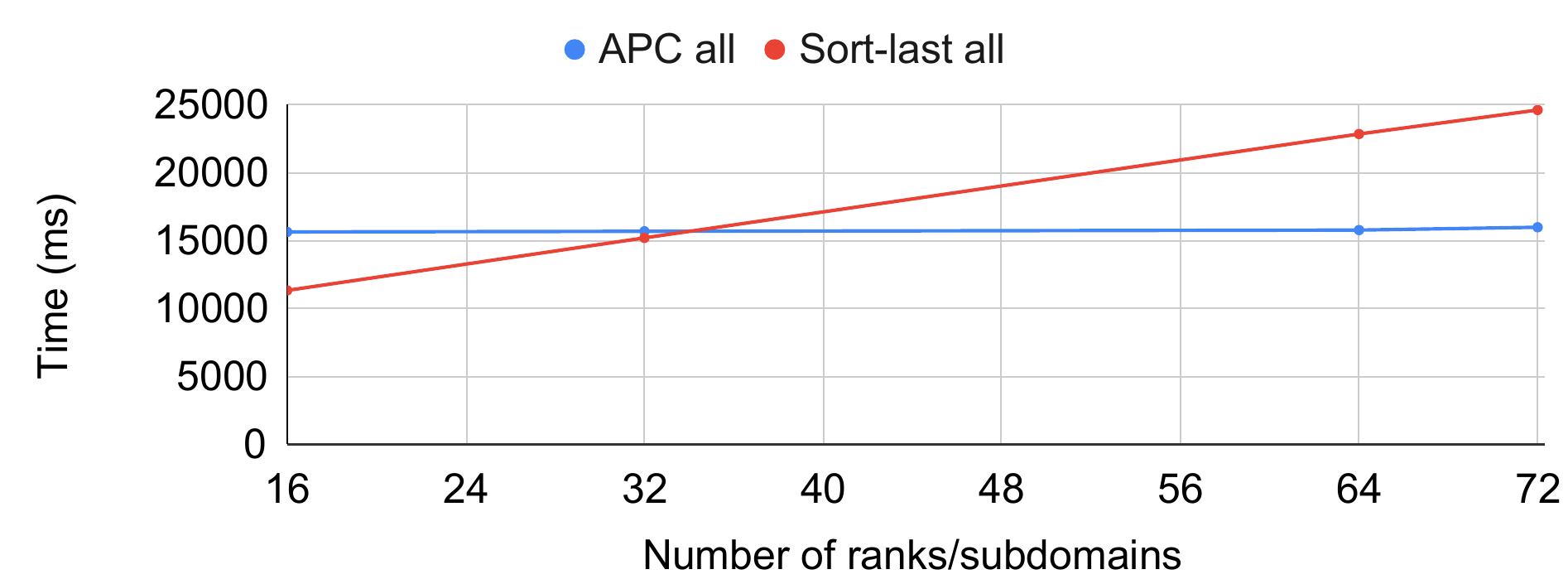}
        \vspace{-1em}
         \caption{All stages}
         \label{fig:FUN3D_all_scale}
     \end{subfigure}
     \begin{subfigure}{0.45\textwidth}
         \centering
         \includegraphics[width=\textwidth]{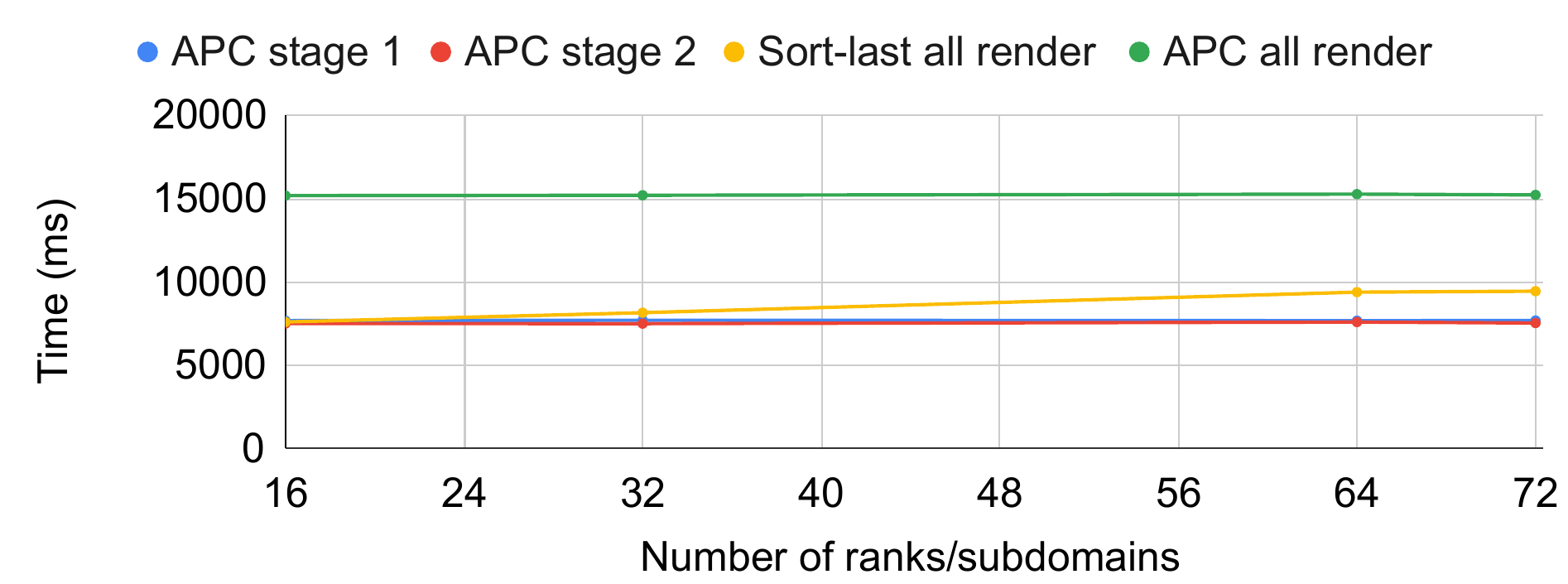}
        \vspace{-1em}
         \caption{Rendering stages}
     \end{subfigure}
     \begin{subfigure}{0.45\textwidth}
         \centering
         \includegraphics[width=\textwidth]{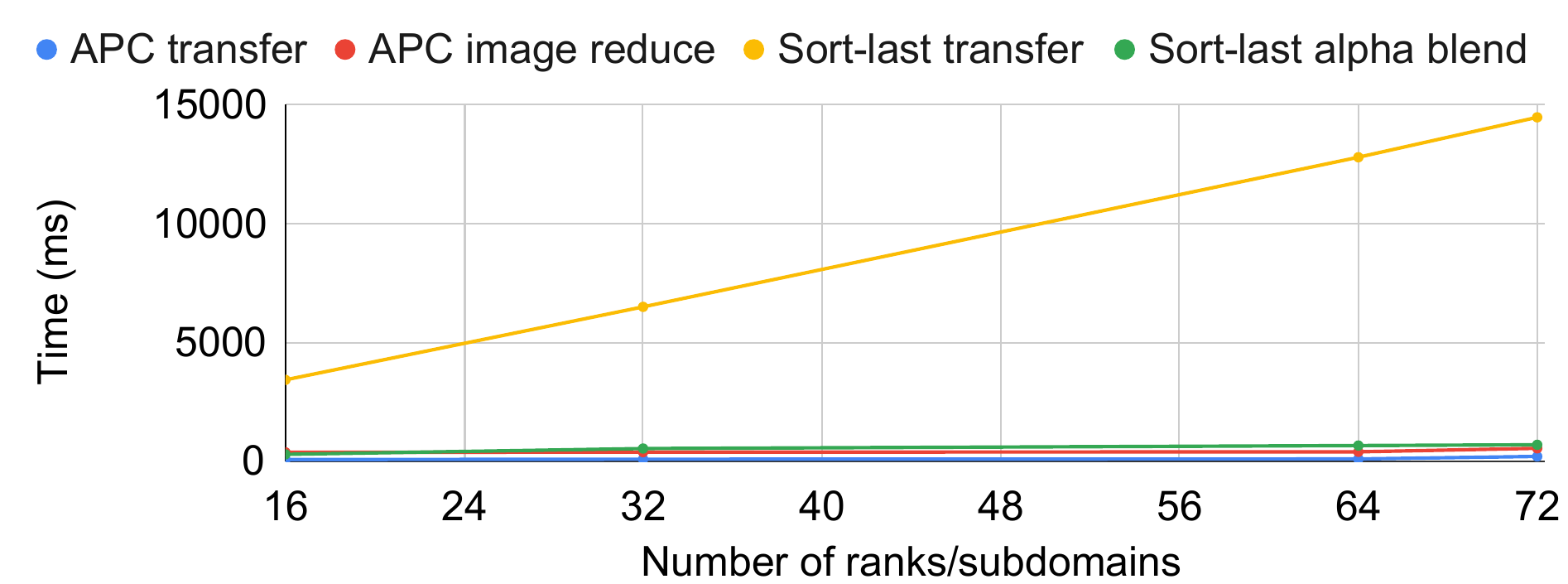}
        \vspace{-1em}
         \caption{\label{fig:FUN3D_scaling_communication}%
         Compositing stages}
     \end{subfigure}
    \vspace{-1em}
    \caption{FUN3D rendering stage performances. Again, for APC both the overall rendering performance and the compositing scale almost constantly. Even though the double rendering pass penalizes overall cost at lower core counts, sort-last becomes disadvantaged with its poor communication scalability. The crossover shown in (a) occurs at about half of the full domain.}
    \vspace{-1.5em}
    \label{fig:FUN3D_scaling_stages}
\end{figure}

\Cref{fig:FUN3D_scaling_cam_orbit} shows rendering performance on the five camera positions, and \Cref{fig:FUN3D_scaling_cam_dist}
shows performance over the different camera distances.
Again, we find that our method does not introduce significant performance overhead at higher core counts overall camera positions and distances.
In~\Cref{fig:FUN3D_scaling_stages} we break down rendering costs into the local rendering and compositing stages.
As in~\Cref{fig:FUN3D_scaling_cam}, 
both algorithms' compositing communication costs (\Cref{fig:FUN3D_scaling_communication}) exhibit similar patterns as in the synthetic case, with APC achieving relatively constant cost and outperforming the sort-last method as the full domain is loaded. The crossover of the two performance curves happens at around half of the domain, as seen in ~\Cref{fig:FUN3D_all_scale}.
Thus, we find that our sorting and redistribution-free approach is well suited to low overhead
rendering of large-scale distributed unstructured meshes.
\new{We also note jumps in rendering times as more mesh partitions are added  due to the uneven
rendering workload each partition incurs.}

\begin{figure}[t]
    \centering
    \begin{subfigure}{0.45\textwidth}
         \centering
         \includegraphics[width=\textwidth]{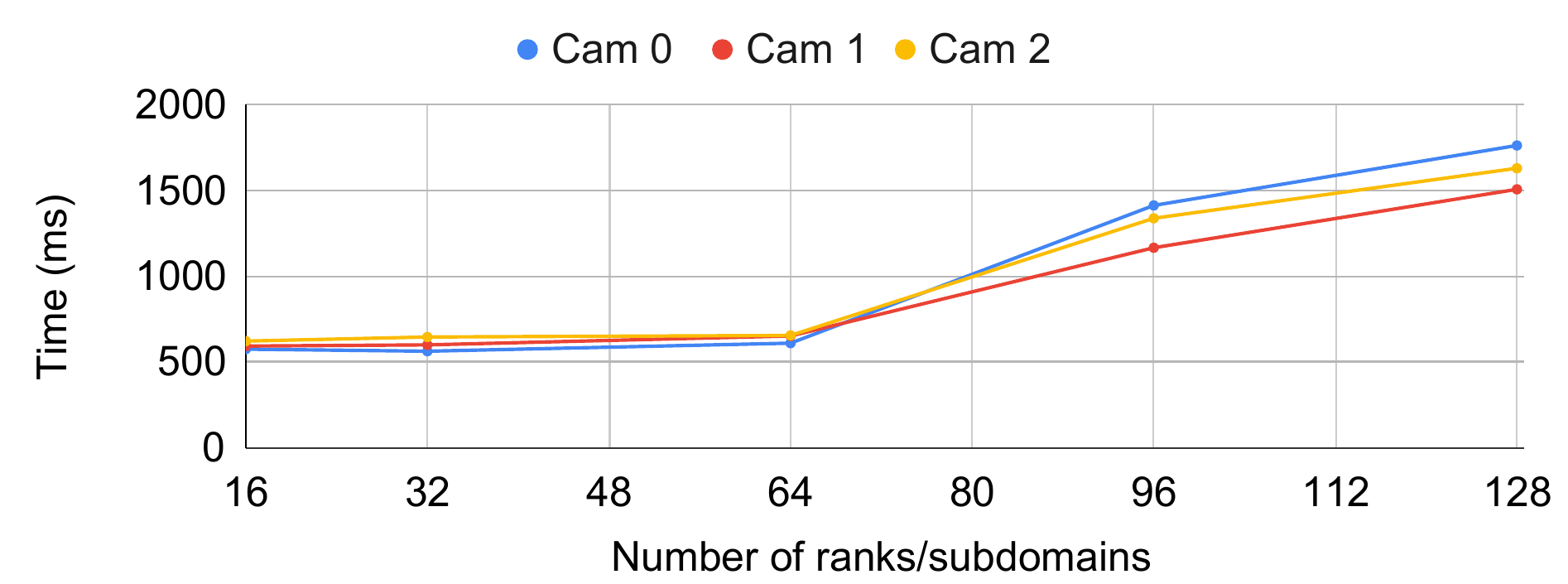}
\vspace{-1em}
         \caption{All cameras}
     \end{subfigure}
    \begin{subfigure}{0.45\textwidth}
         \centering
         \includegraphics[width=\textwidth]{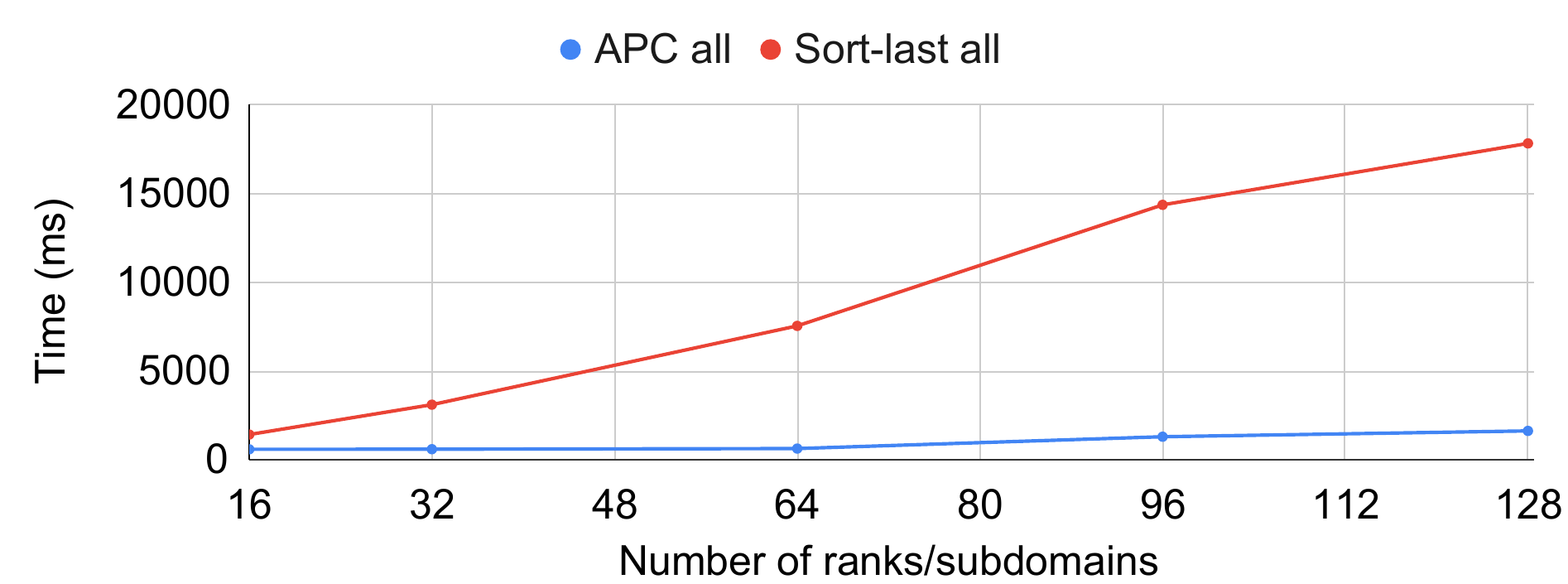}
\vspace{-1em}
         \caption{All stages}
     \end{subfigure}
     \begin{subfigure}{0.45\textwidth}
         \centering
         \includegraphics[width=\textwidth]{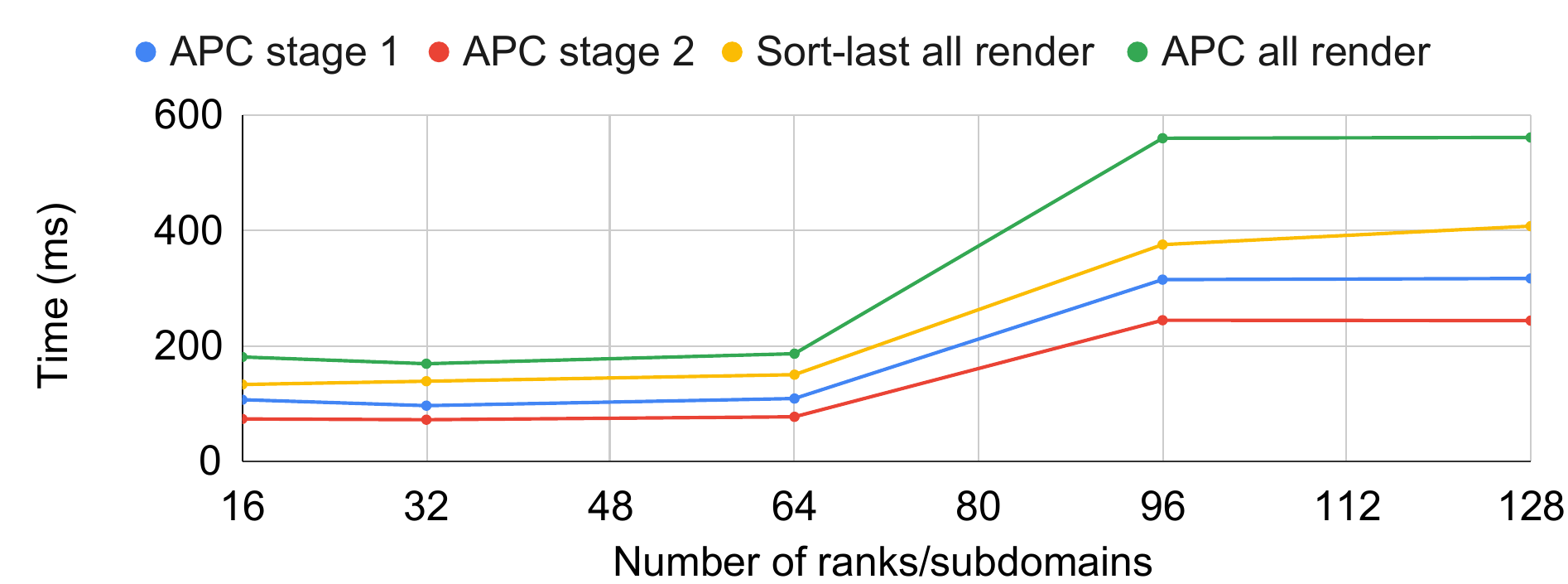}
\vspace{-1em}
         \caption{Rendering stages}
     \end{subfigure}
     \begin{subfigure}{0.45\textwidth}
         \centering
         \includegraphics[width=\textwidth]{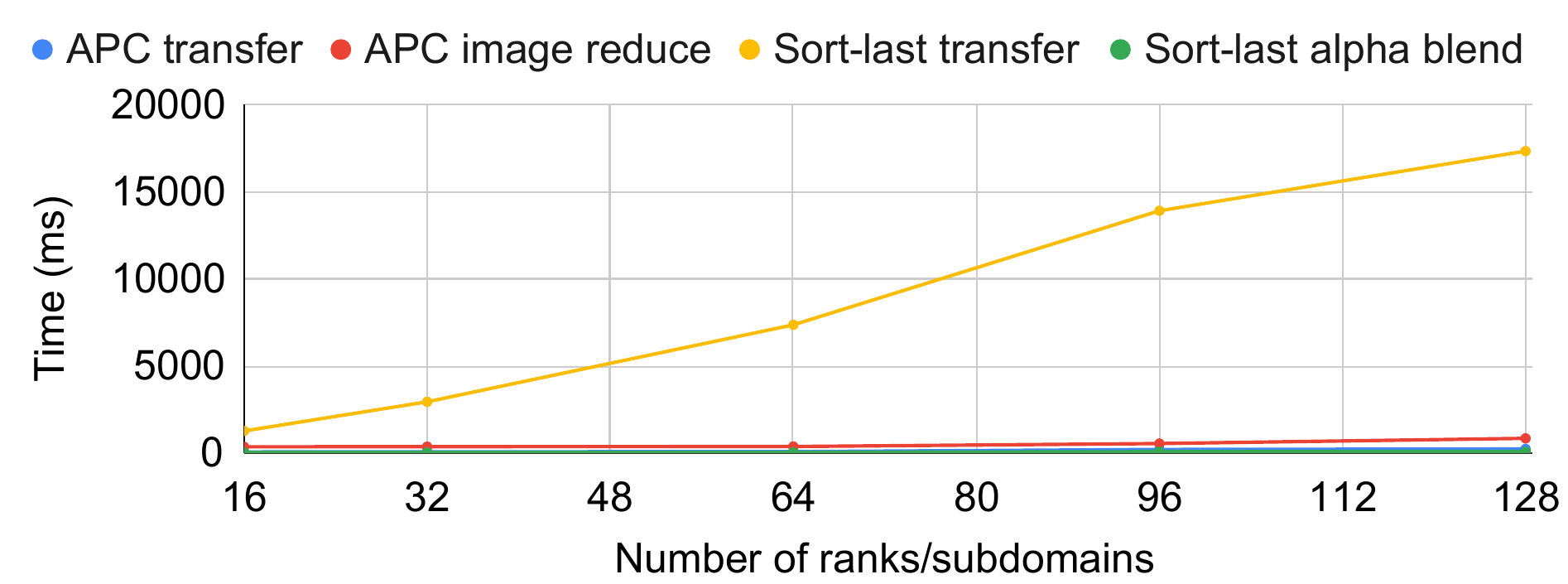}
\vspace{-1em}
        \caption{\label{fig:jet_scaling_communication}%
         Compositing stages}
     \end{subfigure}
\vspace{-1.em}
    \caption{\label{fig:jet_scaling_stages}
    Exajet rendering stage performances up to 128 core counts. (a) Varying cameras. (b) The overall weak-scaling of APC remains flat compared to sort-last. (c, d) show pipeline timing breakdowns. We see similar patterns in local rendering on both methods with divergence in communication scaling, leading to faster end-to-end performance for APC, especially at high core counts.}
\vspace{-2em}    
\end{figure}

The Exajet experiment is run with three camera positions around an orbit, and the results are shown in \Cref{fig:jet_scaling_stages}. Again, for the real-world data benchmark, we vary the number of loaded subdomains with each rank handling its own data. Despite being affected by the more unbalanced rendering loads, the rendering curves for both methods show similar behaviors as in the FUN3D case. As shown in \Cref{fig:jet_scaling_stages}'s break down, 
APC suffers from a second rendering stage but has a near-constant communication curve, whereas sort-last's communication costs grow rapidly with the number of ranks. Furthermore, due to the relatively smaller rendering load and more complex boundaries on each rank, communication dominates the overall performance starting from very low core counts. 

With three datasets of various data distribution scenarios, we have shown that APC is resistant to communication overheads at high core counts for unstructured meshes with unassuming boundary shapes. \new{In particular, when using 4 power moments, the memory requirement is reduced to 16 bytes per pixel to store transmittance information in the first pass and colors for additively blending to a final image. This is in contrast to sort-last segment compositing, which would require 16 bytes$\times \#\text{segments}$ per-pixel.
MBOIT was originally implemented for GPU rendering~\cite{munstermann2018moment}, and thus our method can be easily ported to a GPU use case requiring only 4-8 single-precision values per pixel depending on the moments variant used.}
By trading off a second local rendering stage, our method provides a more scalable solution to large-scale AMR meshes of unpredictable data distributions.

\subsection{Algorithmic Analysis}
\label{sec:eval_algorithmic_analysis}

Finally, we perform an algorithmic comparison against standard compositing algorithms
to evaluate how traditional sort-last compositing techniques may scale when adapted
to support per-segment sorting and compositing.
Traditional sort-last methods, e.g., Binary Swap~\cite{ma94}, rely on constructing
a global sort over the ranks' individual partial images.
When ranks' local data overlaps, it is no longer possible to construct this order.
To support such data distributions, one could consider extending sort-last compositing to support
multiple color segments per rank instead of a single partial image per rank.
Each rank would then produce a color segment for each continuous ray-volume
interval, and pass this set of segments to the compositing pipeline.
Essentially, each rank is treated as multiple virtual ranks per pixel, with
one virtual rank per segment.

Given $n$ ranks and $m$ total segments, APC's communication cost scales with $O(n)$; however,
Direct send compositing and binary swap would scale with $O(m)$.
A traditional sort-last rendering case would have $m = n$, i.e., each
rank produces a single segment per pixel for its local brick of data.
However, for large-scale unstructured datasets
there would be many segments produced on each rank due to the jagged boundaries,
and we would expect $m >> n$. Thus, the linear scaling with $n$ of APC would
lead to better overall performance in practice.

To evaluate the compositing in a real-world scenario, we compute the number of segments per pixel
for the middle distance view of the FUN3D on 64 ranks (\Cref{table:comp_compare}).
This configuration has a maximum of 32 segments per pixel.
We find that, overall, a large number of pixels produce a single segment for this configuration,
resulting in good data transfer costs for direct send and binary swap.
APC's communication costs, on the other hand, are fixed, using just an all-reduce and a reduce.

Besides sorting the entire list, direct send does not consider load-balancing in image space whereas our method ensures an even workload for all pixels with a single image-add operation, and the constant-size reduce/all-reduce communication is well optimized by MPI libraries.
Binary swap provides better scaling than direct-send but requires a large number of pair-wise image swaps and leads to underutilization at the higher levels of the swap tree.
Whereas sort-last methods are unbounded in potential data transfer costs to support such overlapping distributions, our method is bounded by a constant message size, i.e., the number of ranks, regardless of scene complexity.
\new{The main scaling limitation of our approach is the performance of MPI Allreduce and Reduce, which are highly optimized
operations in MPI libraries.}

\begin{figure}[t]
     \centering
     \begin{subfigure}{0.2\textwidth}
         \centering
         \includegraphics[width=\textwidth]{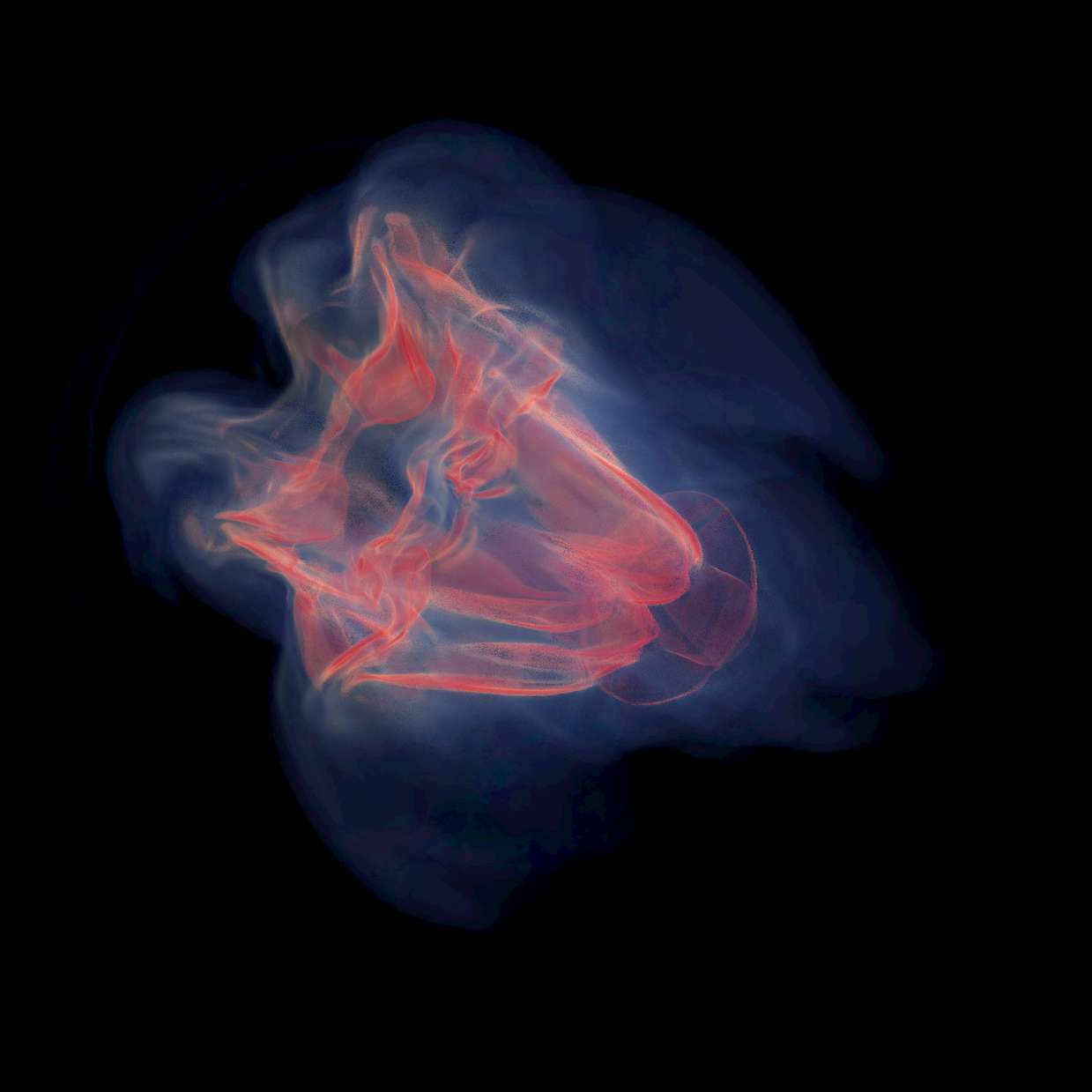}
         \caption{Overview}
     \end{subfigure}
     \begin{subfigure}{0.2\textwidth}
         \centering
         \includegraphics[width=\textwidth]{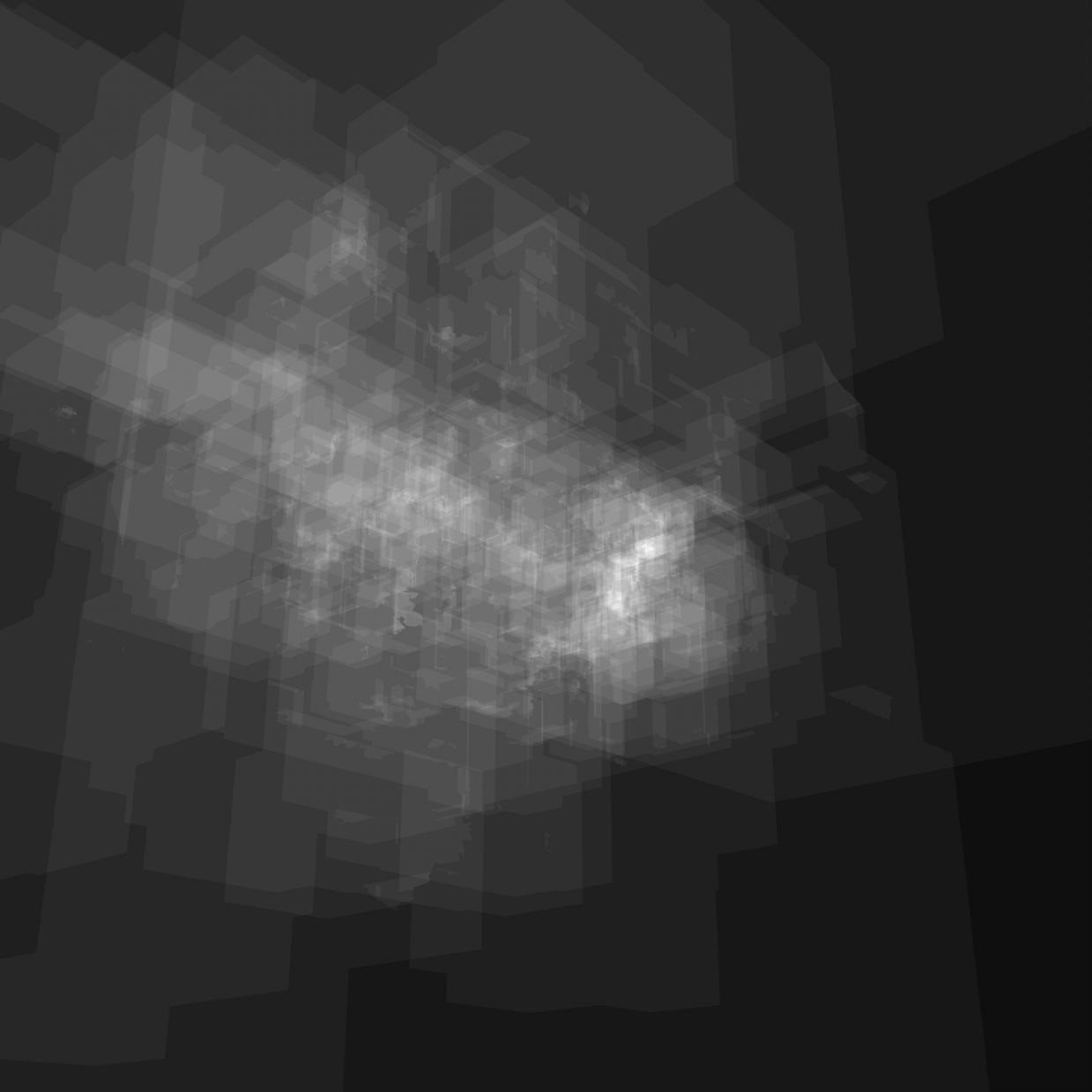}
         \caption{Heatmap}
     \end{subfigure}
     \begin{subfigure}{0.2\textwidth}
         \centering
         \includegraphics[width=\textwidth]{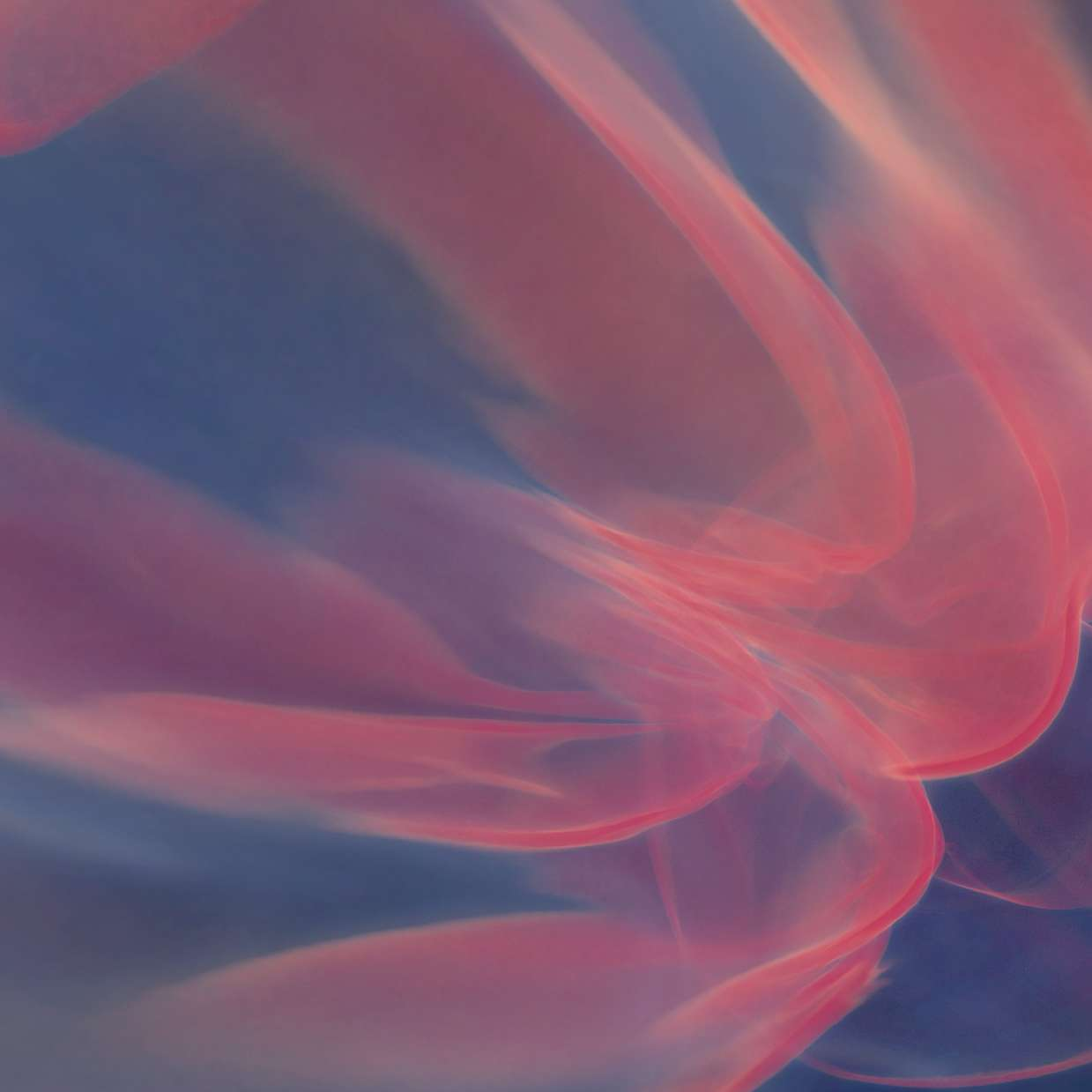}
         \caption{Closeup}
     \end{subfigure}
     \begin{subfigure}{0.2\textwidth}
         \centering
         \includegraphics[width=\textwidth]{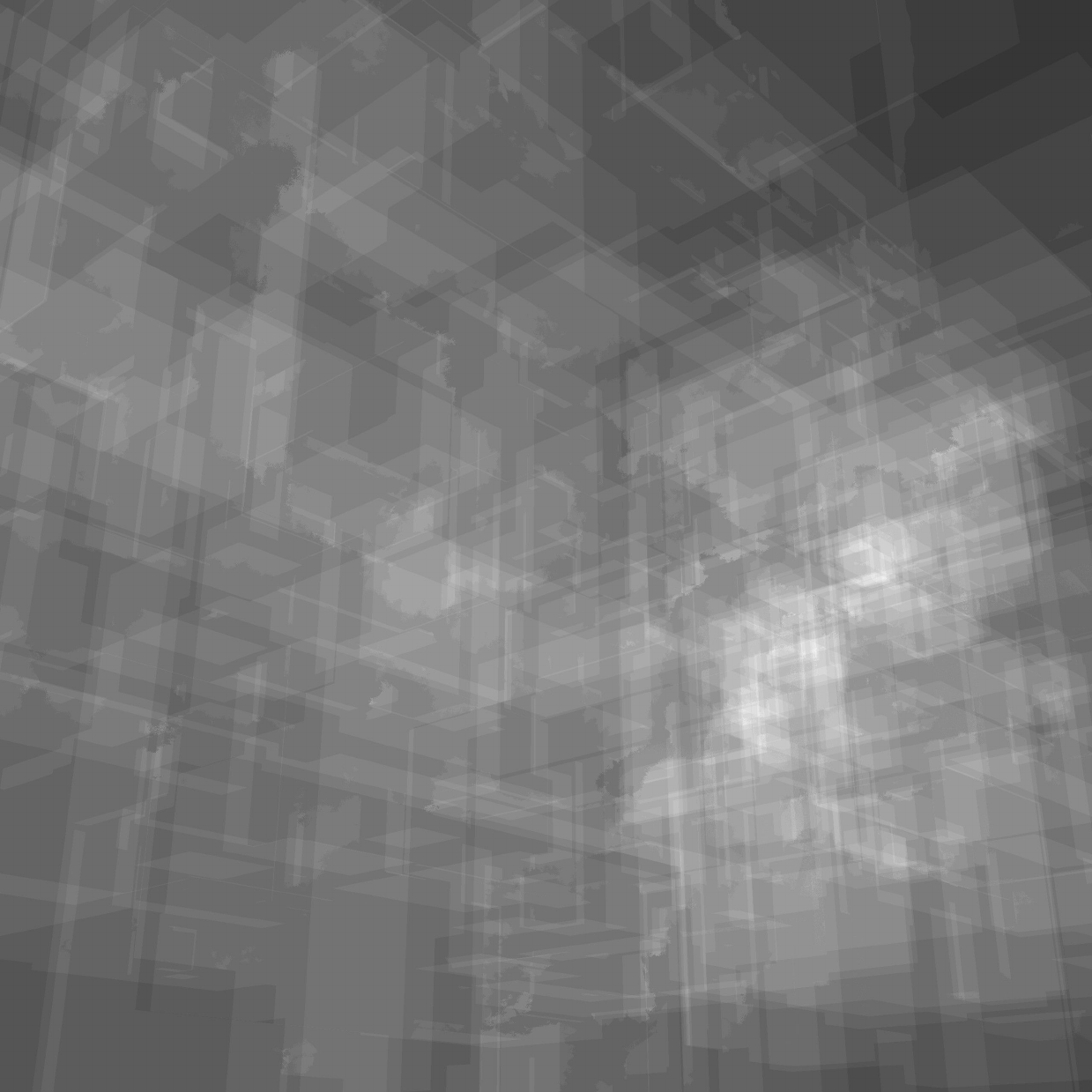}
         \caption{Heatmap}
     \end{subfigure}
     \vspace{-1.5em}
     \caption{Segment heatmap with max number of segments = 28 and 32, grayscale image scaled for visibility. Segment density, within or across ranks, is heavy in certain regions, requiring large amounts of data to be transferred in sort-last compositing.}
     \vspace{-1.em}
     \label{fig:heatmap}
\end{figure}

\begin{table}[t]
\centering
\begin{tabular}{@{}lrr@{}} 
\toprule
    Method & Average & Upper Bound \\ [0.5ex] 
    \midrule
    Ours & 11.7531 & 128 \\
    Direct Send & 6.9848 & $\infty$\\ 
    Binary Swap & 18.4772 & $\infty$\\
    \bottomrule
    \end{tabular}
    \vspace{-0.5em}
    \caption{\label{table:comp_compare}Average number of segments transferred per nonempty pixel on FUN3D overview with n=64, Figure \ref{fig:heatmap} (a). The number of ranks (n=64) is chosen as a power of two for fair comparison with binary swap. Our method guarantees an upper bound of twice the number of ranks, which is 128 in this case. }
    \vspace{-2em}
\end{table}

\section{Conclusion and Limitations}
We have presented a technique for compositing large-scale unstructured mesh data for in situ rendering.
By adopting an order-independent blending technique, our APC pipeline eliminates the need for sorting or ordering partial images across ranks.
Our performance evaluations across synthetic and real-world datasets demonstrate APC's scalability and effectiveness under diverse data boundary conditions.

The compositing performance is discussed in further detail with comparison to the traditional sort-last compositing techniques, and APC shows superior scalability, leading to potentially better overall performance, especially at higher core counts. Even though APC introduced a second local rendering overhead, the message communication steps can both be achieved in small constant sizes through single MPI calls that are optimized by the library.  We also examine the output images by comparing to both single-node MBOIT and single-node sort-last rendering, validating that APC delivers high-precision results.
Our technique enables efficient parallel distributed visualization with a high-quality transparency approximation
for rendering complex data distributions that are not suited to traditional sort-last compositing techniques.

The main limitations of our proposed method come from using order-independent transparency techniques
to eliminate the need for sorting, at the cost of per-sample transmittance accuracy.
As MBOIT is an estimation in the end, APC inherently produces images that are slightly different from those of sort-last alpha blending.
We note that this would be the case for any order-independent transparency method, as all form an approximation of the transparency term in some form. \edit{Eventually, the sort-last technique itself is also one way to approximate real-world light behavior with the advantage of strict fragment ordering. Thus, we believe that an exact color match to the traditional rendering results is not the ultimate goal. The sort-last images are used more as a reference to ground truth in terms of depth perception.} The approximated result effectively preserves object ordering, with the cost of being not entirely energy-conserving and thus may need additional bias adjustments.
The reconstruction method also requires that there is no volume overlapping, as the opacity is not well defined in this case.  
Although the requirement that the data be rendered twice for MBOIT incurs an additional cost, we note that
this workload is entirely local to each rank and thus achieves good scaling by itself.

\acknowledgments{
This work was funded in part by NSF OAC award 2138811, NSF 
CI CoE Award 2127548, NSF OISE award 2330582, the Advanced Research Projects Agency for Health (ARPA-H) grant no. D24AC00338-00, the Intel oneAPI Centers of Excellence at University of Utah, the NASA AMES cooperative agreements 80NSSC23M0013 and NASA JPL Subcontract No. 1685389.
Results presented in this paper were obtained in part using the Chameleon, Cloudlab, CloudBank, Fabric, and ACCESS testbeds supported by the National Science Foundation. This work was performed in part under the auspices of the DoE by LLNL under contract DE-AC52-07NA27344, (LDRD project SI-20-001).

}




\end{document}